\documentclass[12pt]{article}
\usepackage{amsmath, amssymb, epsf, subfigure, rotating}

\usepackage{biophysj}

\usepackage[dvips]{epsfig,graphicx, color}

\usepackage{times}

\newcommand{\be}[1]{\begin{eqnarray}\label{#1}}
\newcommand{\ee}{\end{eqnarray}}

\newcounter{lastnote}

\newcommand{\gl}[1]{Eq.\,\ref{#1}}
\newcommand{\ab}[1]{Figure \ref{#1}}
\newcommand{\ta}[1]{Tab.\,\ref{#1}}
\newcommand{\tas}[2]{Tab.\,\ref{#1} and \ref{#2}}
\newcommand{\tato}[2]{Tab.\,\ref{#1} -- \ref{#2}}
\newcommand{\se}[1]{Section \ref{#1}}
\newcommand{\gln}[2]{Eqs.\,\ref{#1} and \ref{#2}}
\newcommand{\abn}[2]{Figures \ref{#1} and \ref{#2}}

\newcommand{\bc}{$\beta$-cell}
\newcommand{\bcs}{$\beta$-cells}

\begin{document}

\parskip1ex
\parindent0mm

\begin{center}
{\bf\LARGE The electrophysiology of the \bc~based on
single transmembrane protein characteristics}\\[5mm]
{\bf\Large Michael E. Meyer-Hermann}\\[2mm]
Frankfurt Institute for Advanced Studies (FIAS),\\
Max von Laue Str.~1, 60438 Frankfurt/Main, Germany\\
Email: M.Meyer-Hermann@fias.uni-frankfurt.de\\[0.5ex]
phone: +49 69 798 47508; fax +49 69 798 47611
\end{center}

\date{\today}




\noindent
{\bf Keywords:}\\ 
electrical bursting, insulin secretion, calcium dynamics,\\ 
ion-channels, ATPases, mathematical modelling


\noindent
{\bf\large Abstract:}\\
The electrophysiology of \bcs~is at the origin of
insulin secretion. \bcs~exhibit a complex behaviour
upon stimulation with glucose including repeated
and uninterrupted bursting. Mathematical modelling
is most suitable to improve knowledge about
the function of various transmembrane currents
provided the model is based on reliable data. This is
the first attempt to build a mathematical model
for the \bc-electrophysiology in a bottom-up
approach which relies on single protein conductivity
data. The results of previous whole-cell-based
models are reconsidered. The full simulation including
all prominent transmembrane proteins in \bcs~is
used to provide a functional interpretation of
their role in \bc-bursting and an updated
vantage point of \bc-electrophysiology. 
As a result of a number
of {\it in silico} knock-out- and block-experiments
the novel model makes some unexpected predictions: 
Single-channel conductivity data imply that
calcium-gated potassium currents are rather small.
Thus, their role in burst interruption 
has to be revisited. An alternative role
in high calcium level oscillations is proposed
and an alternative burst interruption model
is presented. 
It also turns out that sodium currents
are more relevant than expected so far.
Experiments are proposed to verify these predictions.

\newpage
\section{Introduction}

The electrophysiology of \bcs~is most relevant
for understanding possible regulatory targets of insulin
secretion. Exocytosis of insulin loaded granules is, among others,
governed by intracellular calcium signalling. Calcium
dynamics exhibits very specific patterns in reaction
to increased glucose concentrations which includes
repeated bursting of the membrane potential. Bursting
denotes oscillations of action potential-like depolarisations,
which are regularly interrupted and set in again after
a phase of silence. Thus, oscillations occur on two
distinct time scales: Fast oscillation of the membrane
potential on the scale of $100$ milliseconds and slow oscillations
of intracellular calcium on the time scale of $10$ seconds,
i.e.~in the rhythm of the repetition
of bursting events.

In a minimal model of oscillations of the membrane potential
\cite{chay83} it was found that an activation-delay between voltage-gated
potassium and calcium channels gives rise to this bursting
behaviour. 
Sodium, the major player in
neurons, is generally believed to be of minor importance
\cite{fridlyand03}.
Therefore, modelling work was concentrating on potassium
and calcium currents \cite{sherman88,chay97,bertram04}.

Initiation of bursting events was found to be related
to ATP-driven potassium channels \cite{keizer89,kanno02}
which induce a relevant outflow of potassium
under resting conditions. When glucose -- and via
glucose metabolism -- ATP is increased, the channel
conductivity is inhibited \cite{ashcroft84,cook84}, 
which leads to the initial depolarisation. 
Then voltage-dependent
calcium currents are activated and the delayed
response of voltage-dependent potassium currents
leads to the bursting event. 

Simulations of the \bc-electrophysiology have shown
how, in principle, repeated bursting can happen
\cite{chay83,sherman88,bertram04,fridlyand03}. And
the found mechanisms are in agreement with experimental
results. However, none of the presently existing
models fulfill the following requirements:
(i) All important ions and membrane proteins are
explicitly modeled including their activation
and inactivation dynamics; 
(ii) the \bc~exhibits a steady state and is
stimulated by changes of glucose concentrations;
(iii) the dynamics of the membrane protein activity are
fully derived from protein experiment.
Only if all these requirements are respected, 
the exact role of the different
membrane proteins might be disentangled.

A major problem in developing such a complete
simulation tool is related to the large variety of measured 
whole-cell conductivities in \bcs. In fact the density of the membrane
proteins is itself a dynamical quantity. Their dynamics
vary for different cell-types so that the
data used for modelling are restricted to 
measurements in \bcs. Not all membrane proteins have
been subject to quantitative measurements in \bcs~and
the \bcs~under consideration can be in various
states of protein expression.
Thus, modelling work was restricted to fit the activity-dynamics
of the membrane proteins to the behaviour which was
expected.

In order to overcome this problem, the present simulation
exclusively relies on data of single protein activity.
Such an approach was used in calcium modelling of neurons \cite{erler04}
before and is now applied for the first time to \bc~electrophysiology.
Today there is a rather
complete knowledge about single proteins conductivity and opening
dynamics available which
allows a bottom-up approach starting from
the single protein level. 

To this end the single protein activity dynamics is separated from
the protein density. It is assumed that the single protein dynamics
is cell-independent. In other words a voltage-gated potassium
channel has the same dynamical properties in a neuron and in a
\bc. This assumption might be set in question if cooperation
of membrane proteins with other cell structures would change
the activity dynamics. However, the universality of single
membrane protein properties can be considered to be a good
approximation because the measured characteristics of the cells
show relatively small variations despite them
stemming from different cell-types.

The conductivity and opening properties
of single proteins are exactly implemented into the model and then
are multiplied by the surface density of the respective proteins in
\bcs. The latter are not precisely known and are the free
parameters of the simulation. In comparison to the huge parameter
space in simulations relying on whole-cell conductivity measurements,
this is a rather small set of unknown parameters. This improves
the predictive power of the mathematical model.
The behaviour of the \bc~in terms of dynamics of 
membrane potential and ion-concentrations 
is emerging from the single protein
level, thus, coupling the molecular to the cellular level.

The framework of the simulation is presented in the methods section.
The single membrane protein properties are collected in the 
supplementary material.
These define the current dynamics as used in the simulation. 
The full model as introduced in \se{methods}
contains the model of a most prominent original work \cite{sherman88}.
The bursting behaviour as found in Sherman et al's work
is reproduced as a special case of the full model
and is used as test for the simulation.
Then the full simulation is employed to set up a steady state of the
\bc, and to allow for stimulation of the \bc~with glucose.
The protein densities are determined to find a realistic behaviour
of the cell -- where the single membrane protein properties are
not touched. On the basis of this simulation various 
{\it in silico} experiments are performed and compared to
real experiments if available. 

The {\it in silico} experiments
include knock-out experiments and partial or total block of membrane
proteins. Calcium-driven potassium channels are modeled according to
data that are used for \bc-simulations for the first time.
The measured dynamics leads to the conclusion that this channel
is of minor importance at normal glucose levels, while it is most
important for the \bc-behaviour at supra-large glucose levels.
In this regime calcium-driven potassium channels drive
uninterrupted bursting \cite{kanno02}.
LVA-channels, which have not been considered in other models,
NCX, and PMCA are suggested to be important 
for the interruption of bursting events at normal glucose levels.
Also the importance of a dynamic reversal potential, which was
mostly neglected in other work, is pointed out.
It is found that the role of sodium currents was
underestimated so far, and their impact on \bc~behaviour is 
explained. The paper is concluded with a novel vantage point on the
interplay of the different ion-conducting membrane proteins
during repeated and uninterrupted bursting. Novel experiments
are suggested to verify the new vantage point.


\section{Methods}
\label{methods}

\subsection{Dynamics of ion concentrations and membrane potential}
\label{model}

Rate equations are used to describe the dynamics of
intracellular ion concentrations. External ion concentrations
are assumed constant. The
ions treated explicitly are sodium $N$, potassium $K$ and calcium $C$.
The dynamics depends on the conductivity of various membrane proteins
which each correspond to one term in the following
differential equations.
\be{ion-dynamics}
\frac{dN}{dt}
&=&
-\frac{\xi}{F}
\left(
\rho_{\rm Na,V} I_{\rm Na,V}
+ 2 \rho_{\rm Na,K} I_{\rm Na,K} \alpha_{\rm Na,K}
+ \rho_{\rm NCX} I_{\rm NCX} \alpha_{\rm NCX}
+ J_{\rm Na}
\right)
\nonumber\\
\frac{dK}{dt}
&=&
-\frac{\xi}{F}
\left(
\rho_{\rm K,ATP} I_{\rm K,ATP}
+ \rho_{\rm K,V} I_{\rm K,V}
+ \rho_{\rm K,Ca} I_{\rm K,Ca}
-\frac{4}{3} \alpha_{\rm Na,K} \rho_{\rm Na,K} I_{\rm Na,K} 
+ J_{\rm K}
\right)
\nonumber\\
\frac{dC}{dt}
&=&
-\frac{\xi}{z_{\rm Ca} F \left(1+x_c\right)}
\left(
\rho_{\rm Ca,L} I_{\rm Ca,L}
+ \rho_{\rm Ca,T} I_{\rm Ca,T}
- \frac{1}{3} z_{\rm Ca} \alpha_{\rm NCX} \rho_{\rm NCX} I_{\rm NCX} 
\right.
\nonumber\\
&& \left.
\qquad\qquad\qquad\quad
+ \rho_{\rm PMCA} I_{\rm PMCA}
+ J_{\rm Ca}
\right)
\quad.
\ee
$I_x$ denote the electrical 
single membrane protein currents, where $x$ specifies
the type of membrane protein as defined in \ab{scheme}
(see also \ta{proteins} in the supplement). 
Negative currents are defined to bring positive ions (charges) into the cell.
As the currents are not defined as ion currents but as electrical
currents they have to be weighted by factors denoting the valence
of the ion $z_{\rm Ca}=2$ and the stoichiometry of the protein
$\alpha_x$.
The single membrane protein currents and their dynamics
are discussed in the supplementary material 
\se{opening} and are schematically 
visualised in \ab{scheme}.
\begin{figure}[ht!]
\begin{center}
\includegraphics[height=6.8cm]{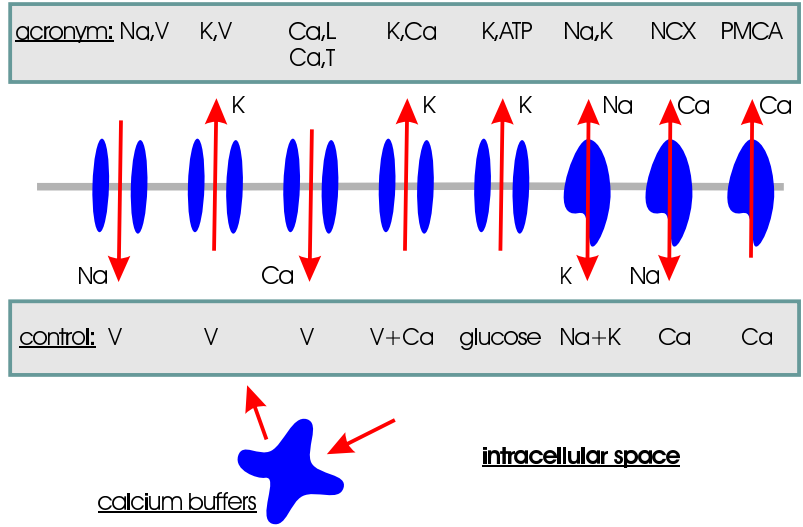}
\caption{\sf {\bf Schematic representation of ion-conducting transmembrane proteins:}
The plasma membrane is visualised as horizontal grey line on which
a set of transmembrane proteins are shown.
The shaded boxes show the {\it acronyms} of the proteins and the
quantities that {\it control} their gating/activity in the model introduced
below. The ions attributed to the arrows denote the direction
of ion flow (assuming normal electrochemical gradients). The {\it calcium
buffers} symbolise all kinds of intracellular calcium binding sites,
including calmodulin and organelles acting as calcium stores.
The acronyms of the membrane proteins are also defined in \ta{proteins}
in the supplement, where also the activity dynamics are explained in detail.}
\label{scheme}
\end{center}
\end{figure}

$\rho_x$ denote the corresponding
surface densities of the membrane proteins. 
$\xi=3$ is a surface to volume factor that relates the current
through the membrane to a change of the ion concentration in
the intracellular volume. The cell geometry is assumed to be
approximately spherical which implies $\xi=3/r$ where
$r$ is the cell radius.

Note that a calcium buffer is included in the equation for
calcium. The calcium buffer is assumed to bind and dissolve
calcium faster than the typical time scale of the ion
dynamics. Thus, the kinetic
equation for the buffer can be solved in the steady
state approximation (rapid buffer approximation).
This implies an additional
factor $(1+x_c)$ in the equation for calcium, where
\be{buffer-ss}
x_c&=&\frac{c_0 K_c}{\left(C+K_c\right)^2}
\quad,
\ee
with $c_0$ the total concentrations of the
buffering calcium binding sites. The most prominent
buffer is calmodulin having more than one calcium
binding site. $K_c$ is the 
dissociation constant of the calcium binding sites.
$C$ is the free calcium concentration.
The total calcium concentration is given by
\be{total-ca}
C_{\rm total}
&=&
C
\left(1+\frac{c_0}{C+K_c}
\right)
\quad.
\ee
The fraction of free calcium is
$f_{\rm Ca}=\frac{C}{C_{\rm total}}$
and adopts 
values between $0.1$ and $1\%$.

$J_{\rm K,Na,Ca}$ denote the leakage currents. They are derived
from the steady state limit (see \gl{leakage_ss} in the supplement)
to guarantee a resting state of the cell. In the resting state
all other currents are balanced by the leakage currents, so that
the net current vanishes and the ion concentrations are constant.

The ions build up the membrane potential $V$.
All non-electroneutral currents introduced in
\gl{ion-dynamics} show up in the equation
for $V$:
\be{voltage}
\frac{dV}{dt}
&=&
- \frac{1}{C_m}
\left(
\rho_{\rm Na,K} I_{\rm Na,K}
+\rho_{\rm K,ATP} I_{\rm K,ATP}
+\rho_{\rm K,V} I_{\rm K,V}
+\rho_{\rm K,Ca} I_{\rm K,Ca}
+\rho_{\rm Na,V} I_{\rm Na,V}
\right.
\nonumber\\
&&
\qquad
+\rho_{\rm NCX} I_{\rm NCX}
+\rho_{\rm PMCA} I_{\rm PMCA}
+\rho_{\rm Ca,L} I_{\rm Ca,L}
+\rho_{\rm Ca,T} I_{\rm Ca,T}
\nonumber\\
&&
\left.
\qquad
+ J_{\rm K} + J_{\rm Na} + J_{\rm Ca}
\right)
\quad.
\ee
No additional leakage current is needed because in the steady
state the right hand side vanishes by construction.
$C_m$ is the membrane capacitance that
relates the electrical currents to changes of the membrane
potential.

\subsection{Reversal potentials}
\label{reversalpot}

Reversal potentials depend on the chemical gradient and
the membrane potential over the cellular membrane. They cannot be
considered to be constant (as assumed in most other models). 
As the reversal potentials linearly
enter Ohm's law 
(which is used to approximate ion current through open pores) this
has to be accounted for. The Nernst-equation is used to calculate
the correct reversal potential during dynamical changes of the
electrochemical constellation:
\be{nernst}
\overline{V_x}
&=&
\frac{RT}{z_x F}\,
\ln\left(\frac{x_{\rm ext}}{x}\right)
\quad,
\ee
where $x={\rm K,Na,Ca}$, and 
$R=8.315 J/(K mol)$ the Rydberg (molar) gas constant.
The temperature $T$ is assumed to be the physiological temperature. 
$x_{\rm ext}$ is the external concentration
of the ion (in thermic bath approximation).
They are calculated from the intracellular ion-concentrations at
rest state to fit the known reversal potential
$V_{\rm K}=-75mV$, 
$V_{\rm Na}=+80mV$, 
$V_{\rm Ca}=+128mV$.

The straight forward application of Nernst equation, which is based
on thermodynamical notions and only involves the electro-chemical
gradient of the ion under consideration, thus, neglecting other
ions and not treating diffusion of ions correctly, is not always
justified. In the case of calcium a more exact theoretical approach
based on the Goldman-Hodgkin-Katz equation leads to a non-linear
I-V-relationship (see e.g.~\cite{hille92}). 
This equation is derived from the theory of electro-diffusion through
a membrane and involves the influence of other ions on the reversal
potential. This more
precise I-V-relationship is approximated by an Ohm's law like
current, which is only justified 
for the linear piece of the curve. A good approximation is to
use a corrected reversal potential of $V_{\rm Ca}=+50mV$. Up to
this rather large membrane potential the I-V-relationship is,
indeed, approximately linear. However, one should keep in mind
that for depolarisation beyond $50mV$ (thus of $\Delta V=120mV$
relative to the resting potential) 
the calcium currents are not correctly
represented. 

\subsection{The basic model assumptions}

The following list summarises the most important features, assumptions,
and limitations of the \bc-model:

{\it Universality of single membrane protein measurements:}\\
It is assumed that the measured single membrane protein
conductivities and opening dynamics are valid irrespective
of the cell-type. This considerably increases the data basis
of the simulation. In view of consistent measurements
of the same protein in different cells this approximation
is reasonable. These parameters are only varied in
order to investigate robustness issues.

{\it Stability of the \bc-resting state:}\\
The leakage currents are assumed to equilibrate all other
currents if all cell properties are at their resting values.
This implies that the cell exhibits a stable steady state
and that the cell returns to this resting state after
stimulation.

{\it Glucose- and not ATP-mediated activation:}\\
Glucose is assumed to directly impact on the conductivity
of the K,ATP-channel. Thus, the \bc~can be stimulated
by changes of the glucose level. However, the dynamics
of ATP \cite{keizer89,fridlyand05} which mediates the effect of 
glucose on K,ATP-channels \cite{hopkins92}
is not considered and left
for future improvements of the simulation.
Also any feedback between ATP and calcium is neglected \cite{detimary98}.

{\it Averaging over the whole-cell:}\\
Inhomogeneities of the cell are averaged out. The ion-concentrations
are assumed to be average quantities over the whole-cell. 

{\it Neglect of calcium-induced-calcium-release:}\\
Calcium-induced-calcium release has been suggested to contribute
to the calcium oscillations in response to glucose-induced 
potassium currents \cite{ammala91}. However, blocking of related
mechanisms does not affect observed oscillations in pancreatic
$\beta$-cells \cite{liu95,gilon99}. Oscillation in the stores
might be orchestrated with oscillations in the cytosol \cite{gilon99}.
The present simulation neglects the effects of 
calcium-induced-calcium-release and assumes that the endoplasmatic
reticulum acts mainly as an additional buffer. 
Thus, the binding site concentration $c_0$ is assumed 
to be larger than expected by calmodulin alone.

{\it Membrane channels follow Ohm's law:}\\
The ion flow through an open pore is assumed to follow
Ohm's law which implies a linear voltage-current-relation.
This is valid for most ion channels. Deviations are found
only for extreme membrane potentials which are not
considered within this investigation.

{\it The \bc~is isolated:}\\
{\it In vivo} \bcs~are in the context of islets and
connected with each others by gap-junctions. This connection
might influence the susceptibility of stimulation. 
Indeed, \bc-stimulation is
more difficult in isolated cells (see e.g.~\cite{kanno02}).
The present simulation neglects effects stemming 
from being embedded into islets.


\section{Results}

\subsection{Evaluate the Sherman et al. 1988 model}

The model of Sherman et al. \cite{sherman88} can be considered
as a classical reference model. It is characterised
by the following properties:
\begin{itemize}
\item
Membrane potential follows a classical ordinary differential
equation with capacity 5.31pF
and three ionic currents (no leakage and no sodium currents).
\item
A voltage-gated potassium channel (corresponding to K,V here)
is described by a sigmoidal function for its asymptotic value
and a differential equation for the opening probability (same as here). 
Inactivation of K,V-channels is not considered.
\item
A calcium-gated potassium channel (corresponding to K,Ca here),
which is described by a Hill function with Hill coefficient 
$n_{\rm K,Ca}=1$.
\item
A calcium channel (corresponding to Ca,L here) including inactivation
and fast activation.
\item
Constant reversal potentials are assumed (no Nernst equation is used).
\item
The steady state of the system is defined as zero current of
the three currents contributing to the voltage equation
(thus no leakage currents are considered). 
\item
An equation for calcium ions including two terms and a constant factor
$f=0.001$ compensating for the buffer (thus, no buffer dynamics). 
One term accounts for Ca,L-currents
the second for calcium removal (corresponding to PMCA here).
\item
PMCA-activity is not described by a Hill-function
but increases linearly with the calcium concentration.
Note that in \cite{sherman88} the calcium extrusion term
is assumed to be electroneutral, which is not the case
as, typically, calcium is removed by PMCA. This pump
is not electroneutral. Thus the PMCA-current also
enters the equation for the membrane potential 
\gl{voltage} in the present model.
\end{itemize}
The full model as introduced in \se{methods}
is reduced to these properties and the results
in \cite{sherman88} are reproduced
(data not shown). In a second step
some parameters are modified to experimentally
known values: capacity, resting concentrations,
reversal potential of calcium.
All these changed values are 
collected in \tato{pars}{densities} in the supplement.
Surprisingly, all modifications together
only moderately alter the results
(see \ab{bc0025}). 
\begin{figure}[ht!]
\begin{center}
\includegraphics[height=6.8cm]{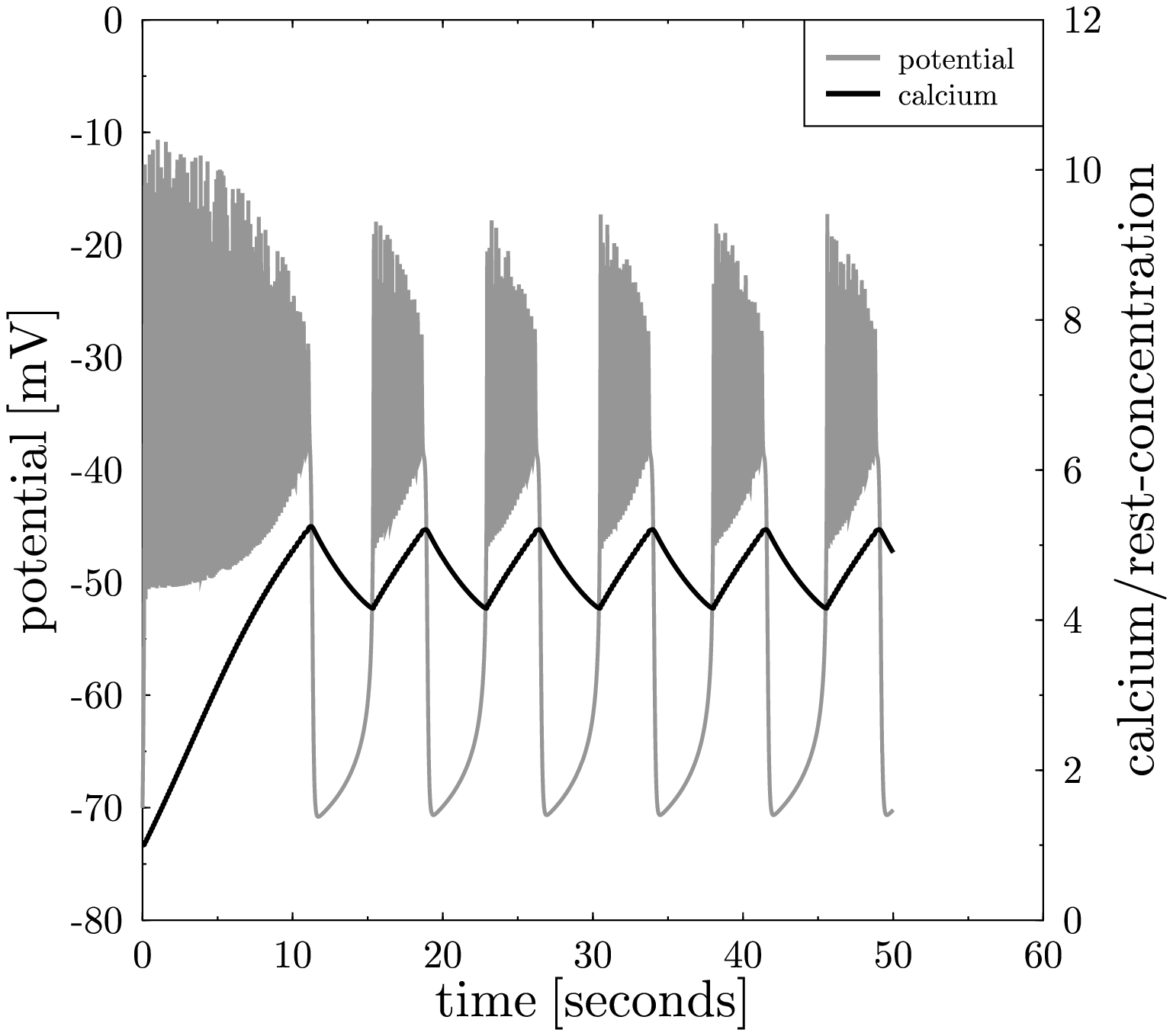}
\includegraphics[height=6.8cm]{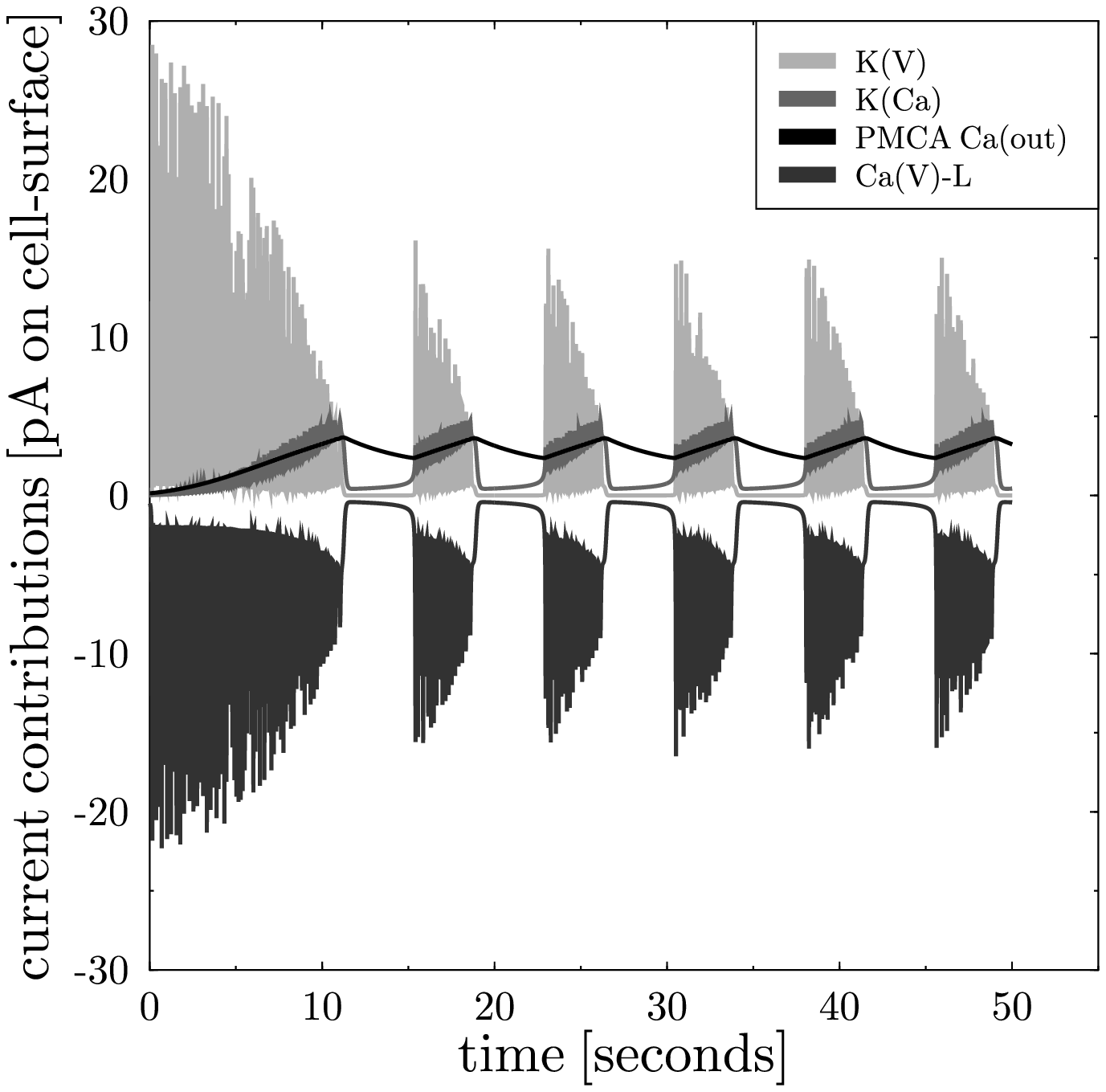}
\caption[]{\sf {\bf The simulation is consistent with Sherman et al's model:}
A simulation of the model of Sherman et al. \cite{sherman88}
with adapted parameters. The membrane potential (left panel,
grey line) exhibits repeated bursting events.
Intracellular calcium concentration is plotted
as ratio to the resting concentration of $0.1\mu M$
(left panel, black full line). It 
rises during bursting events and decreases during the silent
phase. The contributions of the transmembrane currents are
shown in the right panel (the inset lists the currents
in the sequence of vertical appearance). Oscillations occur between 
voltage-gated potassium (K,V right panel, light grey line) 
and calcium channels (Ca,L right panel, dark grey line).
K,Ca- and PMCA-activity follow the calcium level
(right panel, medium grey and black line, respectively.}
\label{bc0025}
\end{center}
\end{figure}

The channels in the initial state of the simulation are
not equilibrated (thus not in a steady state). 
This induces a certain inflow of calcium through Ca,L
which depolarises the cell. As
the K,V current reacts with a delay the potassium outwards
current comes later and allows for an above threshold
depolarisation. This leads to the oscillation on a higher
voltage level. Now, free calcium accumulates in the
cell due to repeated inflow. Note that the largest part
of the incoming calcium is assumed to bind to the buffer, 
such that calcium accumulates much slower
than according to the calcium inflow. 
More calcium induces increasing $g_{\rm K,Ca}$
(note that the K,Ca-currents used in \cite{sherman88}
are not in agreement with single protein experiments)
and PMCA-activity, which in turn increases 
the potassium and calcium outflow, respectively. The
oscillations are suppressed for some threshold K,Ca
and PMCA conductivity. From here the whole process restarts.
The essential properties are the delay between
Ca,L- and K,V-current, the calcium buffering
which induces a slow calcium increase, and
a membrane protein which reacts to the
increased calcium level by dropping oscillations.

The back-reaction
of PMCA activity on the membrane potential was neglected
in \cite{sherman88}. 
If this is included into Sherman et al's model, bursting
disappears.
The back reaction of PMCA
activity on the membrane potentials induces hyperpolarisation
of the cell. This infers that K,V and K,Ca (in view 
of the reversal potential of $\overline{V_{\rm K}}=-75mV$)
but more importantly Ca,L (because of very low opening
probability) are almost inactive. In this situation
PMCA reduces intracellular calcium below its resting state
while further hyperpolarising the cell. This is a problem
related to lack of steady state in Sherman et al's model.
The cell does not go back to resting state after a bursting
process. In conclusion, neglect of a back-reaction
of the process removing calcium from the cell on the membrane
potential turns out not to be a justified approximation.

\subsection{\bc-bursting with calcium, potassium, and sodium currents}

The second simulation includes all membrane proteins as introduced
in \se{methods} and, thus, extends the model \cite{sherman88}. The
characteristics of activation and inactivation are taken
from single protein experiments. The densities are adapted
to find reasonable electrical bursting. 
The resulting set of densities is given in 
in the supplement \ta{densities} (full model).
First, the steady state
is established and then the cell is stimulated by $10mM$ glucose.
The result is shown in \ab{bc0221_10}
and is in good agreement with corresponding
measurements \cite{kanno02,beauvois06}.
\begin{figure}[ht!]
\begin{center}
\includegraphics[height=6.8cm]{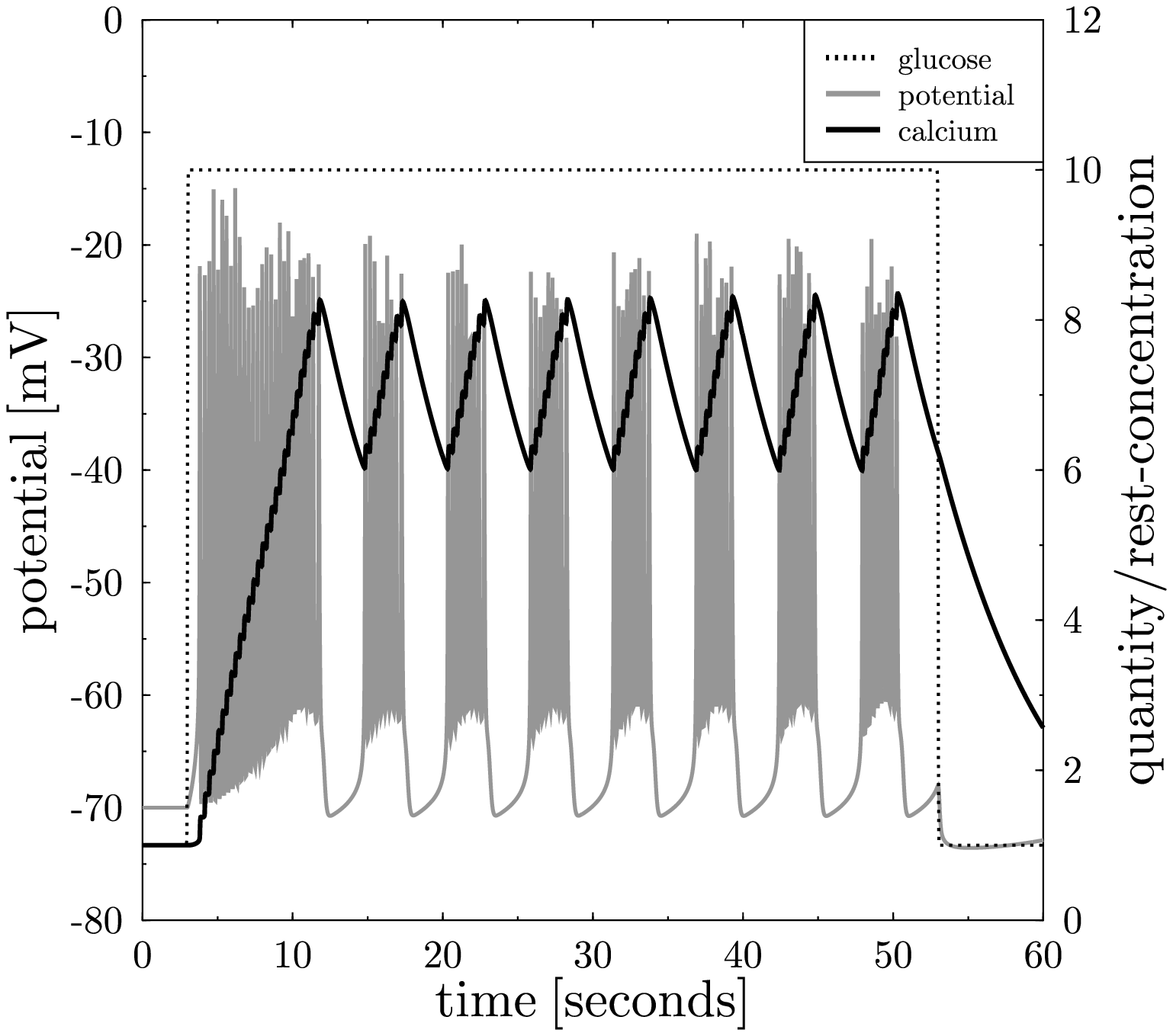}
\includegraphics[height=6.8cm]{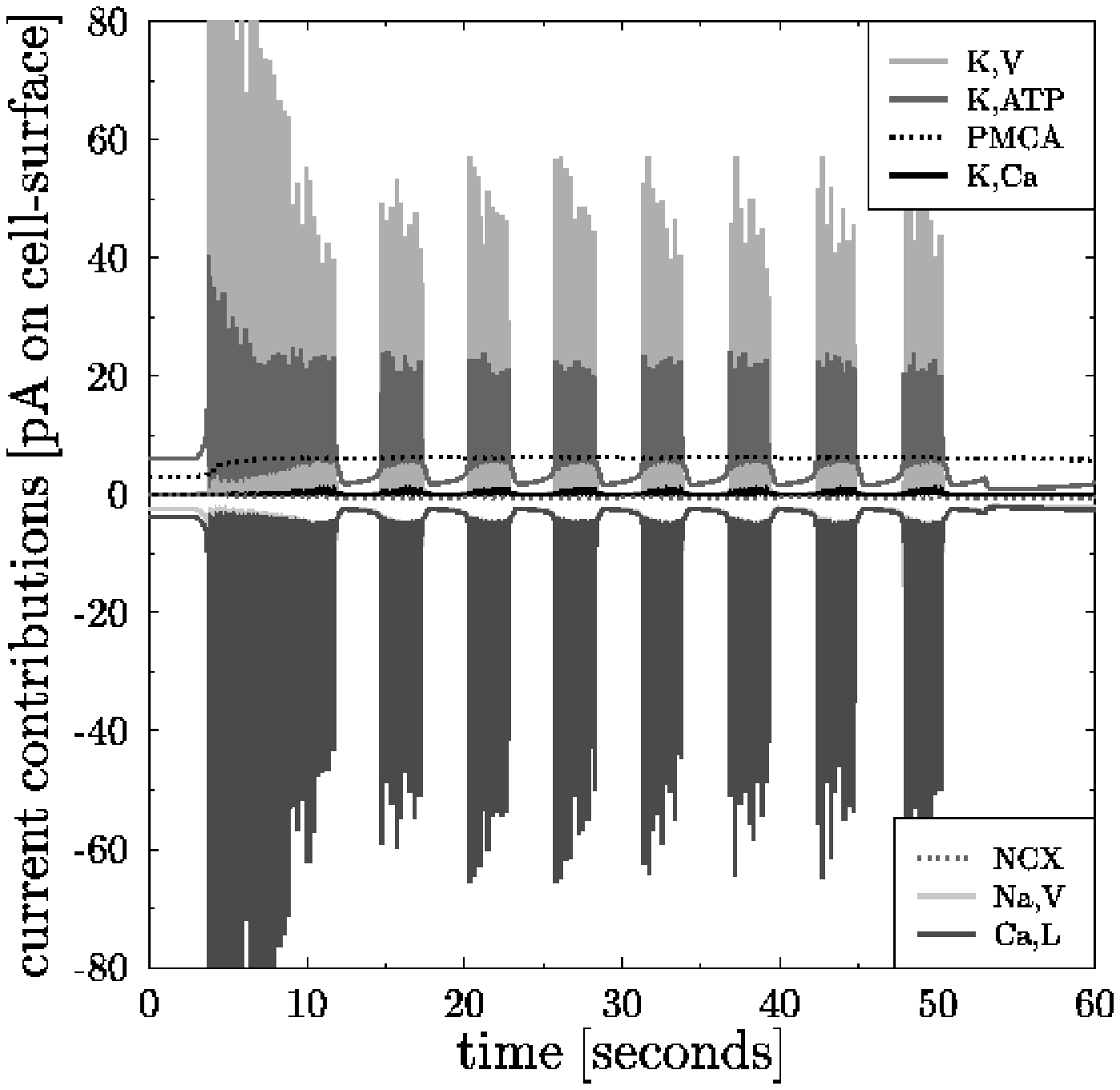}
\caption[]{\sf {\bf Repeated bursting including all transmembrane proteins:}
The full model for the electrophysiology of the \bc~is used
to simulate repeated bursting after activation with glucose.
Before stimulation
the \bc~exhibits a stable steady state.
Activation with $10mM$ glucose at $t=3 s$ 
(left panel, dashed line)
first leads to instabilities.
Then the membrane potential (left panel, grey line)
shows regular repeated bursting activity.
The burst to
silent ratio and the bursting frequency are realistic. 
The intracellular calcium level 
(left panel, black full line)
increases and decreases in cycles corresponding to
the bursting events. 
The whole-cell contributions
of the different currents are shown in the right panel
(the proteins in the inset are given in the sequence
of current contributions).
Note that here and in the corresponding subsequent figures
the electrical currents, not ion-currents, are shown. 
Therefore the NCX-current appears
with negative sign.}
\label{bc0221_10}
\end{center}
\end{figure}
The whole-cell peak ion-currents are $12$, $50$, and $70pA$ for
sodium, calcium, potassium, respectively, which is also
found in experiments. The whole-cell K,ATP-conductivity
is $1nS$ at $1mM$ glucose (derived from \ab{bc0221_10_open}
as $4\pi r^2 \rho_{\rm K,ATP}\overline{g_{\rm K,ATP}}(1-g_{\rm K,ATP})$), 
which compares to $1-3nS$ in experiment \cite{goepel99}.
\begin{figure}[ht!]
\begin{center}
\includegraphics[height=6.8cm]{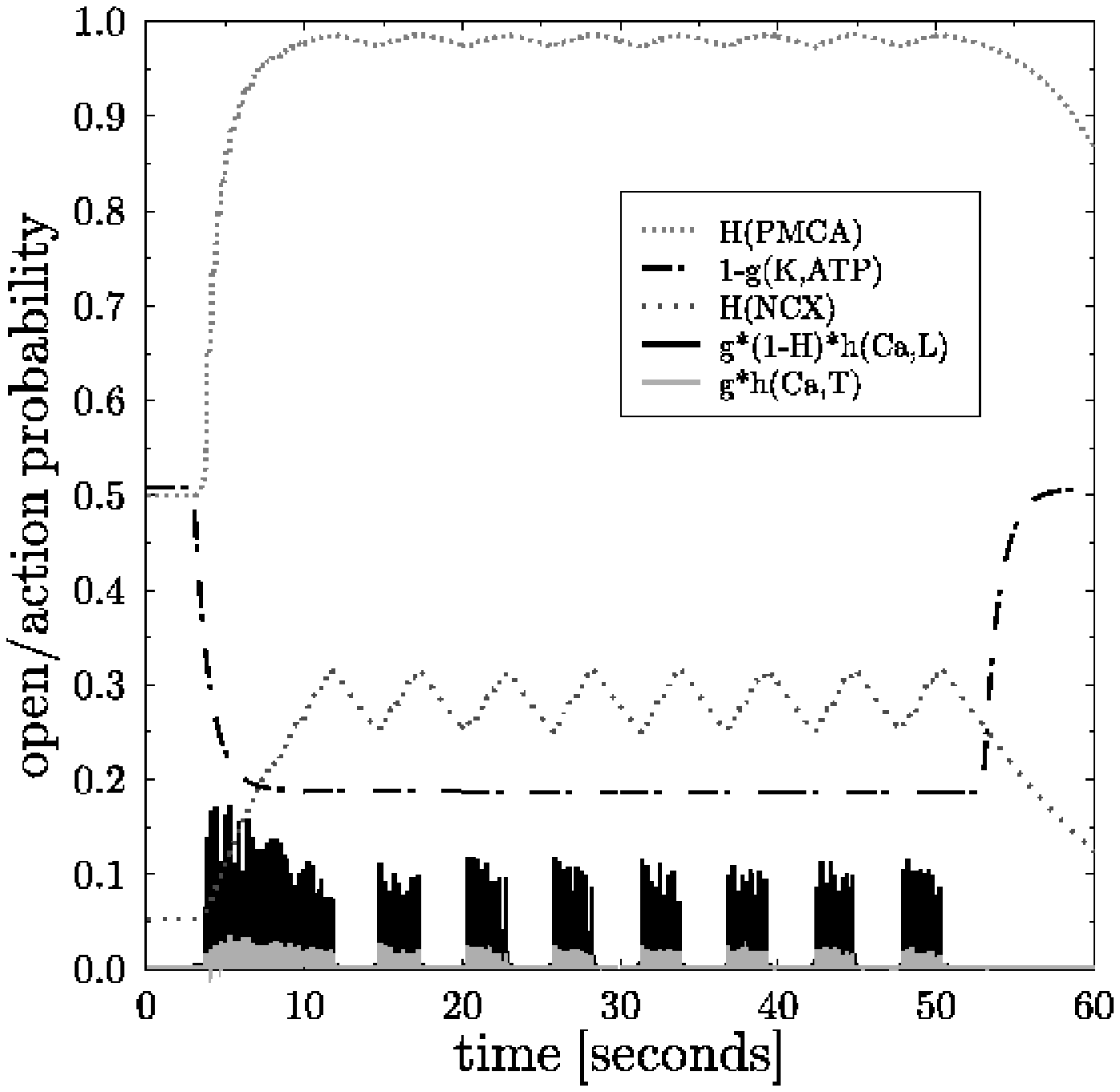}
\includegraphics[height=6.8cm]{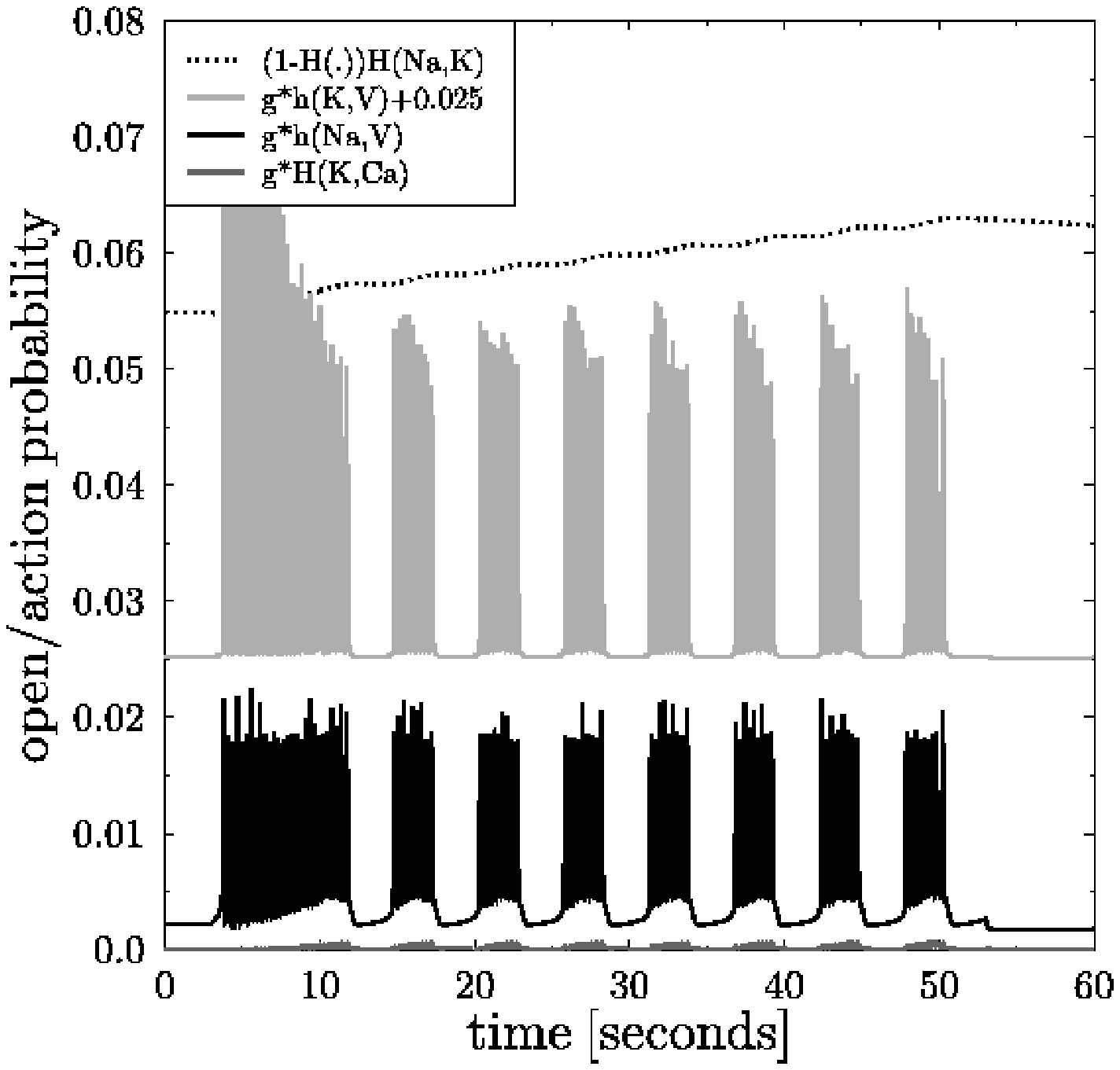}
\caption[]{\sf {\bf Opening/activity states of the membrane proteins:}
The states of the transmembrane proteins during repeated
bursting as given in \ab{bc0221_10} are shown as
probability of activity or opening.
The left panel collects all calcium-relevant proteins
and the K,ATP-channel. The right panel all sodium-relevant
and other potassium-conducting proteins (note the different
scale in the right panel and the shifted open probability
for K,V-channels). The inset lists the probabilities
in the sequence of vertical appearance and gives a hint how
the probabilities are calculated (see supplement
\se{opening} for details).
The simulation
protocol underlying this simulation is explained
in the figure legend of \ab{bc0221_10}.
Note that the PMCA-activity is at its limit in
an actively bursting \bc.}
\label{bc0221_10_open}
\end{center}
\end{figure}
An analysis of the contributions of the different membrane proteins
induces a picture which in parts coincides with the 
one proposed by Sherman et al.
\cite{sherman88}. However, there are also some 
important modifications which
will be analysed and isolated in subsequent knock-out and block
experiments.

\subsubsection{High resolution of bursting events}

A single burst event in \ab{bc0221_10} is analysed in more detail.
To this end a high resolution figure of a burst event is
depicted in \ab{bc0221_10_fine}.
\begin{figure}[ht!]
\begin{center}
\includegraphics[height=6.8cm]{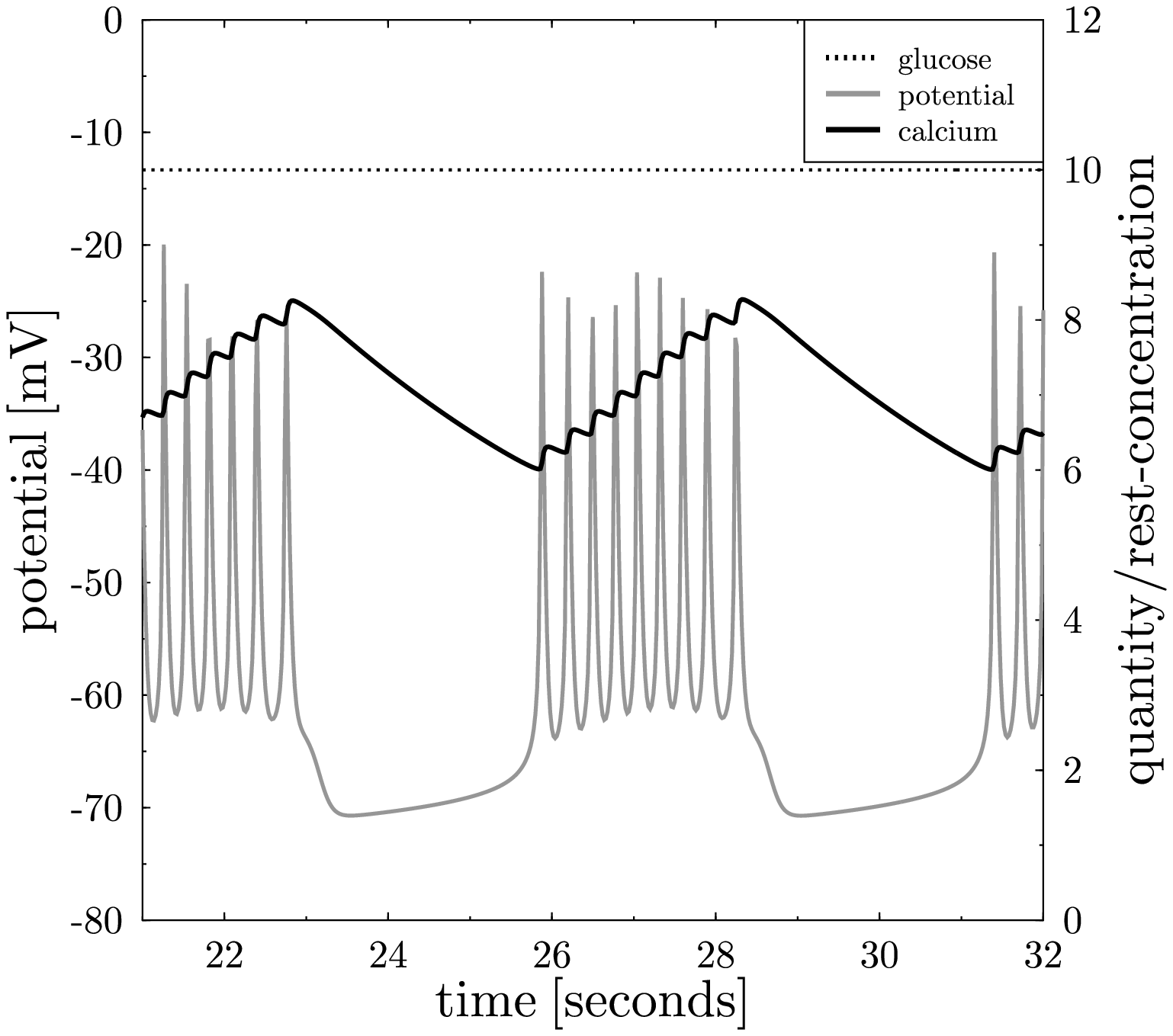}
\includegraphics[height=6.8cm]{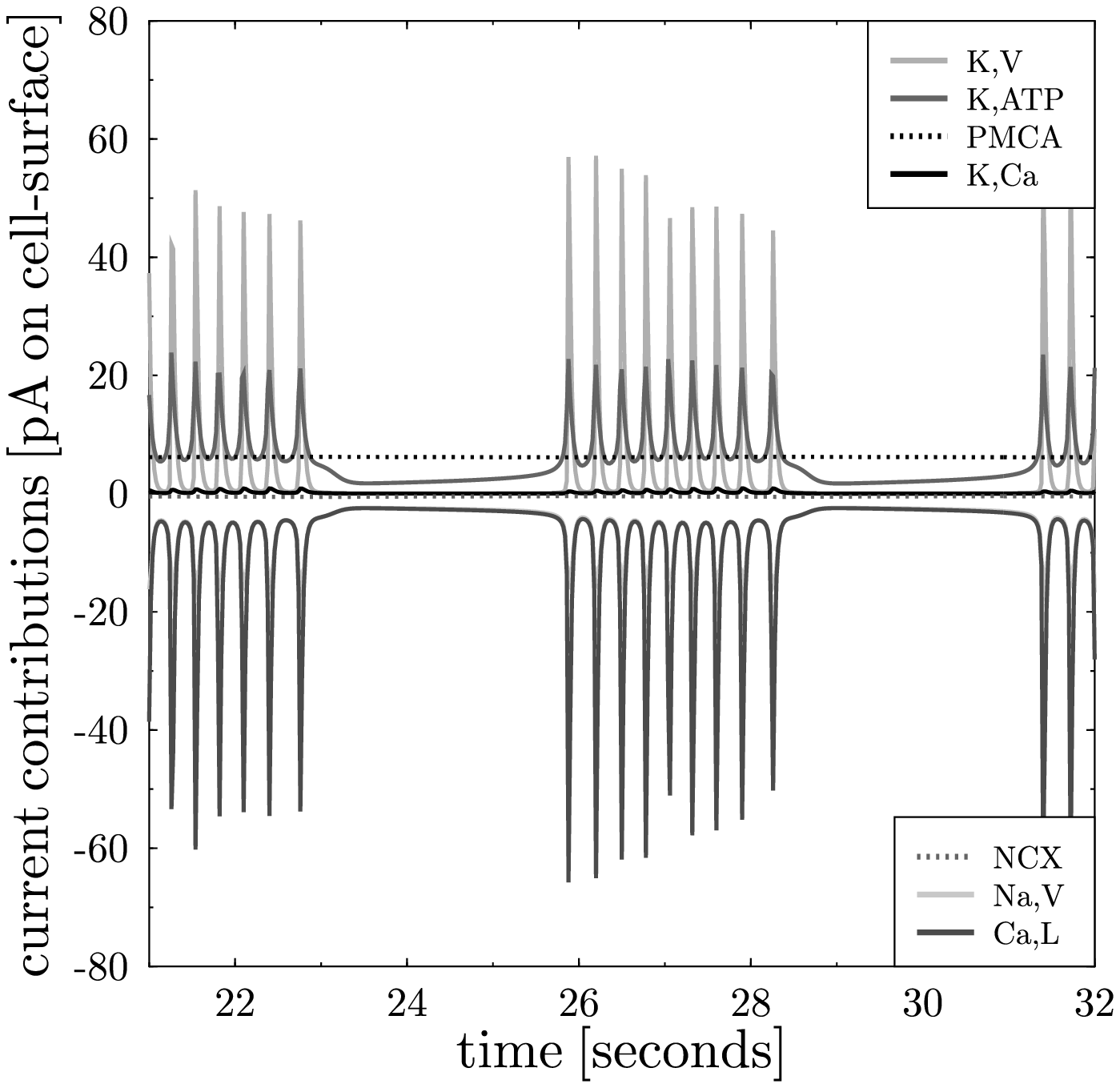}
\caption[]{\sf {\bf High-resolution view on a burst event:}
The third bursting event in \ab{bc0221_10}
is shown in higher resolution. All shown curves
have the same meaning as explained in the legend
of \ab{bc0221_10}.}
\label{bc0221_10_fine}
\end{center}
\end{figure}
The analysis is started when the 
membrane potential has reached its minimum.
It can be clearly seen that the burst event is initialised
by a K,ATP-current. The Ca,L-current quickly follows.
The K,V-currents are suppressed in this phase of the burst
event. This is not related to the delayed activation
(which causes $50ms$ at the most) but to the general
single protein activation properties. These do not
allow for a strong K,V-current near the reversal
potential of potassium.
K,V-currents become relevant only after strong
depolarisations which are induced by strong Ca,L-currents.
The delay of K,V-response with respect to the Ca,L-dynamics
leads to stable oscillations. Calcium levels
increase with each depolarisation spike. This induces
small K,Ca-currents, involving a slightly decreasing
frequency of spikes during each burst event
which is in agreement with experimental data
\cite{kanno02,beauvois06}.

Interruption is mediated by a concerted action of
NCX, PMCA, K,Ca and Ca,T in response to the increasing
calcium concentration. But on top of changed conductivity
the reversal potential plays an important role. For
calcium it changes from $22mV$ at the beginning of
a burst to $0mV$ at the end of a burst. This reduces
the calcium currents through voltage-gated channels
by about $35\%$ (at $V=-40mV$) 
and is a relevant contribution to
the interruption of bursts. Indeed, replacing
the Nernst-equation by constant reversal
potentials (the ones corresponding to the resting
state) leads to uninterrupted bursting (data not shown).

\subsection{Knock-out of K,Ca-protein}
\label{K-Ca-KO}

The calcium-gated potassium channel K,Ca is for the first
time implemented according to single membrane protein
characteristics \cite{barrett82}, which differ strongly
from the values used in other models. 
Currently, no model for
the K,Ca-opening dynamics exists. The measured opening
dynamics depends on the membrane potential and the calcium
concentration in a complicated way.  
A detailed analysis of the measured data (see supplement 
\se{K,Ca-opening})
has revealed that, in addition, the asymptotic opening probability 
for a given calcium concentration also depends on the
membrane potential (see \gl{dynamic_half_C}). 

The novel model
for K,Ca-opening probability is used for the \bc-electrophysiology
and it is predicted that under normal conditions the
K,Ca-currents remain rather small because of a
low opening probability. This finding suggests that
K,Ca might even be unnecessary for
repeated bursting.
K,Ca is believed to abort burst-events when calcium
is sufficiently increased. In principle, this can also be achieved by PMCA
which also directly depends on the calcium level.
K,Ca is switched off ($\rho_{\rm K,Ca}=0$) and
other densities have to be slightly adapted to
$\rho_{\rm K,ATP}=0.13/\mu m^2$ and
$\rho_{\rm PMCA}=1420/\mu m^2$
in order to keep regular bursting
(see \ab{bc0222}).
Note that this simulation does not correspond to
a block of K,Ca. The \bc~is constructed
without K,Ca, thus, this simulation
corresponds to a knock-out experiment.
\begin{figure}[ht!]
\begin{center}
\includegraphics[height=6.8cm]{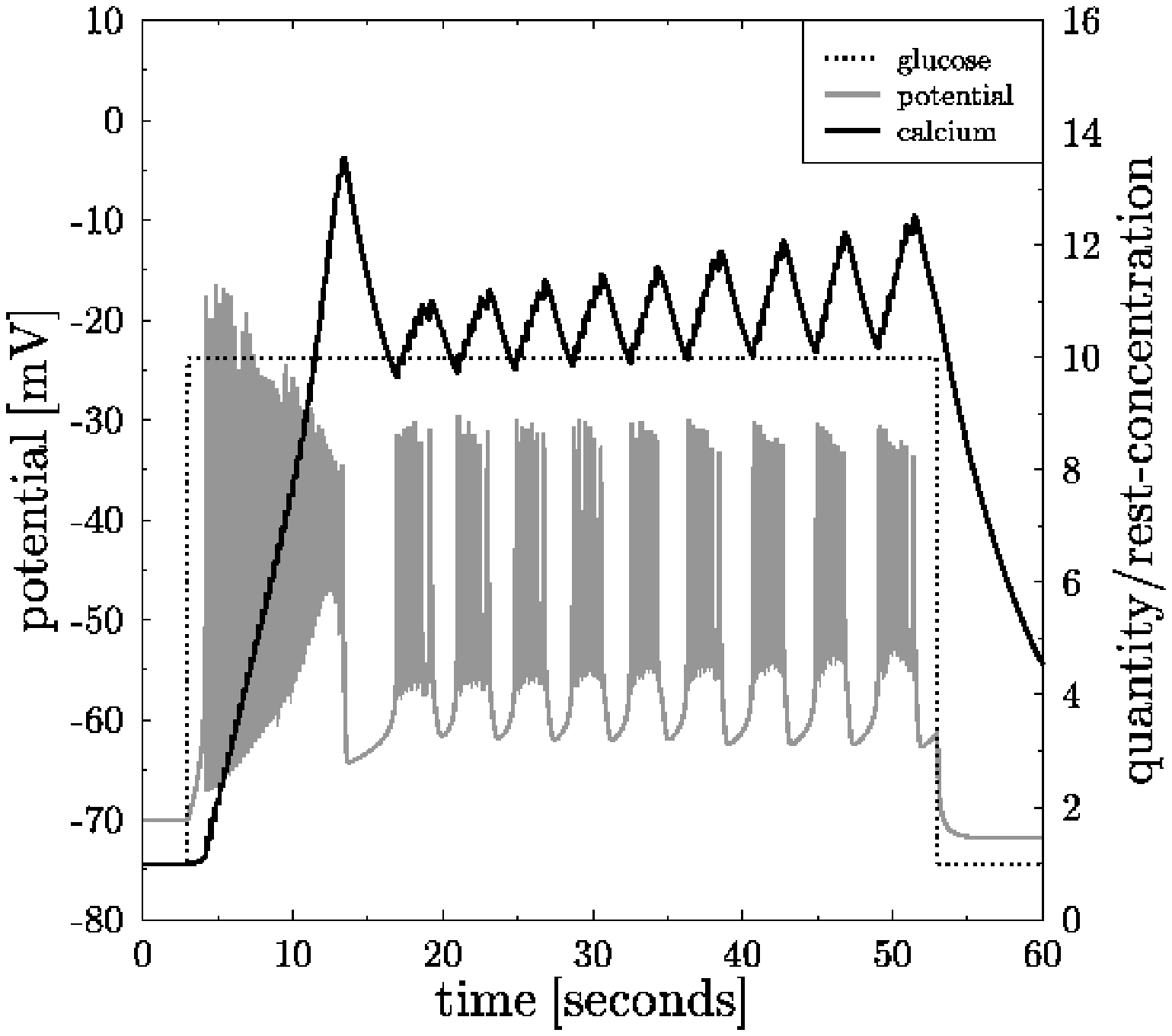}
\includegraphics[height=6.8cm]{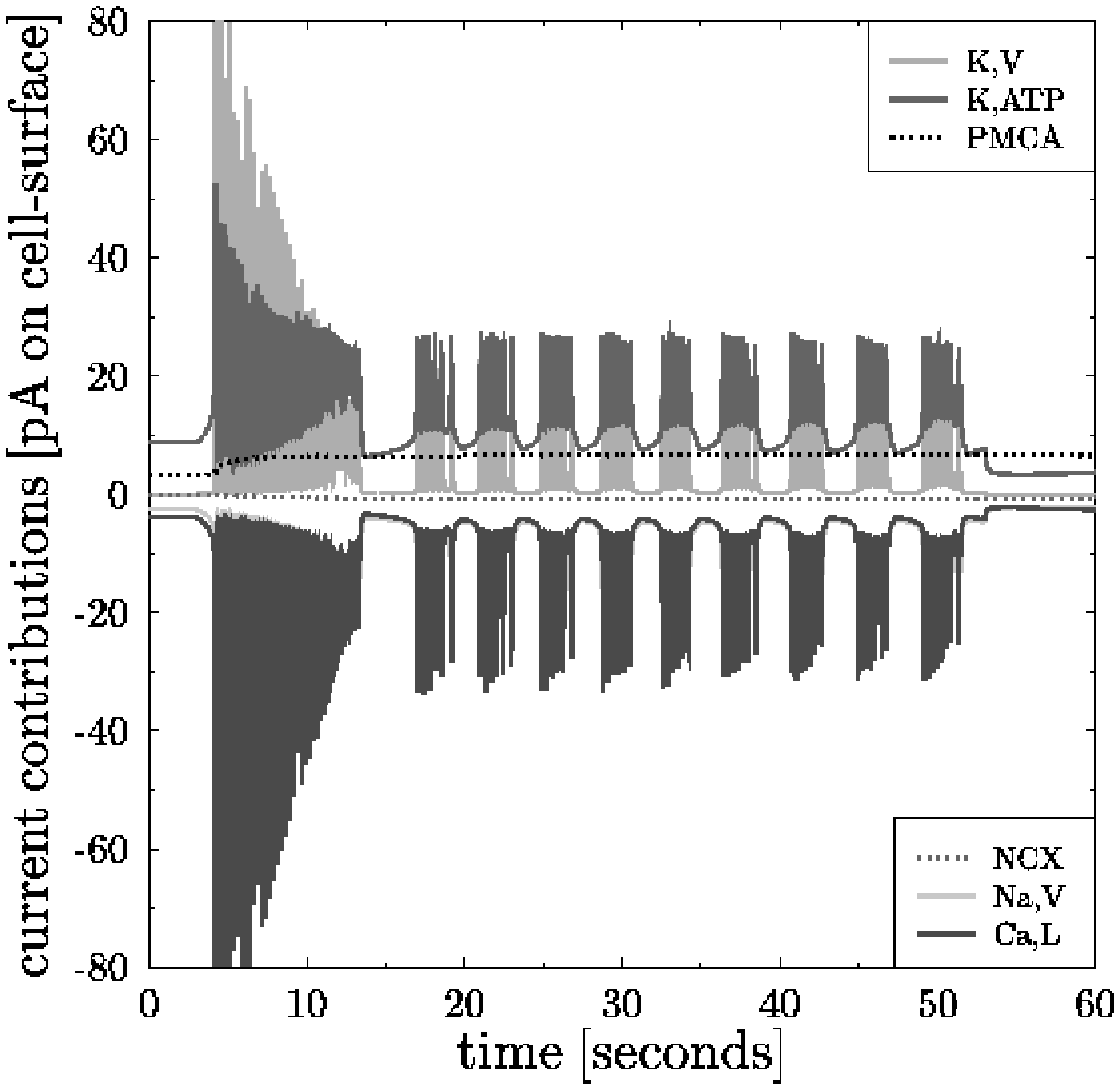}
\caption[]{\sf {\bf \bc-bursting without K,Ca-channels:}
Simulation without K,Ca-channels
and with protein densities adapted to
$\rho_{\rm K,ATP}=0.13/\mu m^2$ and
$\rho_{\rm PMCA}=1420/\mu m^2$.
Bursting frequency and the reached calcium level 
are both higher compared to \ab{bc0221_10}.
Lines are explained in the legend of \ab{bc0221_10}.}
\label{bc0222}
\end{center}
\end{figure}

In the normal \bc~a late increase in K,Ca-activity happens
for high calcium levels (at the end of each burst, see \ab{bc0221_10}
and \ab{bc0221_10_open}) 
leading to
potassium efflux. This counterbalances the calcium influx
and drops depolarisation.
In the present simulation without K,Ca, 
PMCA takes over this function.
However, PMCA-activity is already at its limits during
burst (see \ab{bc0221_10_open}).
This is also found in the K,Ca-knock-out 
experiment (not shown).

The K,Ca-knock-out experiment suggests that
for low calcium concentrations PMCA
is the main interrupter of bursts. 
The interpretation that
K,Ca is important for burst interruption is
only supported for calcium concentration
above $500\mu M$.
In this regime PMCA is not sufficiently flexible 
to further increase its activity and, thus,
to interrupt bursting. 
K,Ca is much more flexible in
this respect. It will be seen later that at high
calcium concentration K,Ca acquires other
functionality.

\subsection{Stimulation by raising external potassium}

The effect of increased external
potassium levels on \bcs~is investigated.
Starting from a \bc~in its resting state at normal
external potassium $K_{\rm ext}=5.7mM$,
external potassium is
increased to different values using a steep
sigmoidal function. 
The \bc~which had adapted to the lower potassium
level is stimulated because the reduced chemical gradient
pushes potassium into the cell and depolarises it.

For moderately increased external potassium the cell first
exhibits a smooth depolarising calcium current.
Larger $K_{\rm ext}$ then leads to K,ATP-induced action 
potential-like events involving potassium and
calcium currents, and finally into a burst-like
activity at $K_{\rm ext}=8mM$. However, this remains a unique
event (data not shown). The burst is interrupted by K,Ca-activity
at a rather high and constant calcium level around $1\mu M$.
Despite a new depolarised equilibrium state which normally
would induce action potentials, 
repeated bursting is prohibited by the constant strong
activity of PMCA and K,Ca.


\subsection{Stepwise increase of glucose levels}

The full model is used to investigate the \bc~behaviour
after stimulation with increased glucose levels. 
This inhibits K,ATP-potassium currents and 
depolarises the cell.
Depending on the target glucose concentration between
the resting value $\gamma_0=1mM$ and $30mM$
this becomes the germ of bursting events.
The \bc~reaction
is shown in \ab{bc0221} for some examples.
\begin{figure}[ht!]
\begin{center}
\includegraphics[height=4.5cm]{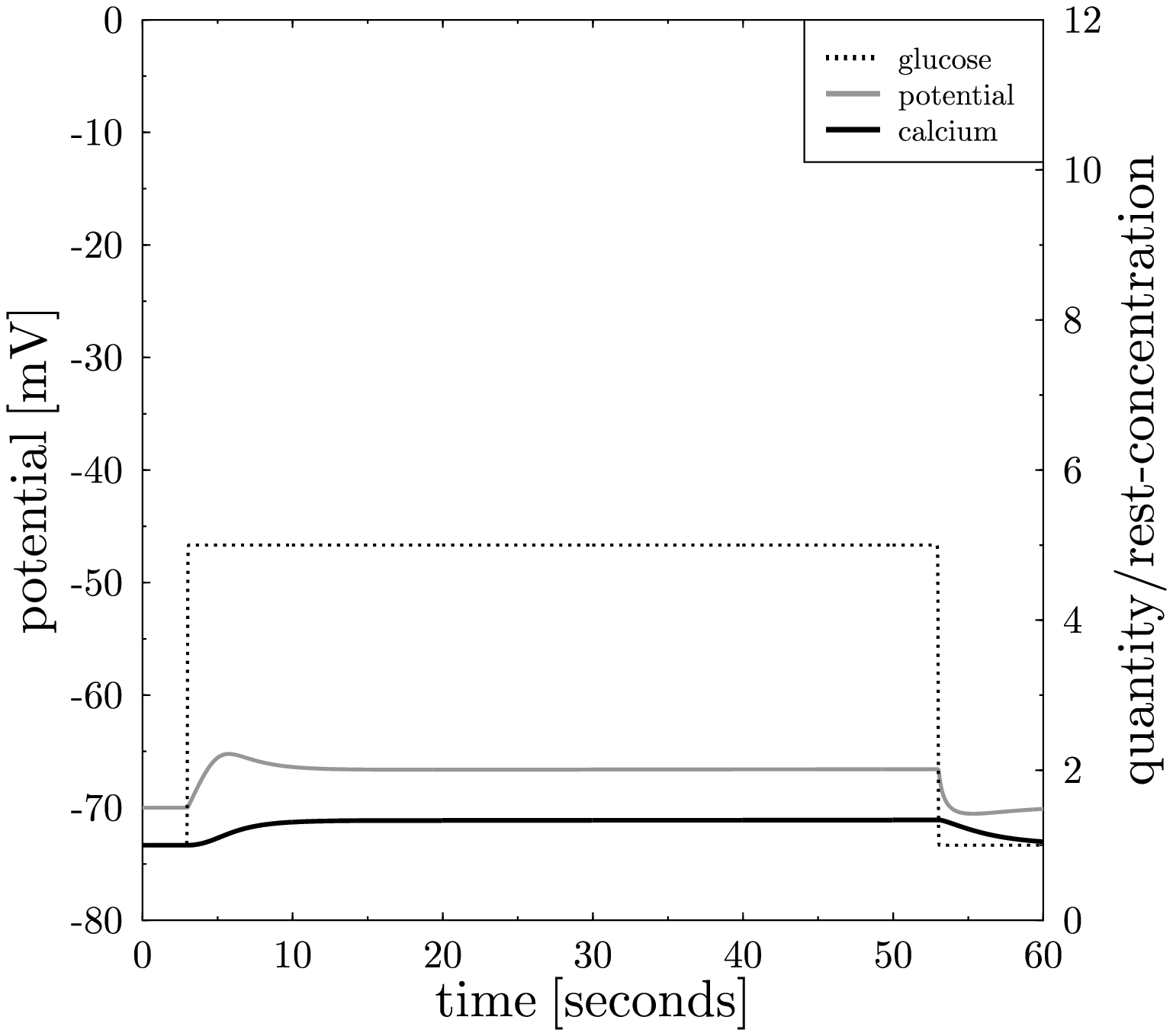}
\includegraphics[height=4.5cm]{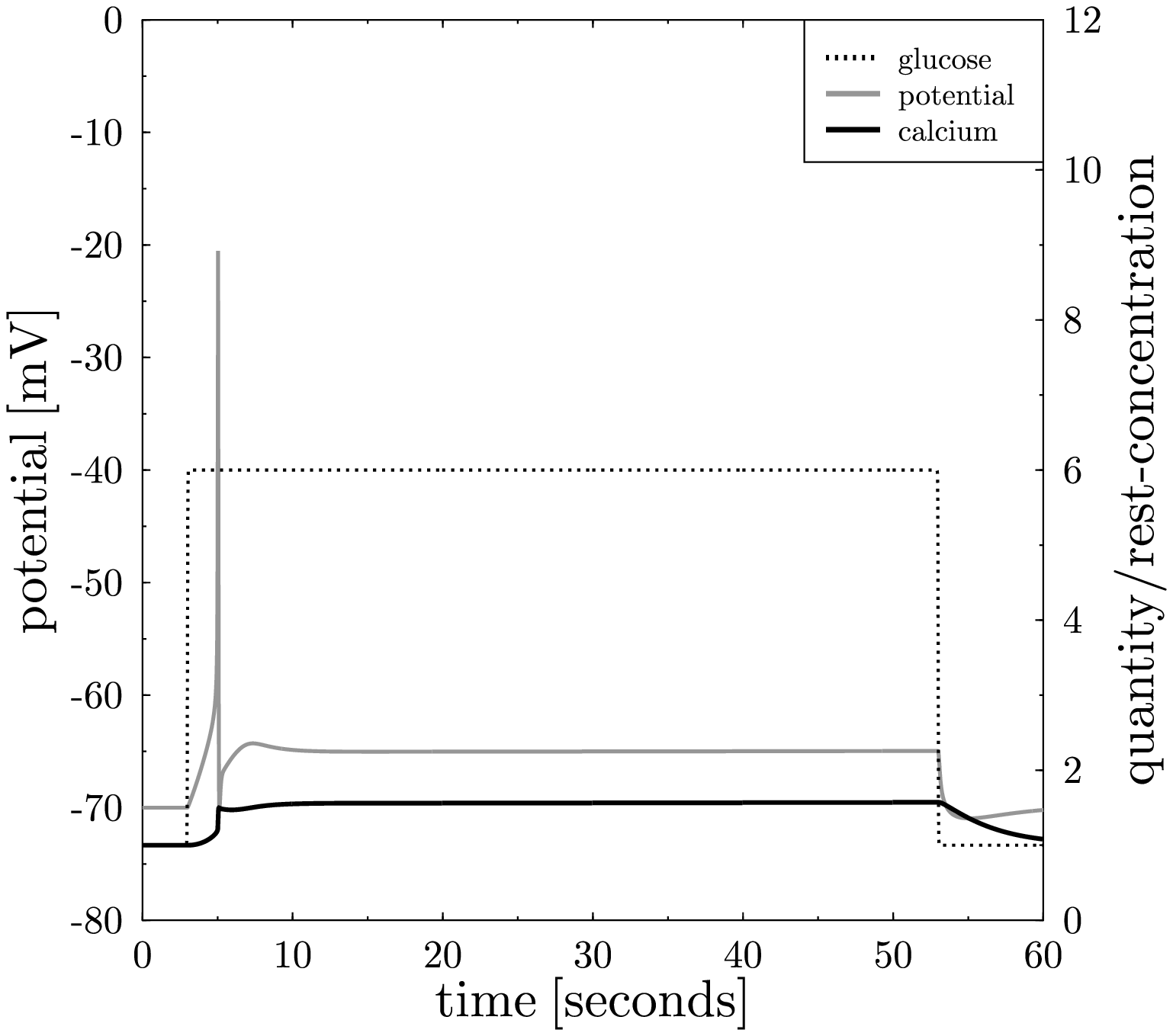}
\includegraphics[height=4.5cm]{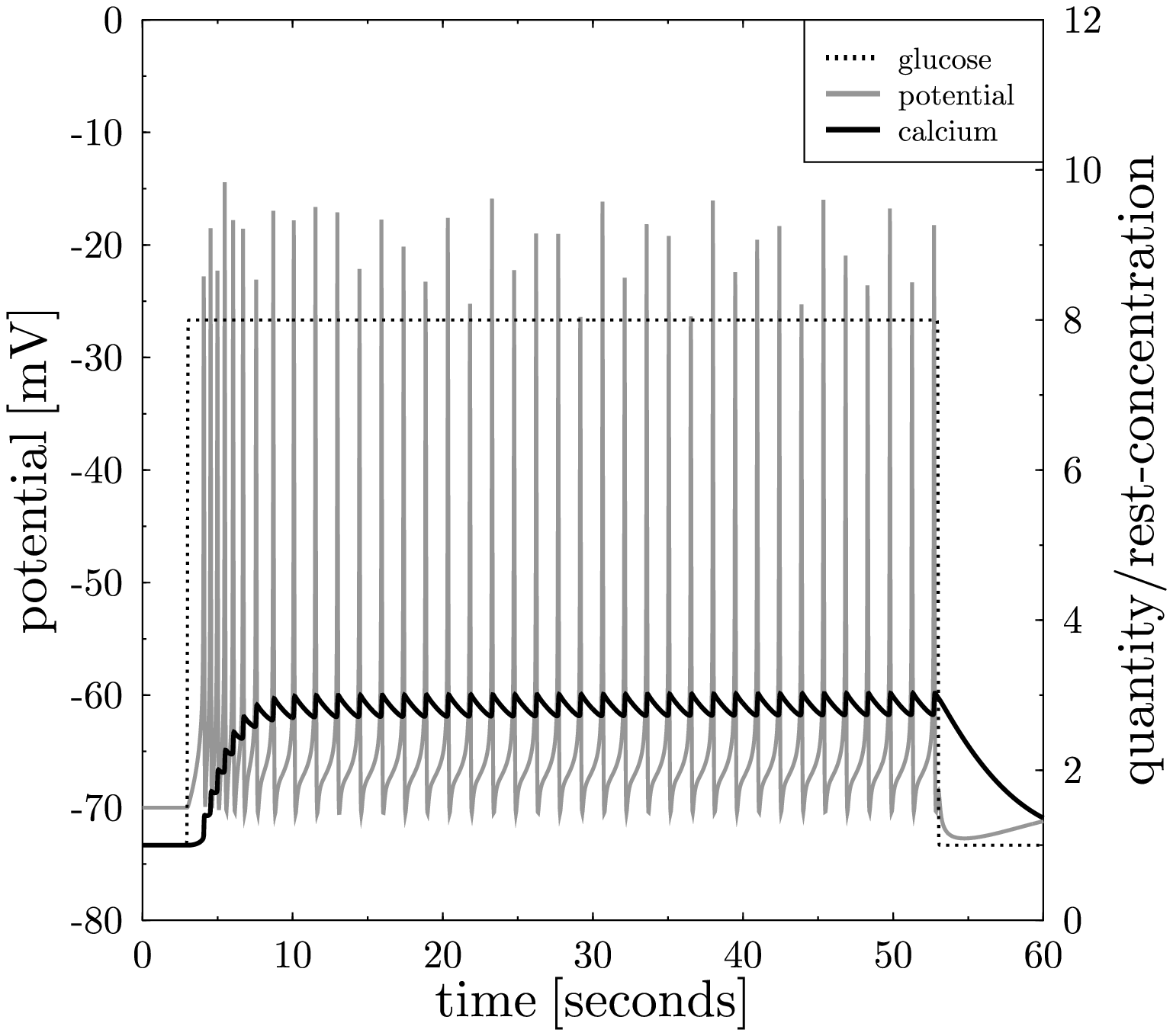}
\includegraphics[height=4.5cm]{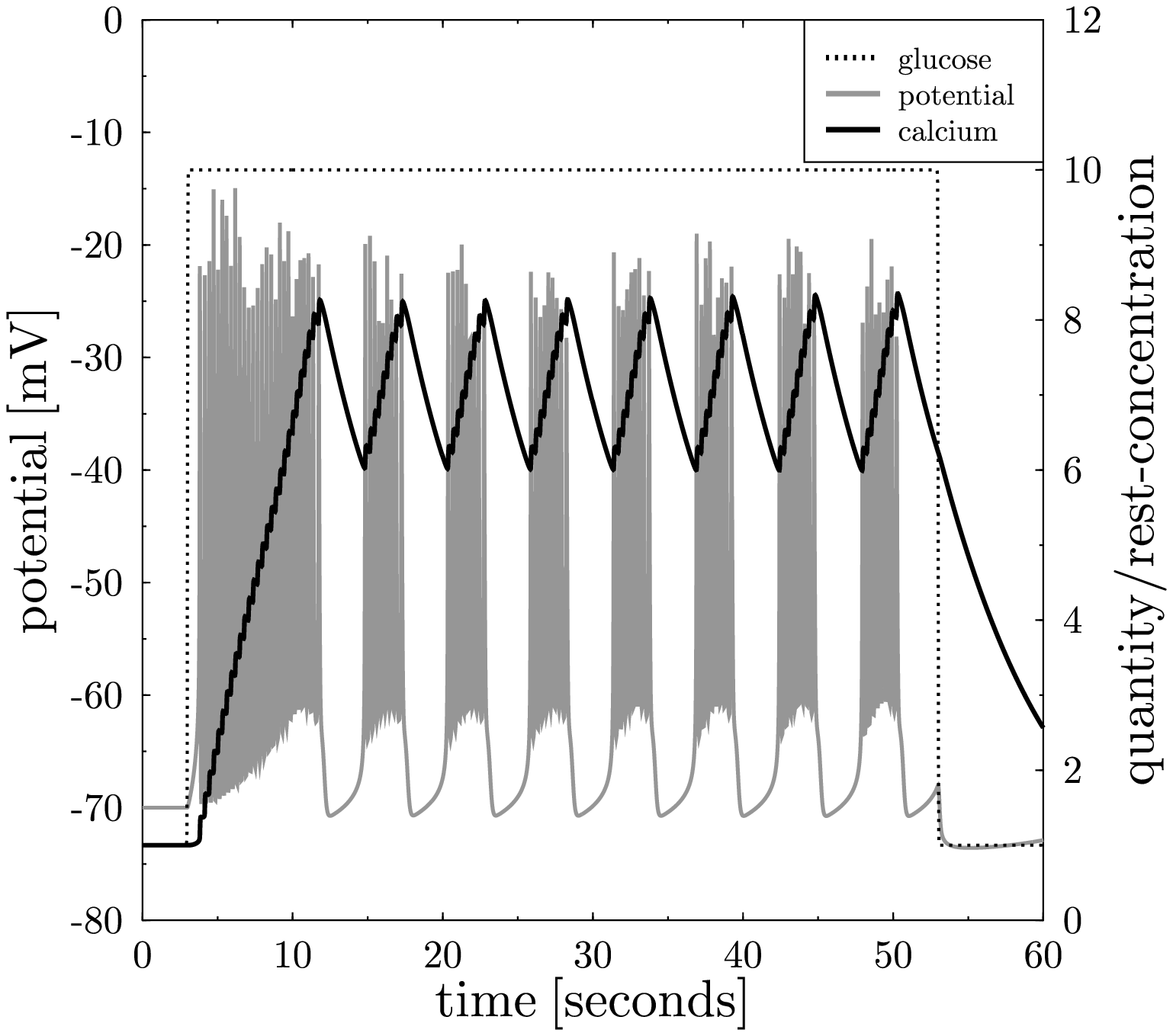}
\includegraphics[height=4.5cm]{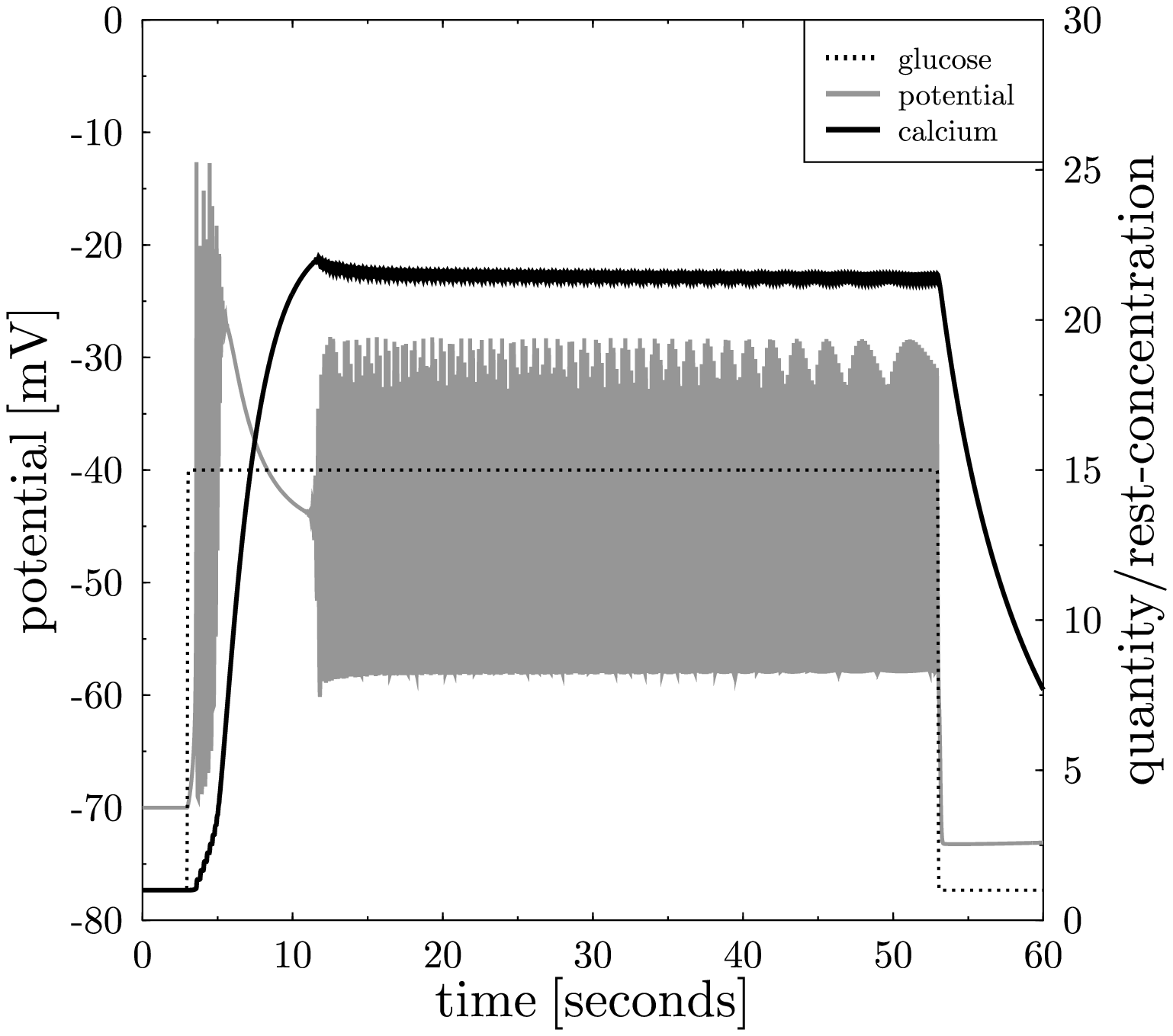}
\includegraphics[height=4.5cm]{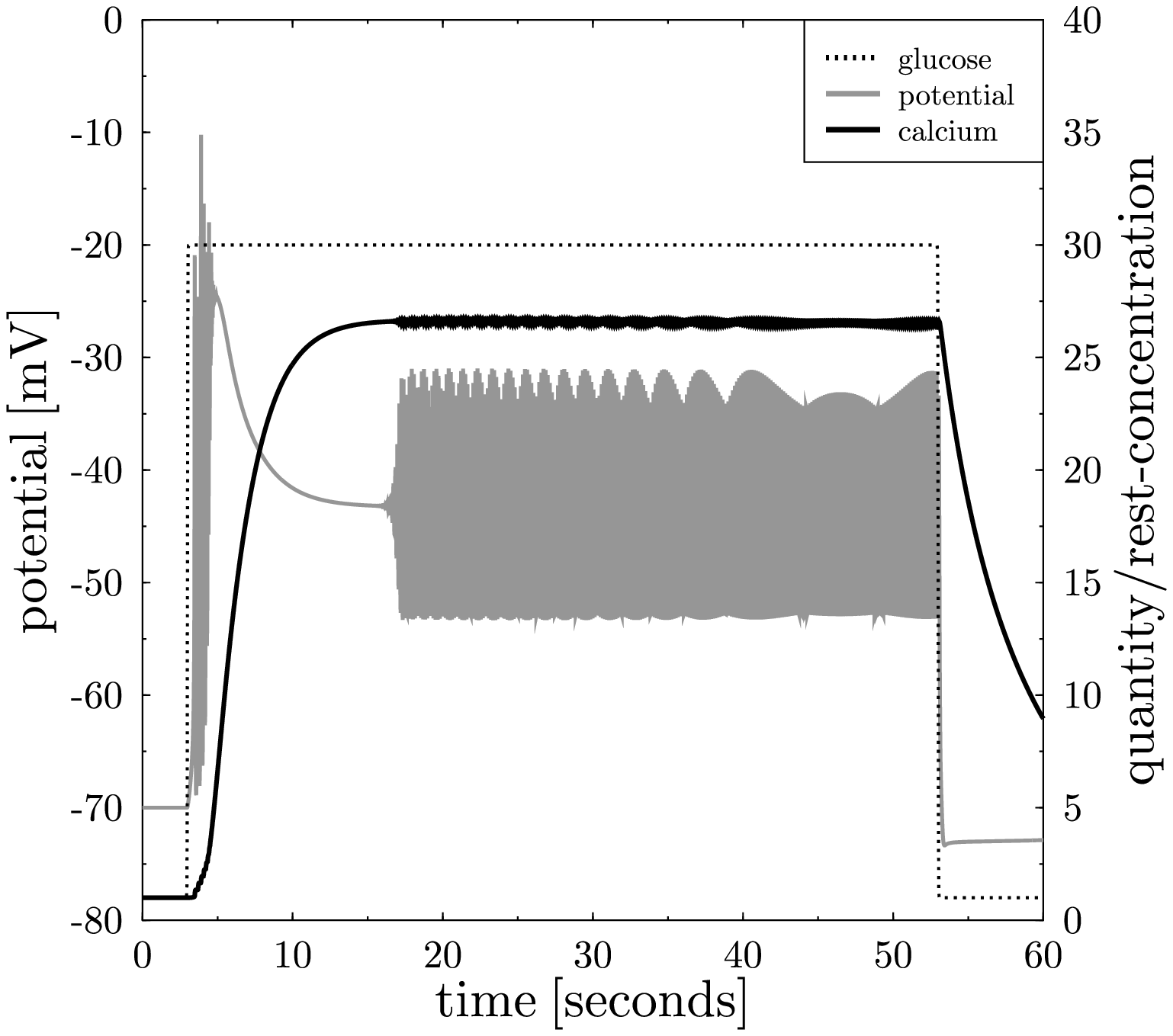}
\caption[]{\sf {\bf The \bc~reaction to increased glucose levels:}
Glucose stimulation is increased in several steps
$\gamma_{\rm stimulation}=5,6,8,10,15,30mM$ (panels in reading
order).  The membrane potential (grey line) 
and the calcium concentration (black line) are shown. Repeated bursting is
found for stimulation with about $10mM$ glucose.
Less glucose only induces single spikes, more
leads to uninterrupted bursting.}
\label{bc0221}
\end{center}
\end{figure}
Up to $5mM$ glucose only a smooth increase of the membrane potential
is found which then stabilises on a higher level. 
A single action potential appears at a threshold of $6mM$
which is in agreement with experiment
\cite{beauvois06}. The spikes then turn
into regular firing with increasing frequency
between $7$ and $9mM$. 
The high-frequency spikes
then are interrupted and replaced by repeated bursting
around $10mM$. 
The frequency of repeated bursting, the burst to silent ratio,
the bursting amplitude, and the bursting baseline
at $\gamma=10mM$ are within the range
of experimental results \cite{beauvois06,kanno02}.
For even larger glucose levels
uninterrupted bursting is found 
which is also in agreement with experiment \cite{kanno02}.

As in islets \cite{beauvois06}
calcium levels constantly increase with increasing glucose levels
(see \ab{bc0221}). This is not the case if the simpler
model for K,Ca-activation is used (data not shown),
which underlines the necessity of using the improved novel
model for K,Ca-dynamics.

\subsubsection{Steepness of glucose increases}

In experiments it is observed that the initial spike or burst
is more intense than the subsequently following ones -- if any.
According to the simulations this is related to the steepness
of the switch to the higher glucose level. With smooth sigmoidal
functions the initial spike can be fully suppressed. 
Similarly, long initial bursts can turn to
a normal duration.
However, the steepness of the glucose increase changes only the initial
action potentials or pre-bursting events. The long-term
bursting behaviour does exclusively depend on the 
asymptotic glucose level irrespective of
how quick it is reached.

\subsubsection{The role of K,Ca-currents at supra-large glucose}

There is an unexpected diversity of model behaviour at high
glucose levels in dependence of the model assumptions.
Uninterrupted bursting for supra-large
glucose levels is only found with the novel model
for K,Ca-activation (see supplement \se{K,Ca-opening}). 
The simpler model for 
K,Ca-activation as used in \ab{bc0025} 
leads to suppression of any spike for large glucose,
which is in contradiction to experiment.
This underlines the
necessity to rely on the measured electrophysiology
of single K,Ca-channels.
A rather weak role of K,Ca-channels during
normal electrical activity in inferred. 
K,Ca gets relevantly active only
at very high calcium concentrations which are associated
with high glucose concentrations. 

A more detailed analysis of \ab{bc0221} reveals that
uninterrupted bursting at supra-large glucose-levels
is based on oscillations in the K,Ca-current. While
at $\gamma=12mM$ the K,V-current is dominant, at
$30mM$ it is the K,Ca-channel which provides the
largest contribution to the potassium current
(see \ab{bc0221_12_30}).
\begin{figure}[ht!]
\begin{center}
\includegraphics[height=6.8cm]{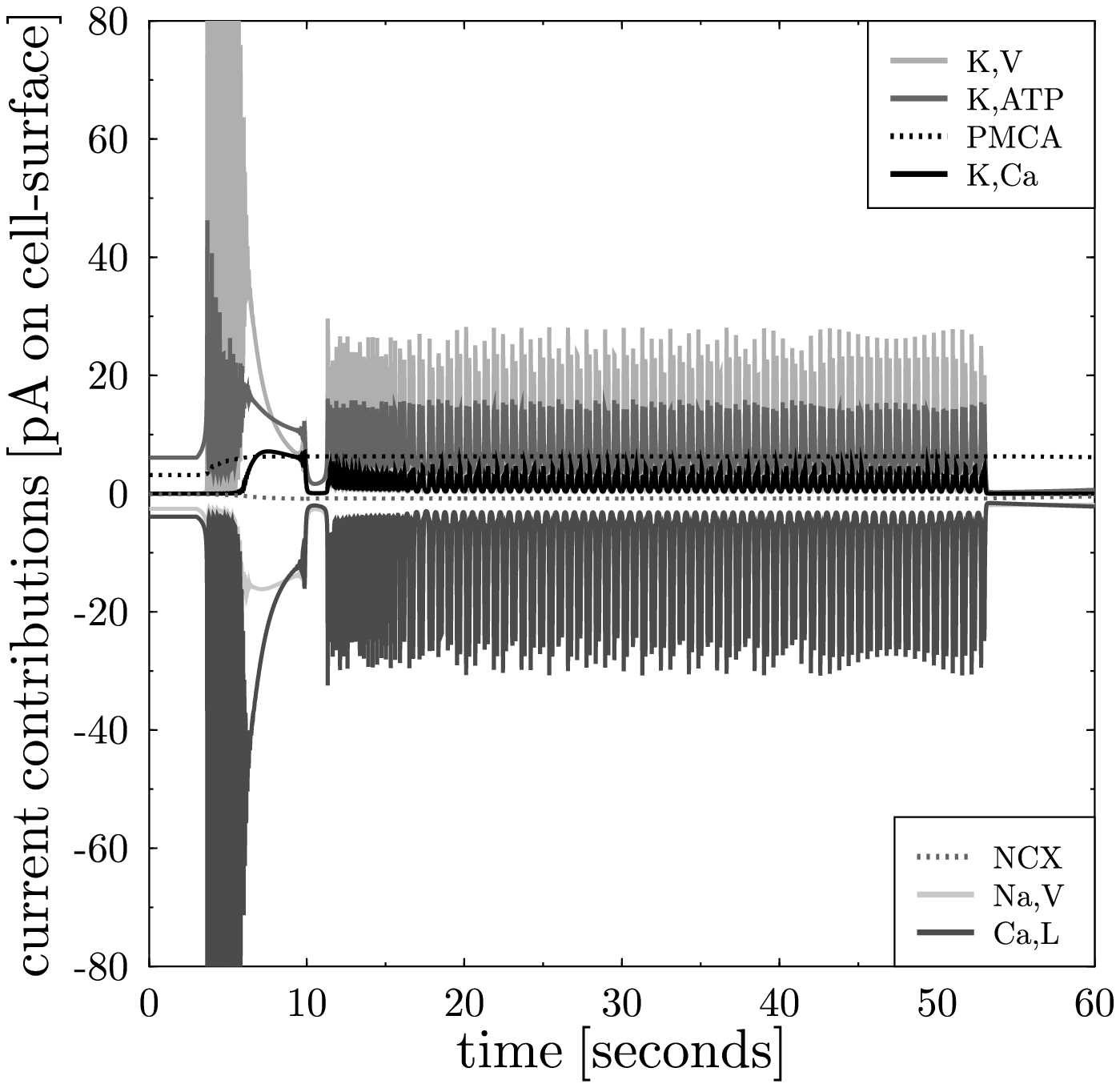}
\includegraphics[height=6.8cm]{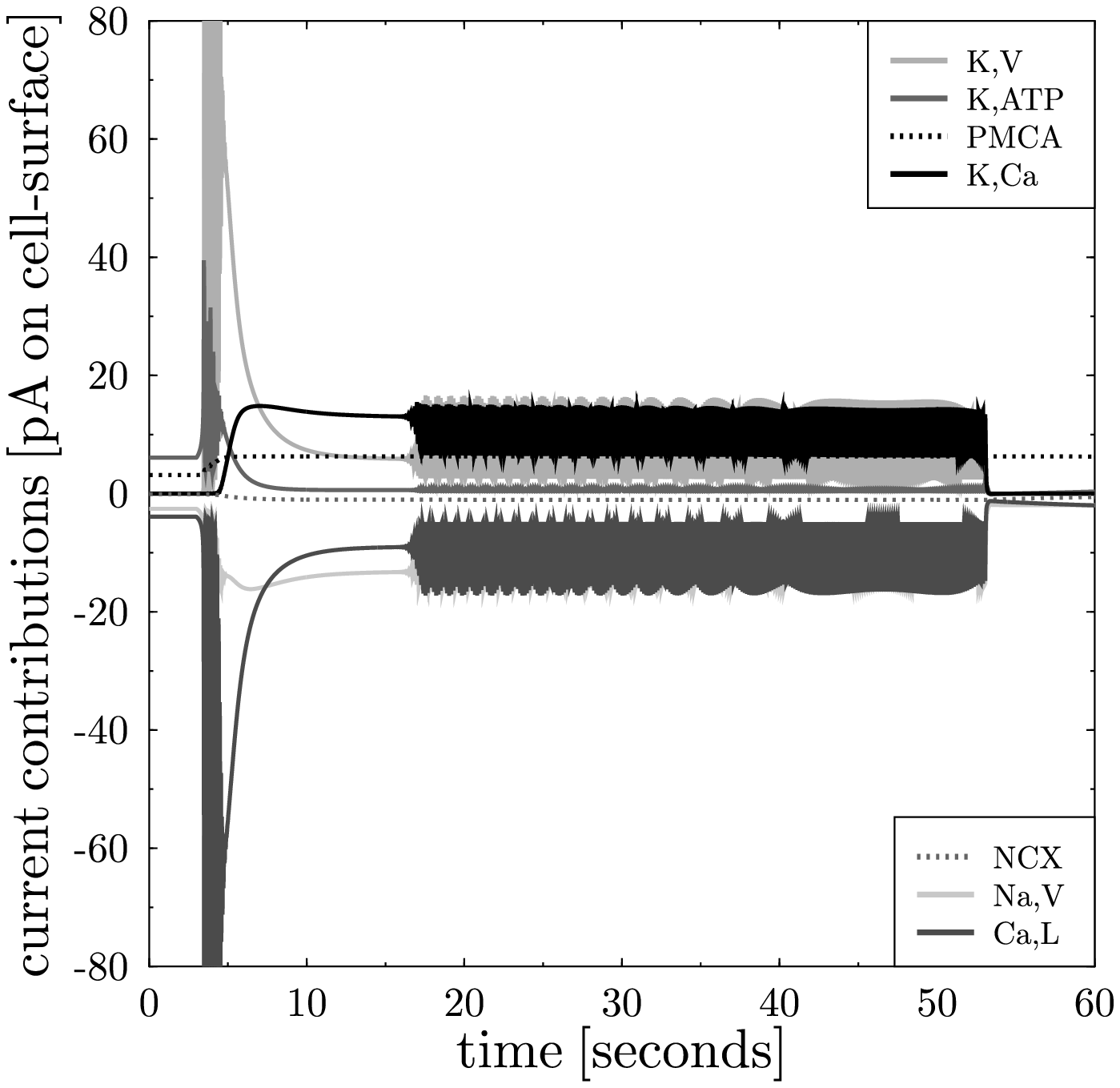}
\caption[]{\sf {\bf K,Ca is the main oscillator at large glucose levels:}
The current contributions 
of K,Ca and K,V to
uninterrupted bursting for glucose
levels of $12$ (left) and $30mM$ (right) are shown.
While at $12mM$ the main contributors to the 
oscillations are K,V (light grey full line) and Ca,L (grey full line), 
these are K,Ca (black full line) and Ca,L at $30mM$ glucose. 
Note also that the contribution
of K,ATP-channels is strongly reduced at supra-high
glucose levels.}
\label{bc0221_12_30}
\end{center}
\end{figure}
Instead of K,V and Ca,L it is now K,Ca and Ca,L
which are swinging. 
This result suggests an additional role of K,Ca-channel
in the electrophysiology of the \bc: It defines
the electrical activity at supra-large glucose and
calcium concentrations.

\subsection{Inhibit SERCA-activity}

Thapsigargin is a SERCA-inhibitor. Its effect is to elevate the
intracellular baseline calcium concentration.
It was found that thapsigargin
can induce continuous firing in \bcs~\cite{kanno02}.
As the present simulation does not include the ER explicitly
SERCA-block is mimicked by induction of its effect on the
calcium baseline. To this end
the calcium leakage current in a cell with established resting state
is modified.

A reduction of the leakage current to
$20\%$, indeed, leads to an increased baseline calcium level. If then
glucose is turned to $\gamma=10mM$ continuous bursting is
found. When the leakage current is turned back to its resting
state value normal repeated bursting is recovered (see \ab{bc0225}).
\begin{figure}[ht!]
\begin{center}
\includegraphics[height=6.8cm]{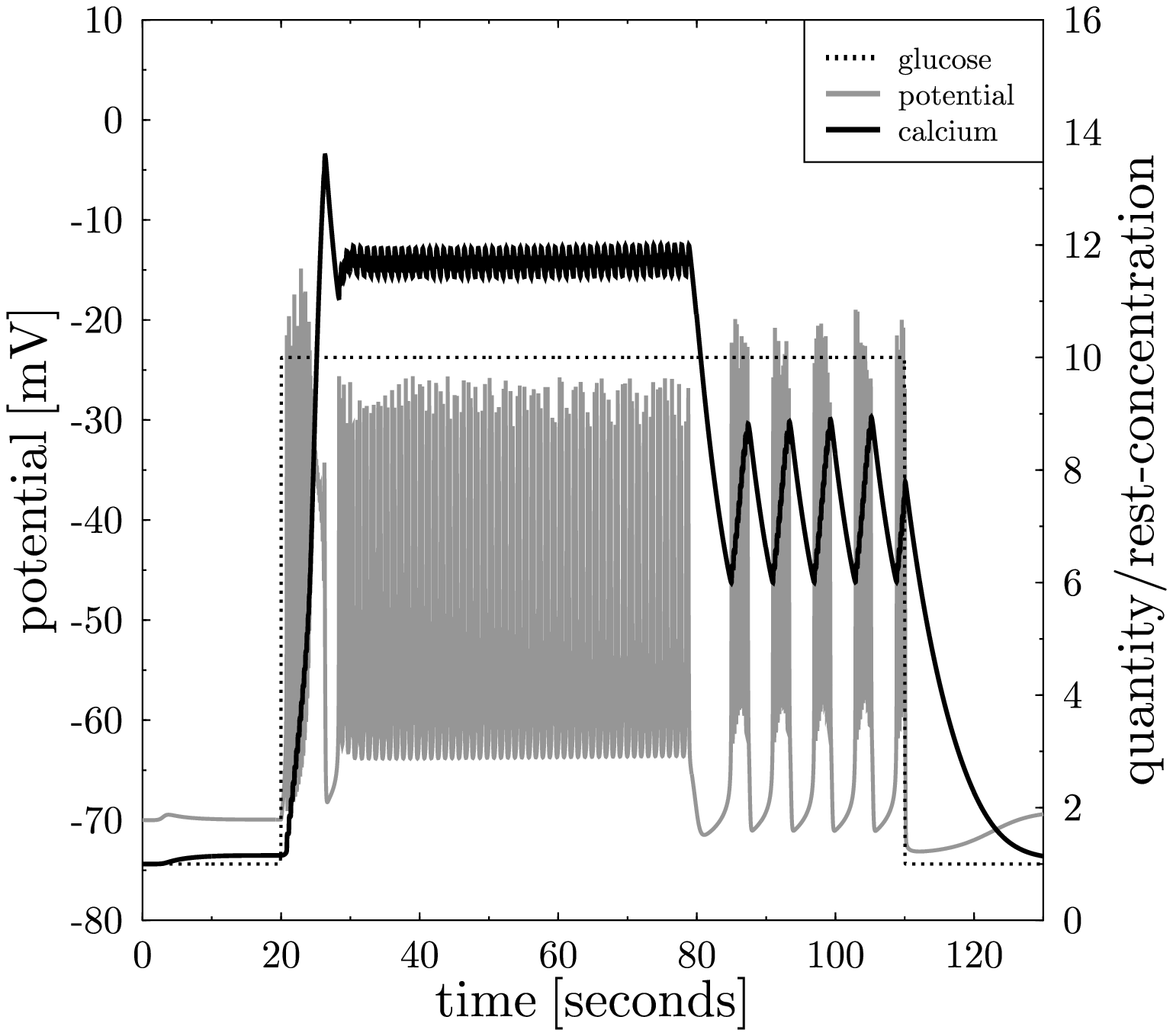}
\includegraphics[height=6.8cm]{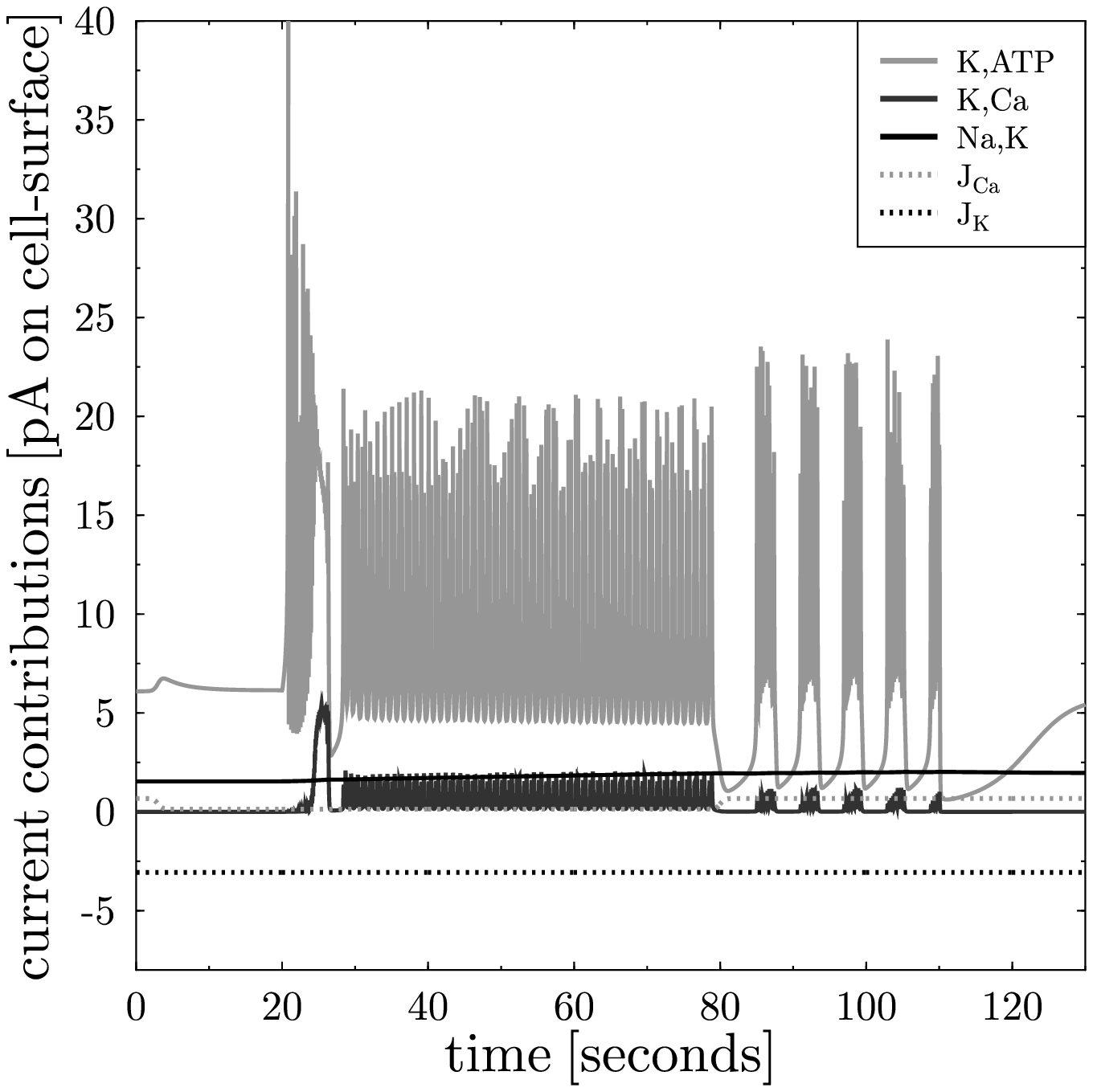}
\caption[]{\sf {\bf Reduction of leakage current to mimic SERCA-block:}
The calcium leakage current is
reduced to $20\%$ at $3 s$ to mimic SERCA-blocking
(right panel, grey dotted line). 
This increases the calcium baseline (left panel, black line). 
At $t=20 s$
glucose is turned to $\gamma=10mM$ (left panel, black dotted line), 
which induces continuous bursting (left panel, grey line). 
At $t=80 s$ the leakage current
is turned back to normal and repeated bursting
(as in \ab{bc0221_10})
is recovered. At $t= 110 s$ glucose is
turned back to its resting value.
The right panel shows all but K,V whole cell potassium currents.
Note that because of its stoichiometry
Na,K-potassium current has negative
sign and double absolute value with respect to
the shown curve (right panel, black full line).}
\label{bc0225}
\end{center}
\end{figure}
Continuous bursting is also found if the leakage current is
reduced after repeated bursting being already established
at high glucose levels. Thus, the sequence of stimulation events
is not important.

From \ab{bc0225} it can be read off that 
the K,Ca-current is increased during continuous
bursting as compared to repeated bursting.
In contrast, the K,ATP-current
is smaller during continuous bursting. 
This is in agreement with the results
found in \cite{kanno02}, Figure 4 therein. However, the
interpretation is different: The authors in \cite{kanno02}
suspected that the reduced consumption of ATP (during SERCA
block) leads to less inhibition of K,ATP-channels.
The simulation suggests
that the increased calcium baseline increases the necessary
PMCA pumping activity in order to establish a steady-state.
In fact, the consumed amount of ATP is more related to
the absolute calcium level in the cell than to the
number of blocked ATPases.
Thus, the 
present simulation suggests that altered calcium concentrations 
and membrane potentials during block of SERCA reduce
K,ATP-currents independent of the intracellular 
ATP-concentration. 

\subsection{Partial blocking of PMCA and NCX}

A very similar result (to \ab{bc0225}) is found when the PMCA-current
is reduced to $90\%$.
The calcium baseline is increased and at $10mM$ glucose
uninterrupted bursting is found.
Substantially stronger inhibition of PMCA-activity
suppresses bursting after an initial bursting event
(data not shown).

A reduction of the NCX-current to $80\%$ also leads
to uninterrupted bursting. However, the characteristics are
different. As before the calcium baseline is
increased. But, instead of an increased calcium level during
uninterrupted bursting, the calcium level is decreased
(data not shown). This points to two different mechanisms
both leading to uninterrupted bursting.

\subsection{Block K,ATP with tolbutamide}

Block of K,ATP-currents 
was investigated in patch-clamp experiments with \bcs~in 
intact islets \cite{kanno02}. {\it In silico} the application
of tolbutamide corresponds 
to a reduction of the K,ATP protein density $\rho_{\rm K,ATP}$ 
in a cell that has established its steady state in the presence
of normal K,ATP-presence.
A reduction of K,ATP-currents to $75\%$ leads to
a result almost identical to \ab{bc0225}. 
An increased calcium baseline is established.
This happens because the
reduced potassium outflow depolarises the cell which,
in turn, partially opens voltage-dependent calcium channels.
The new equilibrium state is, thus, established on a higher
level of calcium influx.

When glucose is increased to $10mM$ uninterrupted bursting is
induced. As in experiment \cite{kanno02} this is reversible:
After release of partial K,ATP-block
normal repeated bursting behaviour is reestablished.
This happens in the
simulation without any additional stimulation.
In \cite{kanno02} the \bc~was depolarised several
times in voltage clamp mode.

Note that an overexpression of K,ATP-channels also leads
to uninterrupted bursting despite the fact that the calcium
baseline is decreased (data not shown). 
Again this points to at least two
mechanisms for uninterrupted bursting. For even more
increased K,ATP-currents regular single firing is observed.

Interestingly, a total block of K,ATP-channels
immediately leads to continuous bursting 
even at the resting glucose concentration $\gamma_0=1mM$
(see \ab{bc0228}).
\begin{figure}[ht!]
\begin{center}
\includegraphics[height=6.8cm]{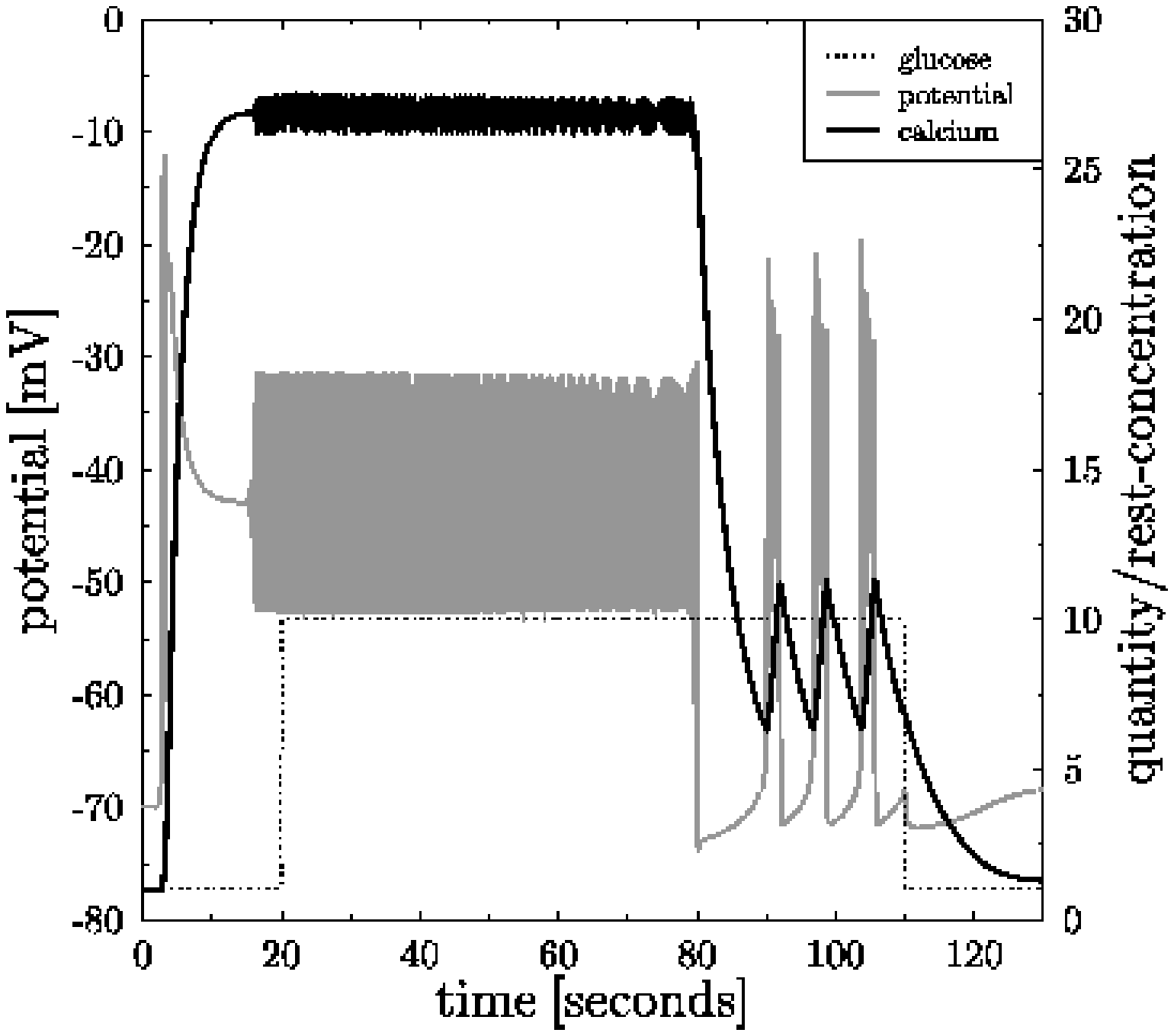}
\includegraphics[height=6.8cm]{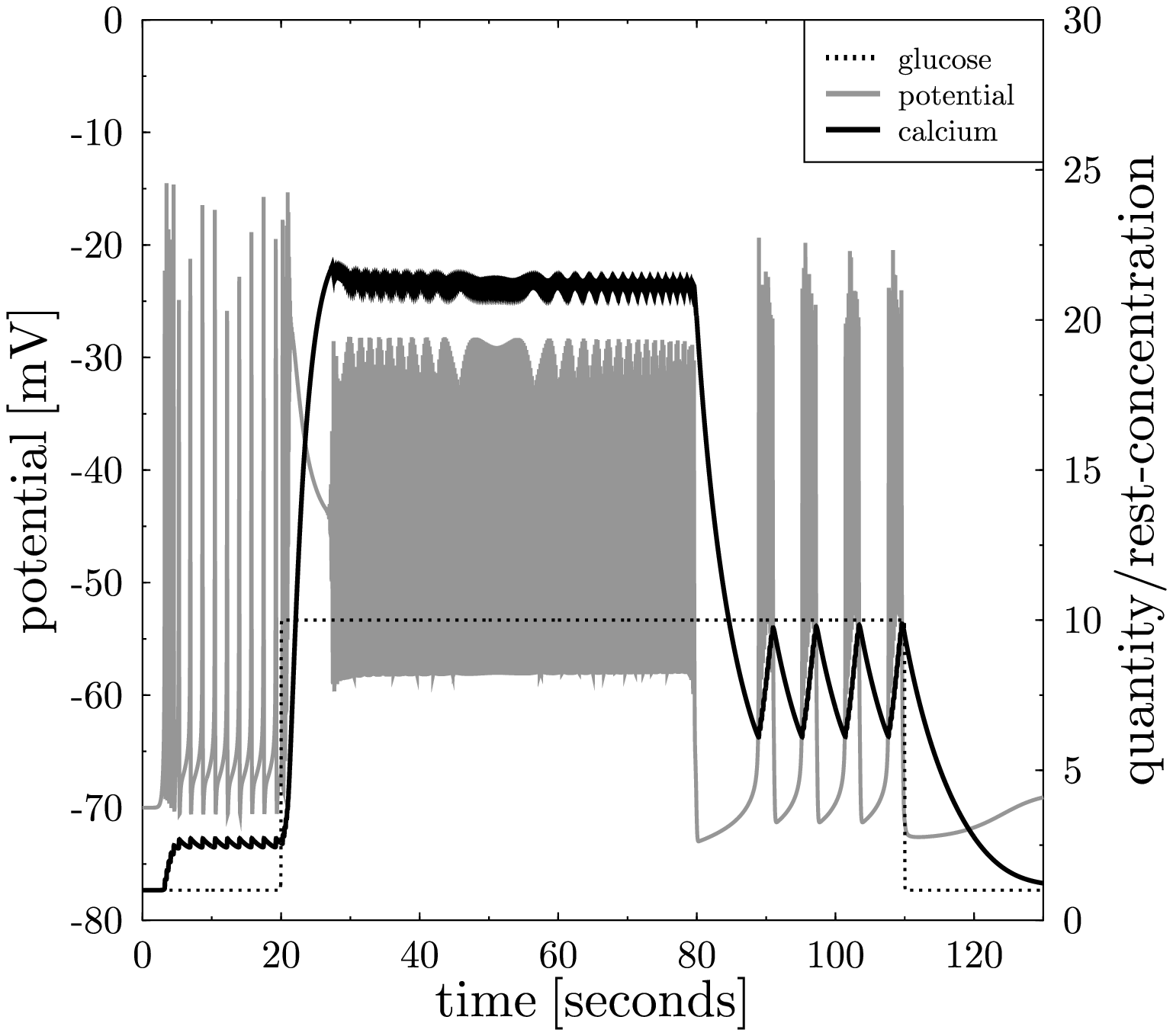}
\caption[]{\sf {\bf Total and partial block of K,ATP:}
K,ATP is fully blocked at $t=3 s$ 
(left panel). This induces strong
depolarisation, calcium inflow and continuous bursting. 
At $t=20 s$
glucose is turned to $\gamma=10mM$, which does not influence
bursting for lack of K,ATP.
At $t=80 s$ the block of K,ATP is released and repeated bursting
is recovered (as found in \ab{bc0221_10}). At $t= 110 s$ glucose is
turned back to its resting value.
The right panel shows the result of partial block
of K,ATP-channels to $50\%$ using the otherwise same protocol.
High frequency repeated bursting is found at resting glucose
levels.}
\label{bc0228}
\end{center}
\end{figure}
This supports that these channels
trigger bursting events by inhibition of potassium
outwards currents.
The observed independence of the increased glucose level
at $t=20 s$ is a direct consequence of the fact
that in the model K,ATP-channels are the only point of effect
of glucose. ATP-ADP ratios are not explicitly
monitored which would also change ATPase activities.

It is possible to stimulate \bcs~by partial block of K,ATP-currents.
Repeated bursting is induced at resting glucose
level (see \ab{bc0228}, right panel, for a $50\%$-block). 
When glucose is increased at $t=20s$ repeated bursting turns into
uninterrupted bursting.

\subsection{Modulation of voltage-gated calcium channels}

While overexpression of LVA calcium channels Ca,T leads
to similar results as shown in \ab{bc0225},
block or knock-out of LVA channels has
only minor effect on the bursting behaviour. Inhibition
only weakly modifies the calcium baseline before stimulation
with glucose. The amplitude
of spikes of the membrane potential is slightly increased
and the reached calcium level is slightly shifted
(data not shown). Thus, according to the present simulations,
LVA have to be considered unimportant for \bc-bursting.

However, this is a weak statement because for a different
set of densities (with a larger fraction of LVA channels)
the role of LVA calcium channels is
more pronounced. In particular, the inactivation of Ca,T
(note that HVA Ca,L-type calcium channels inactivate
only slowly) might play a role for interruption of bursting
events. In these simulations uninterrupted bursting
is found for weak and for total inhibition of LVA-currents.
The present simulation does not yet allow
to make precise statements on the importance of
LVA channels and on their inactivation.
However, in can be stated that if Ca,T-type channels
were expressed to a relevant amount in \bc, their inactivation
properties would have impact on the interruption
of bursting events. This is further supported by the
observation that
$70\%$ of the calcium current in whole-\bc-experiments 
is inactivated after $100ms$ \cite{goepel99},
which is impossible on the basis of Ca,L-channels that
show only little or no inactivation. 
It would be most interesting
to get data about the ratio of LVA and HVA calcium
channels in \bcs.

Moderate inhibited or strengthened Ca,L-currents also
lead to uninterrupted bursting combined with an
increased or lowered calcium baseline, respectively
(data not shown).
For stronger inhibition ($20\%$ remaining)
spiking activity disappears, which is consistent
with the dominant role of the Ca,L-current for \bc-oscillations.

\subsection{Modulation of sodium currents}

The sodium-potassium exchanger Na,K uses ATP to transport both
sodium and potassium against their electrochemical gradient.
Even though this membrane protein is only weakly active
in the resting state, shows little dynamics upon \bc~stimulation 
(variation of intracellular potassium concentration
is relatively small), and induces only small currents, the bursting
behaviour is sensitive to partial blocking of the exchanger.
Already at $70\%$ of the Na,K-current uninterrupted bursting
is found. 
This comes with a comparably small increase in the calcium
baseline.
Uninterrupted bursting is also found for increased
Na,K-currents ($130\%$ or more). Interestingly,
the calcium baseline is decreased in this scenario, and,
thus, cannot be at the origin of uninterrupted bursting.
Note that due to the weak dynamics of Na,K-currents
corresponding effects are not found in Na,K-knock-out experiments.

Similarly the voltage-dependent sodium channel Na,V is generally
not considered to play a major role in \bc~electrophysiology.
At a partial block to $80\%$ (or lower) 
uninterrupted bursting is found together with a slightly
reduced calcium baseline (see \ab{bc0233} left panel).
Uninterrupted bursting is also found if
the Na,V-currents are increased by a factor $1.3$ or
higher together with a slightly increased calcium baseline
(see \ab{bc0233} right panel). 
\begin{figure}[ht!]
\begin{center}
\includegraphics[height=6.8cm]{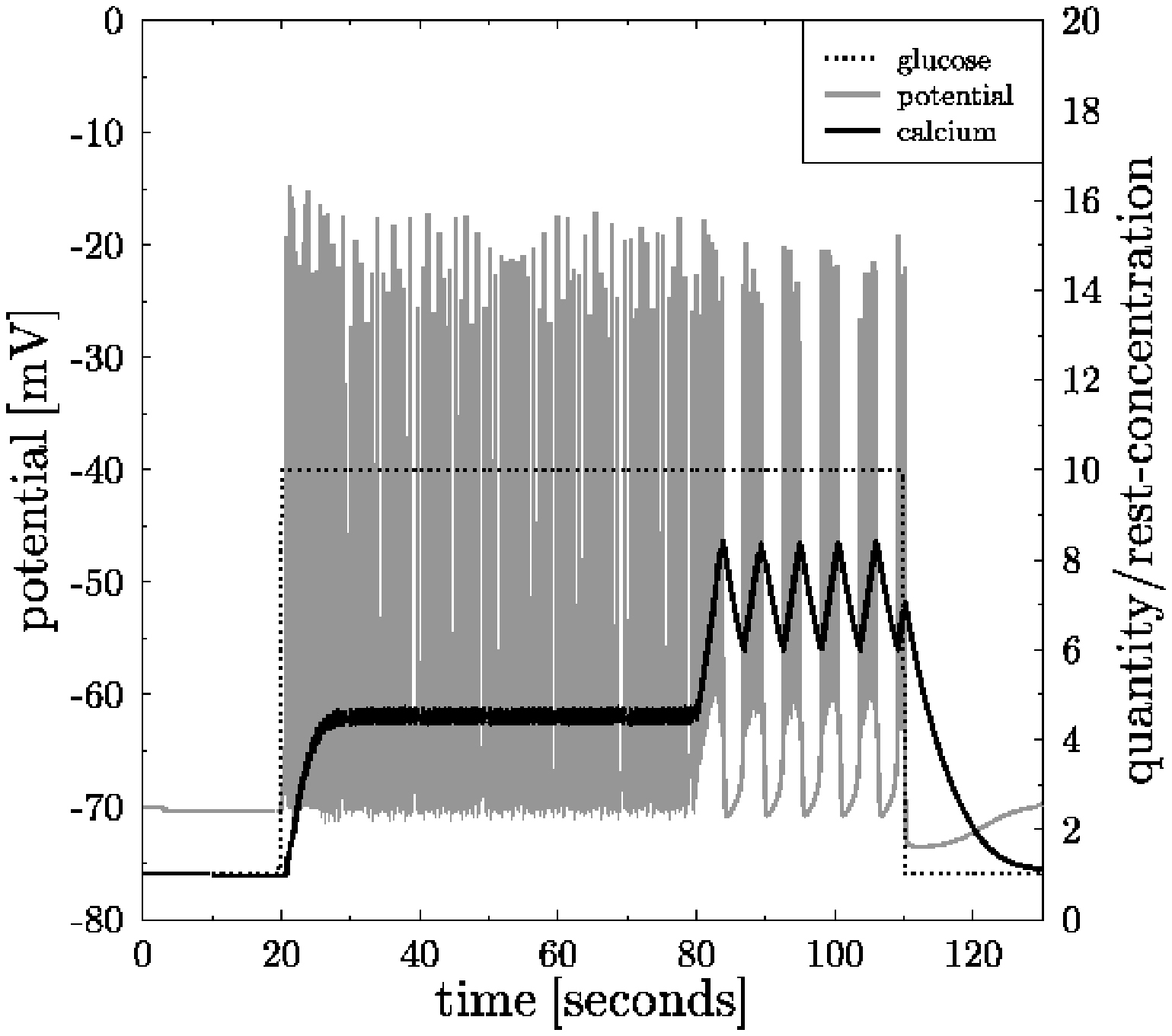}
\includegraphics[height=6.8cm]{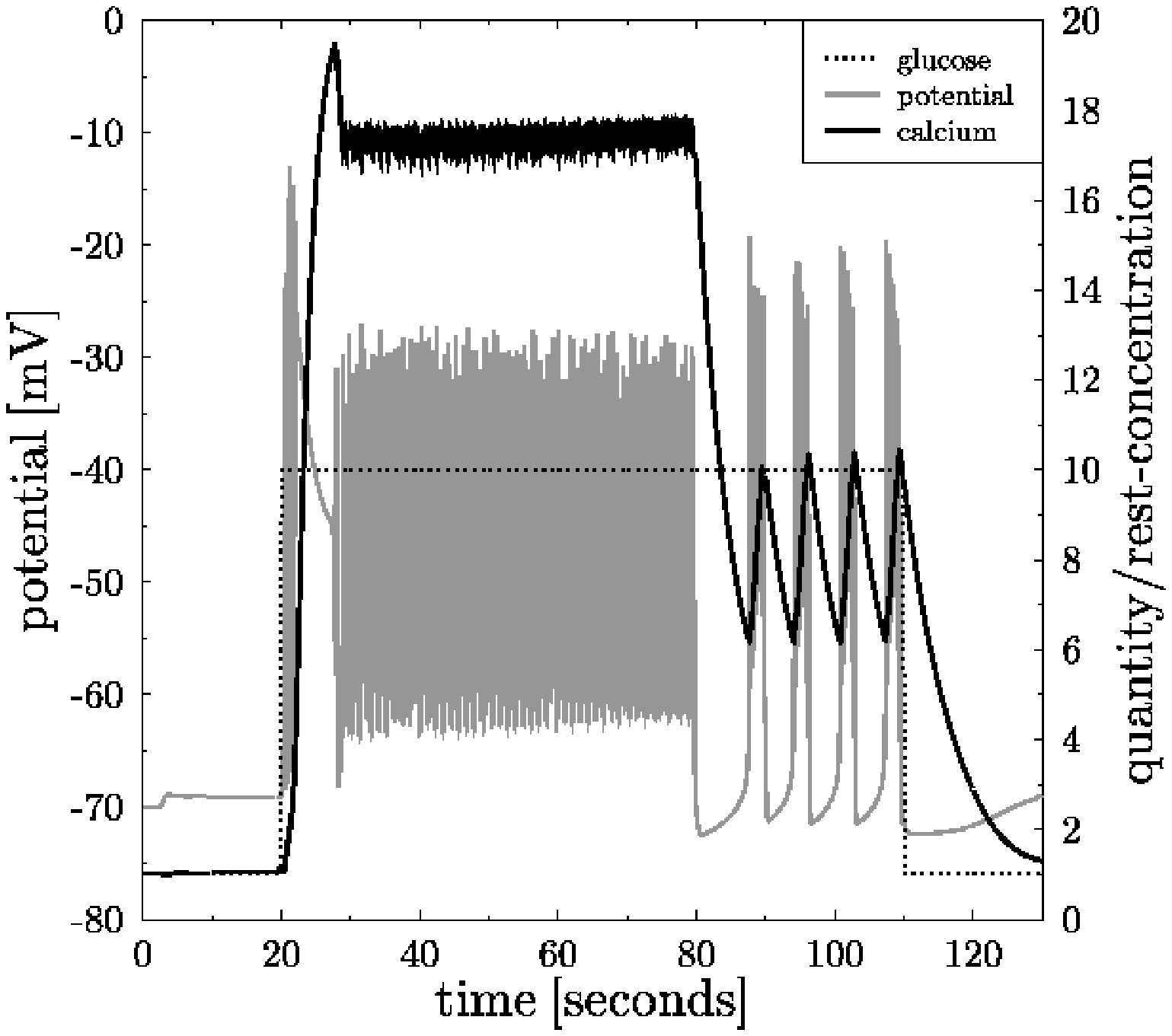}
\caption[]{\sf {\bf Modulating Na,V-currents:}
Na,V-currents are inhibited to $90\%$ (left panel) 
and increased to $130\%$ (right panel) using the same
protocol as in \abn{bc0225}{bc0228}.
The calcium baseline (full black line) is reduced and increased, respectively.
At $10mM$ glucose (see $t>20s$) 
uninterrupted bursting is found in both cases with
differing characteristics. While uninterrupted bursting
with increased calcium baseline is similar to \ab{bc0225},
amplitude and burst baseline are different when coming
together with a reduced calcium baseline. This points to different
mechanisms leading to uninterrupted bursting.}
\label{bc0233}
\end{center}
\end{figure}

Why is uninterrupted bursting found with reduced calcium
baseline when either Na,K-currents are increased or
Na,V-currents are reduced? Both, increased Na,K- and
reduced Na,V-currents polarise the cell because
intracellular sodium levels are reduced. This reduces
the calcium influx through voltage-gated calcium
channels, and, in turn, reduces the calcium baseline.
In the same way intracellular potassium levels 
are increased because of less potassium efflux.
This corresponds to a pre-stimulation in complete
analogy to the inhibition by K,ATP-channels.

Thus, we have to distinguish low and high calcium
uninterrupted bursting.
\bcs~become prone to oscillations either by
increased calcium baseline or by increased
potassium baseline. The equilibrium of both
is regulated by the sodium-currents.
Despite the sodium conducting proteins playing
an unspectacular role in \bc~activity they nevertheless
influence the electrophysiology in an unexpected strength.

\subsection{Modulation of potassium currents}

Inhibition of K,Ca-currents reduces the burst to silence ratio
and the bursting frequency. 
Even at currents of $10\%$ 
repeated bursting is still observed. 
Only when K,Ca-currents are totally blocked, bursting is
fully suppressed.
Increased K,Ca-currents
lead to increased burst to silence ratios and at $150\%$
to uninterrupted bursting (the characteristics correspond
to those as found in \ab{bc0233} left panel).
The calcium baseline and the membrane potential
are not visibly altered by modulation of the K,Ca-currents
(data not shown).

Increased K,V-currents lead to uninterrupted bursting.
A reduction of K,V-currents reduces the burst to silence ratio.
Surprisingly, 
even with totally blocked K,V-currents, oscillations are still
found with almost unaltered calcium baseline. 
The potassium current in the action potentials 
is then taken over by the K,Ca-currents.
This shows that K,Ca-proteins have the potential
to drive the oscillations during burst events --
similar to the scenario found for supra-large glucose
levels.

\subsection{Long-term stimulation}

Stimulation by increased glucose levels not only induce
repeated bursting, the \bc~is also shifted out of its
resting state. If stimulation by glucose is applied
for long durations, and assuming in a thought experiment
all other cell properties unaltered during that period,
the \bc~might
find a novel stable resting state which is in equilibrium
with the high glucose level. 
The simulation in \ab{bc0221_10} is repeated
with $10mM$ glucose applied for one hour.
Due to limited responsiveness of Na,K-exchangers
each burst event (involving Na,V-current oscillations)
increases the intracellular sodium level.
After half an hour sodium reaches a novel resting state 
at $170\%$ of its resting concentration adopted to the
ongoing stimulation (data not shown). 
Similarly the intracellular potassium concentration decreases
to a new equilibrium state.
Now, Na,K-activity and sodium inflow at each bursting
event get into equilibrium. 
Repeated bursting is still present at 
this higher sodium-state. Only the baseline
of bursts is increased by about $20mV$.


\section{Discussion}

{\it A bottom-up approach for the \bc~with quantitative ambitions}\\
A simulation tool for the electrophysiology of \bc~was
introduced. The model is strictly based on experimental
data for single membrane protein activity or
conductivity measurements. This is considered to
be a first step towards a quantitative modelling
of \bc~electrical activity. Previous modelling
work either structurally considered only those elements
necessary to induce repeated bursting or omitted
some essential protein properties.
All previous models relied on whole-cell experiments.
Thus, the present model can be considered to be the
first bottom-up approach of the \bc~starting at the molecular
level and predicting emerging properties on the
cellular scale. 

{\it Improvements with respect to other modelling approaches}\\
The present model, besides on relying on single protein
characteristics, includes several \bc-properties that
have been neglected in previous modelling approaches.
It was found that the back-reaction of the PMCA-currents
on the membrane potential are very important and strongly
alter bursting behaviour as found in \cite{chay83,sherman88}.
The inactivation properties (neglected in \cite{fridlyand03})
of the channels turn out to
be relevant for burst interruption and the stability
of oscillation. Finally, the Nernst-equation was used
to dynamically calculate the reversal potential (which was assumed
constant in \cite{chay83,sherman88,bertram04})
and turned out to substantially change the currents during burst
events.

{\it The endoplasmatic reticulum might change the dynamics}\\
Even though a block of processes related to 
the endoplasmatic reticulum does not affect \bc~oscillations
\cite{gilon99}, calcium-induced-calcium-release might well
modify the dynamics of raising calcium. These modifications
might lead to adaptations of protein densities. 
Also the simple minded
cell geometry has impact on the protein densities.
Thus, even though the resulting whole-cell currents are realistic,
the protein densities used in the simulations
(see \ta{densities} in the supplement)
are not considered as a quantitative prediction in the
present stage of the model. 
Note that calcium pumps in the endoplasmatic reticulum will
in first approximation act as an additional sink of free
calcium which, in the present model, 
is effectively included in a large concentration
of calcium binding sites.

{\it Characteristics of glucose-dependent \bc-stimulation are reproduced}\\
Despite the necessity for further improvement, the
agreement with current experimental data is rather
convincing and strengthens the relevance of the
found results. 
The calcium response is of similar strength, and
whole cell currents are in the right range \cite{goepel99}.
Bursting is found in the same regime
of glucose levels as in experiment \cite{beauvois06,kanno02}.
The glucose threshold for \bc~electrical
response is quantitatively reproduced to be around
$5mM$ \cite{beauvois06}. 
The bursting frequency is in the range of experimental data. 
Uninterrupted bursting is found for
large glucose levels as in experiment \cite{kanno02}.
Thus, it can be stated that the simulation respects 
the relevant characteristics of repeated bursting
in \bcs.

{\it The range of glucose leading to bursting is too small}\\
A more detailed analysis reveals quantitative deviations
between simulation and experiment. The most relevant
one concerns the range of glucose for which
repeated bursting is found. While the simulation
predicts $9$ to $11mM$ in experiment the lower
limit is at $7mM$ \cite{beauvois06} and the upper
limit at $15mM$ \cite{kanno02}. This discrepancy
might be explained by a structural difference between
the experimental set-up and the simulation: The
simulation models single \bcs, while both cited
experiments investigate \bc~in a context of 
intact islets. Repeated bursting is difficult
to be induced in isolated \bcs, which can explain
why the range of glucose is so small in the simulations.
It has to be evaluated in future research 
whether the gap-junction-mediated
contact to other \bcs~can enlarge the glucose regime
for repeated bursting.

{\it A novel model of \bc-bursting}\\
The fact that the main properties of \bc-electrophysiology
are correctly described by the simulation
encourages to draw a novel picture of how \bc-bursting
emerges. This picture is presented as a sequence of events:
\begin{itemize}
\item {\bf Glucose inhibits K,ATP:} Increased glucose only gives
a short pulse of inhibition of K,ATP-currents. This slightly
depolarises the \bc~which is further amplified by
induced sodium and calcium entry through voltage-gated channels
(Na,V; Ca,L; and Ca,T).
\item {\bf K,ATP-currents increase again:}
Even though the opening of K,ATP-channels is inhibited the
current increases upon cell depolarisation. According to
\gl{K,ATP} (see supplement) 
inhibition is over-compensated by the increased
electrochemical gradient.
\item {\bf Ca,L and Na,V cause electrical spiking:}
If the stimulus is over-critical a spike of cell
depolarisation is induced by voltage-dependent calcium
and sodium currents.
\item {\bf Oscillations are induced by delayed K,V-response:}
The potassium current repolarises the cell, but this
happens only with a delay $\tau_{\rm K,V}$. It is well
known that delayed response can lead to stable fast
oscillations.
\item {\bf Calcium rises during the burst:}
During a bursting event calcium and sodium is constantly entering
the cell while potassium is leaving it. While the effect
on potassium and sodium is small, calcium rises because of
a strong electrochemical gradient and the small resting
concentration. The speed of calcium increase is controlled
by the intracellular calcium buffer, and is much slower
than the oscillations.
\item {\bf The burst is interrupted by a concerted action of 
Nernst-reversal potential, PMCA, NCX, K,Ca and Ca,T:}
During bursts calcium levels increase and
the calcium reversal potential is reduced. The latter
inhibits the calcium-currents, while the first increase
repolarising PMCA-, NCX-, and K,Ca-currents, where
K,Ca-currents play a minor role at normal calcium
levels \cite{barrett82}. At a critical
calcium concentration and reversal potential oscillations
are suppressed.
The interruption of bursts is facilitated by the 
inactivation dynamics of LVA calcium channels.
\item {\bf Uninterrupted bursting is driven by K,Ca:}
As also K,Ca acts with a certain delay this membrane
protein has totally analogous potential to drive oscillations
with Na,V- and Ca,L-mediated depolarising currents. According
to the measured activity characteristics \cite{barrett82} 
oscillations are possible at large calcium concentrations only
and can even drive uninterrupted bursting. 
\item {\bf Repolarisation after burst events is governed by PMCA, NCX, and Na,K:}
In complete analogy to neuronal repolarisation, ATPases and
exchangers are needed to reestablish the resting concentrations.
However, sodium and potassium are continuously changing their
level during long stimulation.
\end{itemize}
Note that this model differs from older models in several
aspects which are discussed in more detail in the following.

{\it Sodium currents are important regulators of \bc-electrophysiology}\\
Na,V-currents directly participate in spikes and
bursting events. The activity
of the \bc~is responsive to all properties of the
Na,V-channel. Thereby the half activation potential
$V_{\rm Na,V}$ has the
greatest influence. Na,V-currents facilitate onset
of action potentials due to their fast dynamics
(the reaction time is the fastest of all considered
channels), participate
in the interplay with delayed potassium currents to
induce oscillations, and also facilitates the end
of action potentials by relatively fast inactivation.
The interpretation of a tightly controlled interplay
of Na,V with other membrane proteins is supported
by the observation that uninterrupted bursting
is observed for partially blocked {\it and} 
overexpressed Na,V-proteins. The latter bursting
is characterised by high calcium, the former
by high potassium. It turns out that Na,V-currents
regulate the equilibrium between both ions
in \bc-stimulation.
According to the model results sodium channels are
more relevant for the \bc~electrophysiology than
previously expected.

{\it Long-term stimulation gives rise to a third time scale}\\
The {\it in silico} experiment of long term stimulation
of \bcs~with high levels of glucose has revealed
three time scales of \bc-electrophysiology:
\begin{enumerate}
\item Delay of K,V or K,Ca currents with respect to
calcium and sodium induced depolarisations
($10-100ms$),
\item slow increase in intracellular calcium level
due to intense buffering which keeps the fraction
of free calcium low ($10s$), and
\item even slower modifications of intracellular
sodium and potassium levels ($10min$).
\end{enumerate}
The third time scale should be observable in experiment.

{\it Glucose only induces a tiny inhibition of K,ATP-currents}\\
In the model the only effect of increased glucose levels
is inhibition of K,ATP-currents. However, the inhibition
turns out to be a tiny effect which quickly is hidden
by the increasd K,ATP-currents in response to cell
depolarisation. This opens the possibility of alternative
effects of glucose via ATP-metabolism. Several channels
and transporters rely on ATP-concentrations and may
alter their activity in response to changed glucose
levels. Some might well initiate bursting as already
established for K,ATP. This might explain the recently observed
insulin secretion and calcium dynamis in islets of
mice lacking K,ATP-activity \cite{szollosi07}.

{\it Bursts are interrupted by NCX, PMCA, Ca,T and the dynamic reversal potential}\\
Interruption of bursting only weakly rely on K,Ca-currents
because of small activity at physiological calcium levels.
Instead, interruption is strongly
related to NCX and PMCA activity. Mostly neglected in former
research is the impact of a dynamic calcium reversal potential
(see \gl{nernst}) which turns out to change
the calcium currents by about $35\%$ during a burst.
The inactivation dynamics
of LVA calcium channels Ca,T also supports burst
interruption. The latter statement depends on the
ratio of LVA- and HVA-channels in \bc, which would
be important to be measured. In summary, burst interruption
seems not to rely on a single type of membrane proteins but on
a concerted action of various factors.

{\it Calcium driven potassium channels K,Ca can drive oscillations}\\
For the first time
it is postulated that K,Ca is an active part of the
bursting behaviour as well.
Thereby the unknown K,Ca-time scale of activation
is relevant: If the K,Ca adapts faster to the
membrane potential (thus using a smaller $\tau_{\rm K,Ca}$)
the currents get out of phase and uninterrupted
bursting is destroyed. Therefore, the measurement of
$\tau_{\rm K,Ca}$ is most important to learn about
this predicted functionality of K,Ca-currents.
According to the simulation results K,Ca acquire a major
role for very large calcium concentrations where
it takes over the role of K,V in oscillations.

{\it Falsification of the predictions of the simulation}\\
On one hand an improved version of the simulation model
has to be developed in order to see whether the predictions
of the simulations and the attributed roles of various
membrane proteins for \bc~electrical activity are robust.
The improved model shall include the impact of intracellular
calcium stores and explicitly model the dynamics of
ATP. On the other hand some of the {\it in silico} experiments
were not executed in real systems. The simulation tool might
serve as a guide for the design of very focused and conclusive 
experiments. This is not restricted to the proposed experiments.
On the contrary the author welcomes the challenge of the simulation
model with further suggestions for experimental set-ups.


{\bf Acknowledgments:}
Michael Meyer-Hermann thanks Michele Solimena for pointing me to the
subject. MMH is indebted to Michele Solimena and Marc Thilo Figge
for regular and fruitful discussion and for revising the manuscript.
FIAS is supported by the ALTANA AG. 
MMH is supported by the EC-NEST-project MAMOCELL.

\bibliographystyle{biophysj}
\bibliography{refs_MMH}

\begin{thebibliography}{54}
\providecommand{\natexlab}[1]{#1}

\bibitem{chay83}
Chay, T., and J.~Keizer. 1983.
\newblock {Minimal model for membrane oscillations in the pancreatic
  $\beta$-cell}.
\newblock \emph{Biophys. J.} 42:181--190.

\bibitem{fridlyand03}
Fridlyand, L., N.~Tamarina, and L.~Philipson. 2003.
\newblock {Modeling of Ca2+ flux in pancreatic beta-cells: role of the plasma
  membrane and intracellular stores}.
\newblock \emph{Am. J. Phyisol. Endocrinol. Metab.} 285:E138--E154.

\bibitem{sherman88}
Sherman, A., J.~Rinzel, and J.~Keizer. 1988.
\newblock {Emergence of organized bursting in clusters of pancreatic beta-cells
  by channel sharing}.
\newblock \emph{Biophys. J.} 54:411--425.

\bibitem{chay97}
Chay, T.~R. 1997.
\newblock {Effects of extracellular calcium on electrical bursting and
  intracellular and luminal calcium oscillations in insulin secreting
  pancreatic $\beta$-cells}.
\newblock \emph{Biophys. J.} 73:1673--1688.

\bibitem{bertram04}
Bertram, R., and A.~Sherman. 2004.
\newblock {A calcium-based phantom bursting model for pancreatic islets}.
\newblock \emph{Bull. Math. Biol.} 66:1313--1344.

\bibitem{keizer89}
Keizer, J., and G.~Magnus. 1989.
\newblock {ATP-sensitive potassium channel and bursting in the pancreatic beta
  cell. A theoretical study}.
\newblock \emph{Biophys. J.} 56:229--242.

\bibitem{kanno02}
Kanno, T., P.~Rorsman, and S.~Gopel. 2002.
\newblock {Glucose-dependent regulation of rhythmic action potential firing in
  pancreatic beta-cells by K(ATP)-channel modulation}.
\newblock \emph{J. Physiol.} 545.2:501--507.

\bibitem{ashcroft84}
Ashcroft, F., D.~Harrison, and S.~Ashcroft. 1984.
\newblock {Glucose induces closure of single potassium channels in isolated rat
  pancreatic $\beta$-cells}.
\newblock \emph{Nature}. 312:446--448.

\bibitem{cook84}
Cook, D., and C.~Hales. 1984.
\newblock {Intracellular ATP directly blocks K+ channels in pancreatic
  B-cells}.
\newblock \emph{Nature}. 311:271--273.

\bibitem{erler04}
Erler, F., M.~Meyer-Hermann, and G.~Soff. 2004.
\newblock {A quantitative model for presynaptic free calcium dynamics during
  different stimulation protocols}.
\newblock \emph{Neurocomputing}. 61:169--191.

\bibitem{hille92}
Hille, B. 1992.
\newblock {Ionic Channels of Excitable Membranes}, 2nd Ed. Sinauer Ass. Inc.

\bibitem{fridlyand05}
Fridlyand, L., L.~Ma, and L.~Philipson. 2005.
\newblock {Adenine nucleotide regulation in pancreatic beta-cells: modeling of
  ATP/ADP-Ca2+ interaction}.
\newblock \emph{Am. J. Physiol. Endocrinol. Metab.} 289:E839--E848.

\bibitem{hopkins92}
Hopkins, W., S.~Fatherazi, B.~Peter-Riesch, B.~Corkey, and D.~Cook. 1992.
\newblock {Two sites for adenine-nucleotide regulation of ATP-sensitive
  potassium channels in mouse pancreatic beta-cells and HIT cells}.
\newblock \emph{J. Membr. Biol.} 129:287--295.

\bibitem{detimary98}
Detimary, P., P.~Gilon, and J.~Henquin. 1998.
\newblock {Interplay between cytoplasmic Ca2+ and the ATP/ADP ratio: a feedback
  control mechanism in mouse pancreatic islets}.
\newblock \emph{Biochem. J.} 333:269--274.

\bibitem{ammala91}
\"Amm\"al\"a, C., O.~Larsson, P.-O. Berggren, K.~Bokvist, L.~Juntti-Berggren,
  H.~Kindmark, and P.~Rorsman. 1991.
\newblock {Inositol trisphosphate-dependent periodic activation of a
  Ca2+-activated K+ conductance in glucose-stimulated pancreatic beta-cells}.
\newblock \emph{Nature}. 353:849--852.

\bibitem{liu95}
Liu, Y., E.~Grapengiesser, E.~Gylfe, and B.~Hellman. 1995.
\newblock {Glucose induces oscillations of cytoplasmic Ca2+, r2+ and Ba2+ in
  pancreatic beta-cells without participation of the thapsigargin-sensitive
  store}.
\newblock \emph{Cell Calcium}. 18:165--173.

\bibitem{gilon99}
Gilon, P., A.~Arredouani, P.~Gailly, J.~Gromada, and J.-C. Henquin. 1999.
\newblock {Uptake and release of Ca2+ by the endoplasmatic reticulum contribute
  to the oscillations of the cytosolic Ca2+ concentration triggered by Ca2+
  influx in the electrically excitable pancreatic B-cell}.
\newblock \emph{J. Biol. Chem.} 274:20197--20205.

\bibitem{beauvois06}
Beauvois, M., C.~Merezak, J.~Jonas, M.~Ravier, J.~Henquin, and P.~Gilon. 2006.
\newblock {Glucose-induced mixed [Ca2+]c oscillations in mouse beta-cells are
  controlled by the membrane potential and the SERCA3 Ca2+-ATPase of the
  endoplasmic reticulum}.
\newblock \emph{Am. J. Physiol. Cell Physiol.} 209:C1503--1511.

\bibitem{goepel99}
Goepel, S., T.~Kanno, S.~Barg, J.~Galvanovskis, and P.~Rorsman. 1999.
\newblock {Voltage-gated and resting membrane currents recorded from B-cells in
  intact mouse pancreatic islets}.
\newblock \emph{J. Physiol.} 521:717--728.

\bibitem{barrett82}
Barrett, J., K.~L. Magleby, and B.~Pallotta. 1982.
\newblock {Properties of single calcium-activated potassium channels in
  cultured rat muscle}.
\newblock \emph{J. Physiol.} 331:211--230.

\bibitem{szollosi07}
Szollosi, A., M.~Nenquin, L.~Aguilar-Bryan, J.~Bryan, and J.~Henquin. 2007.
\newblock {Glucose stimulates Ca2+ influx and insulin secretion in 2-week-old
  beta-cells lacking ATP-sensitive K+ channels}.
\newblock \emph{J. Biol. Chem.} 282:1747--1756.

\bibitem{chapman83}
Chapman, J., J.~E., and J.~Kootsey. 1983.
\newblock {Electrical and biochemical properties of an enzyme model of the
  sodium pump}.
\newblock \emph{J. Membr. Biol.} 74:139--153.

\bibitem{owada99}
Owada, S., O.~Larsson, P.~Arkhammar, A.~Katz, A.~Chibalin, P.-O. Berggren, and
  A.~Bertorello. 1999.
\newblock {Glucose Decreases Na+,K+-TPase activity in Pancreatic beta-Cells }.
\newblock \emph{J. Biol. Chem.} 274:2000--2008.

\bibitem{maixent93}
Maixent, J., and I.~Berrebi-Betrand. 1993.
\newblock {Turnover rates of the canine cardiac Na,K-ATPases}.
\newblock \emph{FEBS Lett.} 330:297--301.

\bibitem{bezanilla87}
Bezanilla, F. 1987.
\newblock {Single sodium channels from the squid giant axon}.
\newblock \emph{Biophys. J.} 52:1087--1090.

\bibitem{plant88_PflugArch}
Plant, T. 1988.
\newblock {Na+ currents in cultured mouse pancreatic B-cells}.
\newblock \emph{Pflug. Arch.} 411:429--435.

\bibitem{hiriart88}
Hiriart, M., and D.~Matteson. 1988.
\newblock {Na channels and two types of Ca channels in rat pancreatic B cells
  identified with the reverse hemolytic plaque assay}.
\newblock \emph{J. Gen. Physiol.} 91:617--639.

\bibitem{dunne01}
Dunne, M., C.~Ammala, S.~Straub, and G.~Sharp. 2001.
\newblock {Electrophysiology of the pancreatic b-cell and the mechanisms of
  inhibition of insulin release}.
\newblock \emph{In} {Handbook of Physiology: The Endocrine Pancreas and
  Regulation of Metabolism}. A.~Jefferson, L.S.;~Cherrington, editor. Oxford
  University Press.
\newblock 79--123.

\bibitem{magee95}
Magee, J., and D.~Johnston. 1995.
\newblock {Characterization of single voltage-gated Na+ and Ca2+ channels in
  apical dendrites of rat CA1 pyramidal neurons}.
\newblock \emph{J. Physiol.} 487:67--90.

\bibitem{juhaszova00}
Juhaszova, M., P.~Church, M.~P. Blaustein, and E.~F. Stanley. 2000.
\newblock {Location of calcium transporters at presynaptic terminals}.
\newblock \emph{Eur. J. Neurosci.} 12:839--846.

\bibitem{blaustein99}
Blaustein, M., and W.~Lederer. 1999.
\newblock {Sodium/calcium exchange: its physiological implications}.
\newblock \emph{Physiol. Rev.} 79:763--854.

\bibitem{delprincipe99}
del Principe, F., M.~Egger, and E.~Niggli. 1999.
\newblock {Calcium signalling in cardiac muscle: refractoriness revealed by
  coherent activation}.
\newblock \emph{Nature Cell Biol.} 1:323--329.

\bibitem{gall99}
Gall, D., and I.~Susa. 1999.
\newblock {Effect of Na/Ca exchange on plateau fraction and [Ca]i in models for
  bursting in pancreatic beta-cells}.
\newblock \emph{Biophys. J.} 77:45--53.

\bibitem{dipolo02}
DiPolo, R., and L.~Beaug\'e. 2002.
\newblock {Ionic ligand interactions with the intracellular loop of the
  sodium-calcium exchanger. Modulation by ATP}.
\newblock \emph{Prog. Biophys. Mol. Biol.} 80:43--67.

\bibitem{varadi95}
V\'aradi, A., E.~Moln\'ar, and J.~Ashcroft. 1995.
\newblock {Characterisation of endoplasmic reticulum and plasma membrane
  Ca2+-TPases in pancreatic beta-cells and in islets of Langerhans}.
\newblock \emph{Biochim. Biophys. Acta}. 1236:119--127.

\bibitem{caride01_cell_calcium}
Caride, A., A.~Filoteo, A.~Penheiter, K.~Paszty, A.~Enyedi, and J.~Penniston.
  2001.
\newblock {Delayed activation of the plasma membrand calcium pump by a sudden
  increase in Ca2+: fast pumps reside in fast cells}.
\newblock \emph{Cell Calcium}. 30:49--57.

\bibitem{carafoli91}
Carafoli, E. 1991.
\newblock {The calcium pumping ATPase of the plasma membrane}.
\newblock \emph{Annu. Rev. Physiol.} 53:531--547.

\bibitem{caride01_JBiolChem}
Caride, A., A.~Penheiter, A.~Filoteo, Z.~Bajzer, A.~Enyedi, and J.~Penniston.
  2001.
\newblock {The plasma membrane calcium pump displays memory of past calcium
  spikes. Differences between isoforms 2b and 4b}.
\newblock \emph{J. Biol. Chem.} 276:39797--39804.

\bibitem{graupner05}
Graupner, M., F.~Erler, and M.~Meyer-Hermann. 2005.
\newblock {A theory of Plasma Membrane Calcium Pump stimulation and activity}.
\newblock \emph{J. Biol. Phys.} 31:183--206.

\bibitem{elwess97}
Elwess, N., A.~Filoteo, A.~Enyedi, and J.~Penniston. 1997.
\newblock {Plasma membrane Ca2+ pump isoforms 2a and 2b are unusually
  responsive to calmodulin and Ca2+}.
\newblock \emph{J. Biol. Chem.} 272:17981--17986.

\bibitem{rorsman03}
Rorsman, P., and E.~Renstrom. 2003.
\newblock {Insulin granule dynamics in pancreatic beta cells}.
\newblock \emph{Diabetologia}. 46:1029--1045.

\bibitem{kelly91}
Kelly, R., R.~Sutton, and F.~Ashcroft. 1991.
\newblock {Voltage-activated calcium and potassium currents in human pancreatic
  beta-cells}.
\newblock \emph{J. Physiol.} 443:175--192.

\bibitem{vinet99}
Vinet, R., and F.~F. Vargas. 1999.
\newblock {L- and T-type voltage-gated Ca2+ currents in adrenal medulla
  endothelial cells}.
\newblock \emph{AJP - Heart and Circulatory Physiology}. 276:H1313--H1322.

\bibitem{smith93}
Smith, P., F.~Ashcroft, and C.~Fewtrell. 1993.
\newblock {Permeation and gating properties of the L-type calcium channel in
  mouse pancreatic beta cells}.
\newblock \emph{J. Gen. Physiol.} 101:767--797.

\bibitem{plant88}
Plant, T. 1988.
\newblock {Properties and calcium-dependent inactivation of calcium currents in
  cultured mouse pancreatic B-cells}.
\newblock \emph{J. Physiol.} 404:731--747.

\bibitem{villarroya99}
Villarroya, M. e.~a. 1999.
\newblock {Voltage inactivation of Ca2+ entry and secretion associated with N-
  and P/Q-type but not L-type Ca2+ channels of bovine chromaffin cells}.
\newblock \emph{J. Physiol.} 516:421--432.

\bibitem{hoefer97}
Hoefer, G., K.~Hohenthanner, W.~Baumgartner, K.~Groschner, N.~Klugbauer,
  F.~Hofmann, and C.~Romanin. 1997.
\newblock {Intracellular Ca2+ incativates L-type Ca2+ channels with a Hill
  coefficient of 1 and an inhibition constant of 4$\mu$M by reducing channel's
  open probability}.
\newblock \emph{Biophys. J.} 73:1857--1865.

\bibitem{atwater78}
Atwater, I., B.~Ribalet, and E.~Rojas. 1978.
\newblock {Cyclic changes in potential and resistance of the beta-cell membrane
  induced by glucose in islets of Langerhans from mouse}.
\newblock \emph{J. Physiol.} 278:117--139.

\bibitem{solimena06}
Solimena, M. 2006.
\newblock {private communication}.

\bibitem{straub04}
Straub, S., G.~Shanmugan, and G.~Sharp. 2004.
\newblock {Stimulation of Insulin Release by Glucose Is ssociated With an
  Increase in the Number of Docked Granules in the "s-Cells of Rat Pancreatic
  Islets}.
\newblock \emph{Diabetes}. 53:3179--3183.

\bibitem{gentet00}
Gentet, L., G.~Stuart, and J.~Clements. 2000.
\newblock {Direct Measurement of Specific Membrane Capacitance in Neurons}.
\newblock \emph{Biophys. J.} 79:314--320.

\bibitem{goepel04}
Goepel, S., Q.~Zhang, L.~Eliasson, X.~Ma, J.~Galvanovskis, T.~Kanno, A.~Salehi,
  and P.~Rorsman. 2004.
\newblock {Capacitance measurements of exocytosis in mouse pancreatic alpha-,
  beta- and delta-cells within intact islets of Langerhans}.
\newblock \emph{J. Physiol.} 556:711--726.

\bibitem{smith88}
Smith, P. 1988.
\newblock {Electrophysiology of B-cells from pancreatic islets of Langerhans.
  Ph.D thesis}. University of East Anglia.

\bibitem{wasserman81}
Wasserman, W., and L.~Smith. 1981.
\newblock {Calmodulin triggers the resumption of meiosis in amphibian oocytes}.
\newblock \emph{J. Cell. Biol.} 89:389--394.

\end{thebibliography}


\clearpage

\section{Supplementary material}

\subsection{Conductivity of membrane proteins}
\label{opening}

The (electrical charge) currents entering \gl{ion-dynamics} and \gl{voltage} are themselves
dynamic quantities. Channels are gated in dependence on
the membrane potential, on the presence of ATP, and on glucose- and ion-
concentration. The resulting current depends
on the electrochemical gradient over the membrane. So every
membrane protein has its own characteristics and the
conductivity of passive or active ion flow depends on
different other parameters. In the following the different
membrane proteins (as listed in \ta{proteins})
\begin{table}[!ht]
\begin{tabular}{|l|l|} \hline
Na,K & sodium-potassium exchanger (Na$^+$ outwards, K$^+$ inwards)\\
Na,V & voltage-gated sodium channels\\
NCX & sodium-calcium exchanger (Na$^+$ inwards, Ca$^{2+}$ outwards)\\
PMCA & Plasma-membrane-calcium-ATPase (Ca$^{2+}$ outwards)\\
K,ATP & ATP-driven (and glucose-dependent) potassium channels\\
K,V & voltage-gated potassium channels (also called delayed rectifier)\\
K,Ca & voltage- and calcium-gated potassium channel\\
Ca,L & L-type voltage-dependent calcium channels\\
Ca,T & T-type voltage-dependent calcium channels\\
\hline
\end{tabular}
\caption[]{\sf {\bf Acronyms of transmembrane proteins:}
Acronyms for the different plasma membrane proteins
contributing to ion flow through the membrane are defined. 
{\it Inwards} and {\it outwards}
specifies in which direction the ion flows, i.e.~in or out of the
cell.
\label{proteins}}
\end{table}
and their gating/activity properties 
are described in more detail. Currents through pores
are generally approximated by a linear voltage-current relation,
which infers currents proportional to the
electrical potential gradient (terms with $V-\overline{V}$ with
$\overline{V}$ the reversal potential of the ion under consideration).
For some channels and for extreme depolarisations
there might be deviations from this approximation.

The activation dynamics of active carrier proteins 
relies on the Hill-function 
\be{Hill}
H(x,x_h,n)&=& \frac{x^n}{x^n+x_h^n}
\quad,
\ee
which can be derived from
chemical kinetics.
$x$ denotes some concentration, $x_h$ is the corresponding
concentration of half activation, and $n$ is the Hill-coefficient.
Large $n$ make the switch between inactive and 
active states steeper.

The asymptotic state of activation and inactivation is
described by two functions of sigmoidal shape
\be{sigmoidal}
\sigma_{\rm act}(x,x_{\rm h},\kappa)
&=&
\frac{1}{1+\exp\{(x_{\rm h}-x)/\kappa\}}
\nonumber\\
\sigma_{\rm inact}\left(x,x_{\rm h},\lambda\right)
&=&
\frac{1}{1+\exp\{(x-x_{\rm h})/\lambda\}}
\quad,
\ee
Such sigmoidal functions are frequently used in literature
and the parameters of these functions are determined
in experiments.
$x_h$ is the concentration of half-maximum value
and $\kappa$ and $\lambda$ regulate the activation and
inactivation steepness, respectively.
Inactivation is a property of ion channels which will
be handled in complete analogy to activation. In a first approximation
ATPases driving ions against their chemical gradient are
inactivated. Only the
channels Na,V; K,V; K,Ca; Ca,L; and Ca,T 
(see \tas{pars_NK}{pars_Ca}) are subject to inactivation.


\subsubsection{The sodium-potassium exchanger Na,K}

The sodium potassium exchanger transports sodium out of the cell
and potassium into the cell. Both transport processes are against
the chemical gradient, so that this process has to consume energy
which is provided by ATP. The current is modelled as 
\be{Na,K}
I_{\rm Na,K} 
&=& 
\overline{I_{\rm Na,K}}
\left(1 - H(K,K_{\rm Na,K},n_{\rm Na,K})\right) H(N,\tilde{K}_{\rm Na,K},\tilde{n}_{\rm Na,K})
\quad,
\ee
where $\overline{I_{\rm Na,K}}$ denotes the
maximum current. The quantities used for the
Hill-function are explained after \gl{Hill}. The first term
describes a decreasing activity with increasing potassium concentration.
The second one an increasing activity with increasing sodium
concentration (see \cite{chapman83}). 
This approach follows approximations frequently adopted in
other models: 
The adaption of Na,K-activity to sodium and potassium levels is 
assumed quick such that the steady state approximation is
justified. 
A direct impact of glucose on Na,K-activity in \bcs~\cite{owada99} 
and a voltage-dependence of Na,K-activity \cite{chapman83} is neglected.
As ATP-dynamics are not modelled, the ATP-level does not
influence Na,K-activity.

{\bf Activation dynamics:}
The used parameters follow those found in \cite{chapman83}
to fit experimental activation dynamics. Good agreement
is found with
$n_{\rm Na,K}=2$ and $K_{\rm Na,K}=33mM$ as well as
$\tilde{n}_{\rm Na,K}=2$ and $\tilde{K}_{\rm Na,K}=20mM$
for dependence on potassium and sodium, respectively.

{\bf Stoichiometry:}
The stoichiometry of the exchanger is 3:2, thus 3 sodium ions are
exchanged against 2 potassium ions entering the cell,
i.e.~$\alpha_{\rm Na,K}=1.5$ in \gln{ion-dynamics}{voltage}.
As both ions have the valence 1 this corresponds to
$1$ positive charges leaving the cell for 2
potassium ions entering the cell. 
Thus $\overline{I_{\rm Na,K}}>0$
enters the different equation with different factors:
$+1$ for $V$ \gl{voltage}, $-2$ for $K$, and $+3$ for $N$ 
\gl{ion-dynamics}.

{\bf Turnover rate:}
The turn over rate of one ATPase
measured as the consumption of ATP-molecules per
second is 
$\overline{I_{\rm Na,K}}=200 ATP/sec\approx 0.00003pA$
in canine cardiac neurons \cite{maixent93},
where the last equality assumes one exchange of 3 sodium
and 2 potassium per ATP.


\subsubsection{The voltage gated sodium channel Na,V}

The voltage gated sodium channel opens when the cell
is depolarised. The sodium ions then follow their
electrochemical gradient to flow into the cell. The
current is therefore composed of two factors, one
describing the open-probability of the channel, the
other describing the flow through the open channel:
\be{Na,V}
I_{\rm Na,V}  & =& 
h_{\rm Na,V} 
g_{\rm Na,V} 
\overline{g_{\rm Na,V}}
\left(V-\overline{V_{\rm Na}}\right)
\quad.
\ee
$\overline{V_{\rm Na}}$ is the sodium reversal potential, and
$\overline{g_{\rm Na,V}}$ is the maximum conductivity.
The open and inactivation probabilities $g,h_{\rm Na,V}$ 
are assumed to exponentially reach their asymptotic values
and, thus, follow the kinetic equations 
\be{g_Na,V}
\frac{dg_{\rm Na,V}}{dt}
&=&
\frac{\sigma_{\rm act}(V,V_{\rm Na,V},\kappa_{\rm Na,V})-g_{\rm Na,V}}{\tau_{\rm Na,V}}
\nonumber\\
\frac{dh_{\rm Na,V}}{dt}
&=&
\frac{\sigma_{\rm inact}(V,W_{\rm Na,V},\lambda_{\rm Na,V})-h_{\rm Na,V}}{\theta_{\rm Na,V}}
\quad.
\ee
The parameters are explained after \gl{sigmoidal}.
It is $\tau_{\rm Na,V}$ that defines the time constant of adaptation
of the sodium channel (the speed of conformity change) in
response to a changed membrane potential, while $\theta_{\rm Na,V}$
determines the time scale of channel inactivation.

{\bf Conductivity:}
The single channel conductivity of the voltage-gated sodium
channel is $\overline{g_{\rm Na,V}}=14pS$ as found in the
squid axon \cite{bezanilla87}.
This value is consistent with findings in other
systems. 
Taking the results in mice \bcs~\cite{plant88_PflugArch}, 
one has a single channel current
of $1pA$ at $-30mV$ which (using $V_{\rm Na}=80mV$)
corresponds to $9pS$. 

{\bf Activation properties:}
Activation is observed between $-60mV$ and $-10mV$ \cite{plant88_PflugArch}. 
Thus, a realistic approximation for the activation dynamics is
$V_{\rm Na,V}\approx -35mV$ and $\kappa_{\rm Na,V}\approx 8mV$.

{\bf Inactivation properties:}
The role of voltage-gated sodium channels in insulin
secretion and in membrane depolarisation is controversial.
Depending on species, the half-maximal inactivation 
is found at $-109mV$ in mouse \bcs~\cite{plant88_PflugArch} 
(where $2.6mM$ calcium were used, and at less calcium the inactivation
point was shifted to even lower voltage)
and total inactivation is found at $-40mV$ \cite{hiriart88,dunne01}.
Thus, one may estimate $W_{\rm Na,V}\approx -100mV$ with a steepness
of $\lambda_{\rm Na,V}\approx 20mV$ (in order to get the observed total
suppression of inactivation at $-150mV$, and total inactivation at $-40mV$).
This very low half inactivation potential if fully
consistent with measurement in islet-\bcs~\cite{goepel99}
and suggests an unprobable
major role in the electrophysiology of \bcs~\cite{plant88_PflugArch},
which is not confirmed in the present simulations.

{\bf Time scales:}
According to the sodium currents as observed in squid giant axons
and to the classical description by Hodgkin-Huxley the time
constant of activation depends on the membrane potential
\cite{hille92}. This is
represented by 
\be{tau_Na,V}
\tau_{\rm Na,V}
&=&
\frac{c}{\exp\{(V-V_0)/a\}+\exp\{(V_0-V)/b\}}
\ee
with $c=11.5ms$, $a=40mV$, $b=50mV$, and $V_0=-70mV$.
Similar $\tau_{\rm Na,V}$ are found in
islet-\bcs~\cite{goepel99}.
%
%
Note that the same simulation results are found with
the constant average value of $\tau_{\rm Na,V}=3ms$.
Inactivation happens on a time scale of $\theta_{\rm Na,V}=4.6ms$
which was found by two pulse experiments \cite{hille92}.

{\bf Calcium permeability of sodium channels:}
Sodium channels in neurons are in general 
permeable for calcium ions. Therefore, Na,V-channels are explicitly blocked
with TTX in experiments to investigate the opening dynamics of voltage-gated
calcium channels in order to separate these from the
currents through sodium channels (see e.g.~\cite{magee95}).
Calcium permeability of Na,V is not considered in the present approach.


\subsubsection{The sodium-calcium exchanger NCX}

The sodium-calcium exchanger takes advantage of the electrochemical
gradient of sodium (pushing sodium into the cell)
to transport calcium ions against its electrochemical gradient.
This exchanger is important to reestablish the resting state calcium
concentration after an excitable calcium influx into the cell.
As in the case of the sodium-potassium exchanger a fast adaption
of the activity to the calcium concentration is assumed. The
current is derived from a corresponding Hill-function:
\be{NCX}
I_{\rm NCX}
&=& 
\overline{I_{\rm NCX}} H(C,C_{\rm NCX},n_{\rm NCX})
\quad.
\ee
Typical values for the single membrane protein
parameters can be found in \cite{erler04}:
$\overline{I_{\rm NCX}}=-0.0005pA$ \cite{juhaszova00} and
$C_{\rm NCX}=0.0022mM$ \cite{blaustein99}. The Hill coefficient
is assumed to be $n_{\rm NCX}=1$. This is
justified by the linear relationship between the observed current
and the calcium concentration as observed for low calcium concentrations
in cardiac cells \cite{delprincipe99}. However, the
NCX-Hill-coefficient is controversial \cite{gall99,fridlyand03}.

{\bf Stoichiometry:}
The stoichiometry of NCX is 3:1, i.e.~$\alpha_{\rm NCX}=3$
in most tissues \cite{blaustein99},
thus three sodium ions have to
enter the cell in order to expel one calcium ion. The net current
of NCX is one positive charge getting into the cell for every
expelled calcium ion. This infers that $\overline{I_{\rm NCX}}<0$.
$I_{\rm NCX}$ enters the different equations with 
factor $+1$ for the equation for the membrane
potential \gl{voltage}, a factor $-1$ in the calcium equation, and
a factor $\alpha_{\rm NCX}$ in the sodium equation \gl{ion-dynamics}.

{\bf Impact of sodium-concentration:}
Note that in the present description \gl{NCX} the sodium concentration
does not enter, even though NCX activity also depends on
the sodium concentration gradient. Even though the sodium dynamics
is modelled, it is 
assumed that the sodium concentration gradient is approximately
constant. This turns out to be a reasonable approximation. 
In contrast,
to neurons, in \bcs~peaks in the potential are 
dominantly generated by calcium and not by sodium.

{\bf Impact of ATP:}
A dependence of the NCX activity on intracellular ATP concentration 
has been observed \cite{dipolo02}. This might be included in improved
versions of this model that treat ATP dynamically.


\subsubsection{PMCA}

PMCA is an ATP-driven calcium pump extruding calcium
from the cell to the extracellular medium, which has been
characterised in \bcs~\cite{varadi95}. In a first attempt
the dependence on the ATP concentration is ignored and the
ATP-concentration is assumed to be large enough to make the pump
work optimally. Then the activity is mainly dependent on the
calcium concentration in the cell. A suitable modelling
approach is
\be{PMCA}
I_{\rm PMCA}
&=& 
\overline{I_{\rm PMCA}} H(C,C_{\rm PMCA},n_{\rm PMCA})
\quad.
\ee
The maximum current $\overline{I_{\rm PMCA}}$ is positive, 
as it carries calcium out of the cell. 
A delayed activation of PMCA by increased
calcium has been observed \cite{caride01_cell_calcium}
which is neglected in the present model (using the
steady state approximation).

{\bf Stoichiometry:}
One calcium ion is transported per used ATP molecule
leading to the stoichiometry of $\alpha_{\rm PMCA}=1$
\cite{carafoli91}. This is surprising in view of the
Hill-coefficient which was determined to be
$n_{\rm PMCA}=2$ \cite{caride01_JBiolChem}, which
would suggest a corresponding stoichiometry
(see also the discussion in \cite{graupner05}).

{\bf Turnover rate:}
The turnover rate of single PMCA is in the order of 30 Hertz
\cite{juhaszova00} corresponding to an activity rate of $k_a=0.03/ms$
(this value was also used in \cite{sherman88}).
This turnover rate can be translated into an electrical
current: Every pumping event corresponds
to a flow of $2$ electrical charges $e$ leading to
$\overline{I_{\rm PMCA}}=z_{\rm Ca} e k_a= 60\cdot 1.6\cdot 10^{-19} C/s
\approx 10^{-17} A = 10^{-5} pA$.

{\bf Half activation concentration:}
Typical values for the half activation calcium
concentration are $C_{\rm PMCA}=0.1\mu M$ 
(see e.g.~\cite{elwess97} Figure 3).
Note that the values of half activation are found
to depend on the calmodulin concentration and the
isoform under consideration \cite{elwess97} which
is neglected in the present approach. The 4b-isoform
has a slightly larger 
half activation concentration of $0.16\mu M$
\cite{elwess97}.

{\bf ATP consumption:}
If modelling ATP concentration dynamics PMCA has to be
considered as a major consumer of ATP. Note that up to
$40\%$ of the ATP consumed to pump calcium out of the cell
is used by SERCA-ATPase, thus, not in the plasma membrane
but at the endoplasmatic reticulum \cite{detimary98}.


\subsubsection{ATP-driven potassium channel K,ATP}

The ATP-driven potassium-selective channel 
(also called G-channel \cite{ashcroft84})
uses ATP to open the channel
and to release potassium from the cell. This process is,
however, also regulated by the presence of glucose which
is a source for ATP-metabolism in mitochondria. At normal
glucose levels the channel has a resting state open probability,
which is counteracted by the activity of the sodium-potassium
exchanger. 
Note that about
50\% of the potassium current is mediated by K,ATP
channels \cite{kanno02}.
At increased levels of glucose the opening of this
channel is inhibited. The single protein current
is modeled by (see also \gl{Na,V}):
\be{K,ATP}
I_{\rm K,ATP}
&=& 
(1-g_{\rm K,ATP}) \overline{g_{\rm K,ATP}}
\left(V-\overline{V_{\rm K}}\right)
\quad,
\ee
with
\be{g_K,ATP}
\frac{dg_{\rm K,ATP}}{dt}
&=&
\frac{\sigma_{\rm act}(\gamma,\gamma_{\rm K,ATP},\kappa_{\rm K,ATP}) - g_{\rm K,ATP}}
{\tau_{\rm K,ATP}}
\quad.
\ee
Here, $\gamma$ is the glucose concentration. All other parameters
are in full analogy to the voltage-gated sodium channel Na,V.

{\bf Ohmic conductivity and calcium dependence:}
In the resting state the current is dominantly flowing through this
channel \cite{dunne01} which explains why the resting potential
is near of the potassium reversal potential.
It is controversial whether
the current-voltage relation is ohmic (\cite{dunne01} versus
\cite{cook84,ashcroft84}). The latter authors show an ohmic
relation in the relevant range of membrane potentials which
justifies the present model 
(see \cite{cook84} Fig.~2, \cite{ashcroft84} Fig.~1b).
The opening frequency is observed to depend neither on
the membrane potential nor on the calcium level
\cite{dunne01,cook84} which justifies the present approximation
\gln{K,ATP}{g_K,ATP}. 

{\bf Conductivity:}
The conductivity of single open channels is
in the order of $50pS$ for relevant membrane 
potentials \cite{ashcroft84} where also 
the dependence on the glucose level was investigated.
This is consistent with the finding that
with increasing depolarisation of the cell increasing outflow
of potassium is observed with a single channel conductivity of 
$\overline{g_{\rm K,ATP}}=54pS$ \cite{cook84} (this value is
used throughout all present simulations). 
An approximately linear current-voltage-relationship 
is found \cite{cook84}.

{\bf Activation time scale:}
The steady state approximation was used in two other models 
\cite{chay97,bertram04}. This points to a relatively short
$\tau_{\rm K,ATP}$. However, in
an effective modelling approach 
(ignoring ATP concentrations)
the time constant does not represent 
the time needed for the channel to adapt to a novel
ATP-concentration but the time needed for the full metabolism
process from glucose through ATP generation up to the changed
open probability. 
The whole process from increased glucose levels up to exocytosis
of insulin carrying granules take 1 minute \cite{rorsman03},
which defines an upper limit for $\tau_{\rm K,ATP}$.
Thus, $\tau_{\rm K,ATP}$ is chosen between this time and
the time scale

{\bf Glucose-driven deactivation:}
As K,ATP provides the dominant part of potassium current 
in resting state it is assumed that the 
half-deactivation-concentration of K,ATP is
around the resting level of glucose ($\gamma_{\rm K,ATP}=1.2mM$).
The steepness of the dependence on glucose is
estimated as $\kappa_{\rm K,ATP}=6mM$.

{\bf Modelling ATP-dynamics:}
The present simulation might be improved by inclusion
of ATP-dynamics as done in \cite{keizer89,fridlyand03}.
The functional relationship is well supported by experiment
\cite{hopkins92}. This will be subject of further research.

{\bf Other ATP-driven potassium channels}
Other, novel, ATP-activated potassium channels have been observed in mice and
humans, however, they differ in characteristics. They are still
poorly defined but they are also believed to be involved in
calcium oscillations.
One additional channel found in human \bcs~has a single-channel
potassium conductivity of $30pS$ \cite{dunne01}.


\subsubsection{The voltage-gated potassium channel K,V}

The voltage gated potassium channel (also called delayed rectifier
potassium channel) increases the potassium
efflux from the cell once the cell is depolarised.
The potassium ions follow their electrochemical gradient.
The model for this channel is in complete analogy
to the voltage-gated sodium channel \gl{Na,V}:
\be{K,V}
I_{\rm K,V}  
&=& 
h_{\rm K,V} g_{\rm K,V} \overline{g_{\rm K,V}}
\left(V-\overline{V_{\rm K}}\right)
\quad,
\ee
with
\be{g_K,V}
\frac{dg_{\rm K,V}}{dt}
&=&
\frac{\sigma_{\rm act}(V,V_{\rm K,V},\kappa_{\rm K,V})-g_{\rm K,V}}{\tau_{\rm K,V}}
\nonumber\\
\frac{dh_{\rm K,V}}{dt}
&=&
\frac{\sigma_{\rm inact}(V,W_{\rm K,V},\lambda_{\rm K,V})-h_{\rm K,V}}{\theta_{\rm K,V}}
\quad.
\ee
This is largely analogous to other models of
\bc~bursting \cite{chay97,fridlyand03,bertram04}.
However, the activation characteristics are not
consistent in these models and inactivation is
not considered
(\cite{chay97} includes a different inactivating
channel called $I_{\rm fast}$ therein).
The present model is based on
single protein properties excusively 
derived from experiment.

{\bf Conductivity:}
Open channels have a single channel conductivity of
$\overline{g_{\rm K,V}}=10pS$ \cite{dunne01}.

{\bf Voltage-dependent activation:}
K,V-channels in human \bcs~were activated with a $400ms$
depolarisations to different membrane potentials \cite{kelly91}. Then the
membrane potential was reduced to -50mV and the current
after this reset was monitored. The authors find a sigmoidal
shape, centered at $V_{\rm K,V}=1mV$ and width
$\kappa_{\rm K,V}=8.5mV$. 

{\bf Time scale of activation:}
The activation time constant is assumed to depend on
the potential in \cite{sherman88} according to
\be{tau_K,V}
\tau_{\rm K,V}
&=&
\frac{c}{\exp\{(V-\overline{V_{\rm K}})/a\}+\exp\{-(V-\overline{V_{\rm K}})/b\}}
\ee
with $c=60ms$, $\overline{V_{\rm K}}=-75mV$, $a=65mV$, and $b=20mV$.
This function is in agreement with experiments where the
time constant was found to vary between $8$ and $37ms$ 
\cite{kelly91}. 

{\bf Slow channel inactivation:}
For long depolarisations inactivation of the channels
is observed.
The characteristics are $W_{\rm K,V}=-25mV$ and width
$\lambda_{\rm K,V}=7.3mV$ \cite{kelly91}.
However, inactivation is not observed within $400ms$ so
that we must conclude that $\theta_{\rm K,V}>400ms$.
The value of $400ms$ is used in all simulations.


\subsubsection{Calcium-gated potassium channel K,Ca}
\label{K,Ca-opening}
The calcium-gated potassium channel opens in dependence
on the membrane potential and the intracellular calcium level.
The dependence of the open probability on calcium
is assumed to be in steady state and 
described by a Hill-function. Voltage-gating is 
described by on Ohm-like approach with dynamic gating
(see also \gl{Na,V}).
\be{K,Ca}
I_{\rm K,Ca}
&=&
g_{\rm K,Ca} \overline{g_{\rm K,Ca}}
\left(V-\overline{V_{\rm K}}\right)
H(C,C_{\rm K,Ca},n_{\rm K,Ca})
\quad,
\ee
with
\be{g_K,Ca}
\frac{dg_{\rm K,Ca}}{dt}
&=&
\frac{\sigma_{\rm act}(V,V_{\rm K,Ca},\kappa_{\rm K,Ca})-g_{\rm K,Ca}}{\tau_{\rm K,Ca}}
\ee
All parameters are in analogy to other proteins described before.
Other models that include detailed opening dynamics of K,Ca
do not exist. The models \cite{chay97,fridlyand03,bertram04}
differ from the present one by very large Hill-coefficients
(between $3$ and $5$) in a single Hill-function. 
As this is the first model of the K,Ca-opening dynamics
it is explained in more detail.

{\bf Ohmic conductivity:}
The calcium-gated potassium channel is rapidly activated
upon depolarisation and show a linear current-voltage
relation \cite{dunne01}. This justifies usage of an
Ohm's law current.

{\bf Conductivity:}
The conductivity of single-channels 
$\overline{g_{\rm K,Ca}}=220pS$ is rather large
\cite{dunne01}. This is also confirmed
in rat muscle cells with values of up to $300pS$
depending on the temperature \cite{barrett82}. 

{\bf Opening dynamics:}
Experiments with rat muscle cells \cite{barrett82}
show a dynamic activation in dependence
on the membrane potential and of the calcium concentration
(see also \cite{dunne01}). 
A sigmoidal opening probability is found for 
constant membrane potential in dependence of different
calcium concentrations (see \cite{barrett82} Fig.~6),
but also for constant calcium
with variable membrane potential (see \cite{barrett82} Fig.~8).

In a first approximation the opening dynamics is 
modelled by a product of a sigmoidal function and a Hill-function
as shown in \gln{K,Ca}{g_K,Ca}.
However, this approach turns out to be in contradiction to
the data in \cite{barrett82}.
In order to reproduce these measurements
more accurately, a dependence of the half activation
calcium concentration $C_{\rm K,Ca}(V)$ on the voltage
has to be assumed (see \ab{C_K_Ca_figure}):
\be{dynamic_half_C}
\frac{dC_{\rm K,Ca}}{dt}
&=&
\frac{C^\infty_{\rm K,Ca}(V)-C_{\rm K,Ca}}{\tau_{\rm K,Ca}}
\quad,
\nonumber\\
\mbox{\rm with}\qquad
C^\infty_{\rm K,Ca}(V)
&=&
\exp\left\{\frac{a-V}{b}\right\}
\ee
with $a=45mV$ and $b=30mV$, and the same time delay than
for the direct voltage-dependence (assuming related
mechanisms). 
\begin{figure}[ht!]
\begin{center}
\includegraphics[height=6cm]{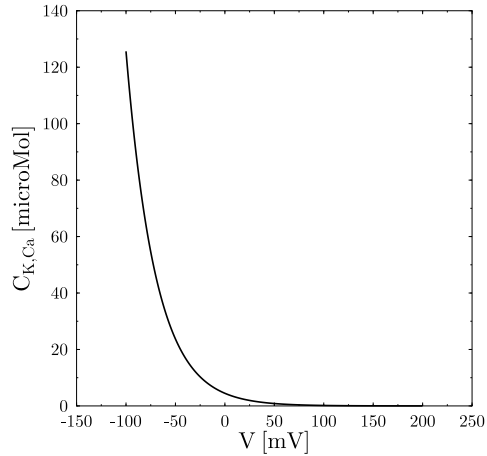}
\caption[]{\sf {\bf Voltage-dependence of half activation:}
The steady-state voltage-dynamics of the half activation calcium
concentration $C^\infty_{\rm K,Ca}$
following \gl{dynamic_half_C} with
$a=45mV$ and $b=30mV$. Note that the dynamics
is relevant at physiological membrane potentials.}
\label{C_K_Ca_figure}
\end{center}
\end{figure}
This is combined with a
Hill-coefficient of $n_{\rm K,Ca}=2$ and the sigmoidal
function for the voltage-dependent activation
with $V_{\rm K,Ca}=-40mV$ and $\kappa_{\rm K,Ca}=25mV$. 
We then get the result shown in
\ab{K_Ca_var_half}.
\begin{figure}[ht!]
\begin{center}
\includegraphics[height=6.8cm]{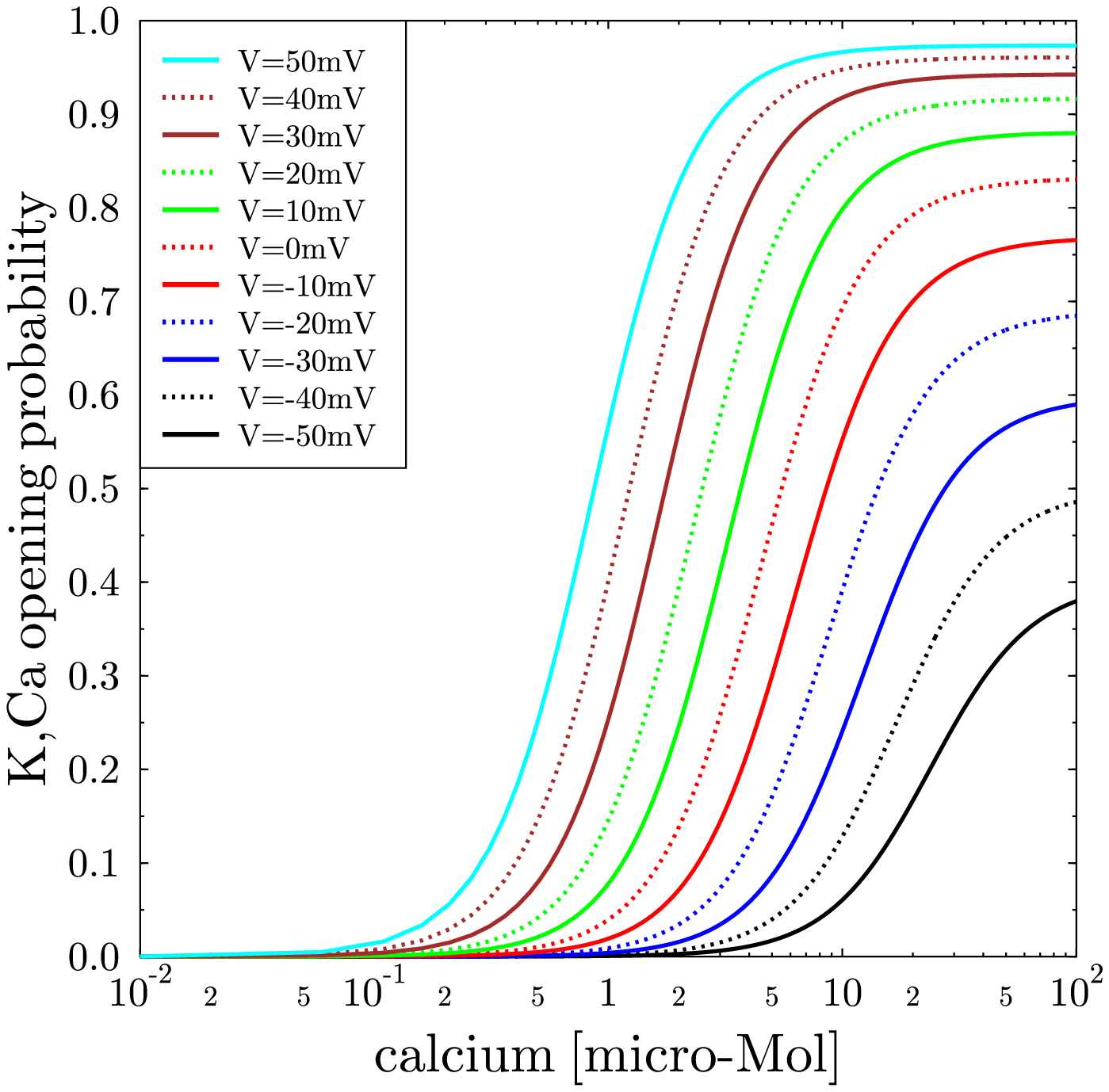}
\includegraphics[height=6.8cm]{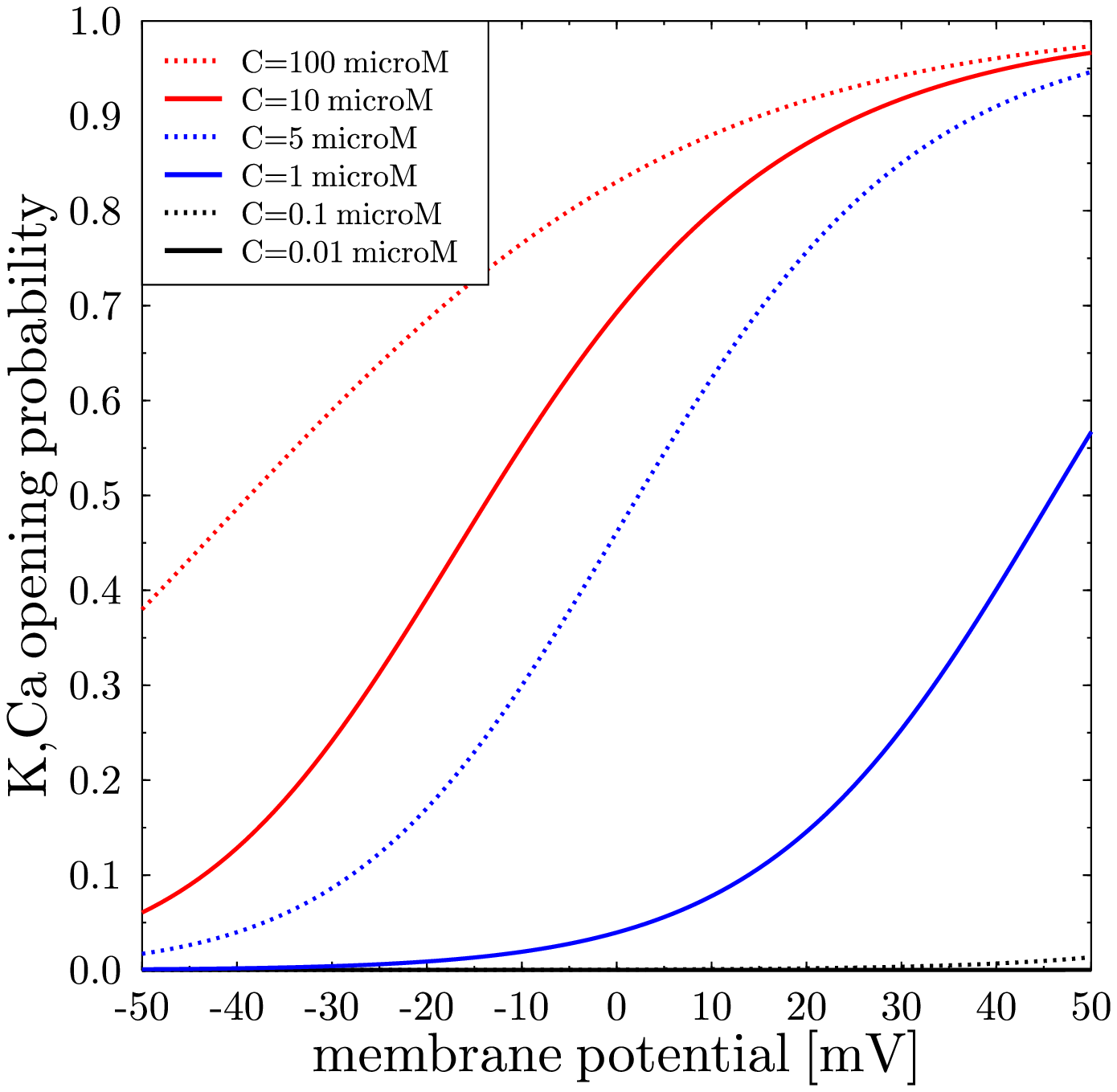}
\caption[]{\sf
Opening probability of K,Ca using a product
of a sigmoidal function (for $V$) and a Hill-function
(for calcium). The calcium concentration of half activation
is assumed to depend on the membrane potential according
to \gl{dynamic_half_C}. The results are shown varying
calcium (left panel) or the membrane potential (right panel).
There is good agreement with the data in Fig.~8 \cite{barrett82}. 
Parameters are 
$a=45mV$, $b=30mV$, $n_{\rm K,Ca}=2$, 
$V_{\rm K,Ca}=-40mV$, and $\kappa_{\rm K,Ca}=25mV$.}
\label{K_Ca_var_half}
\end{center}
\end{figure}
A physiological interpretation of \gl{dynamic_half_C} is
not obvious. Basically this equation says that less calcium
is needed to reach the same level of activation when the cell
is more depolarised. If the dependence of the opening
probability on voltage and calcium level are supposed
to correspond to two different molecular gating-mechanisms,
then this equation says that either depolarisation acts
on both mechanisms or the mechanism influenced by depolarisation
acts back on the calcium-dependent mechanism. A concise
interpretation is, however, lacking, and has to be
clarified in further experiments.

{\bf Inactivation:}
Inactivation was only rarely observed. In rat muscle cells
the channels remained open for several hundred milliseconds
\cite{barrett82}. Thus, inactivation of these channels
is neglected in the model.


\subsubsection{Voltage-gated calcium channels Ca,L/T}

Voltage-gated calcium channels induce the inflow of 
calcium, following the electrochemical gradient of calcium,
after an initial depolarisation of the cell.
There are different subtypes of these channels which
differ in their responsiveness. The dominant types in
\bcs~resemble L- and T-type channels \cite{dunne01}
which can be classified as HVA- and LVA-channels,
respectively.
The equations
for these channels are in analogy to the voltage-gated
sodium channels \gl{Na,V}
\be{Ca,L}
I_{\rm Ca,L}  
&=& 
h_{\rm Ca,L} 
\left(1-H(C,C_{\rm Ca,L},n_{\rm Ca,L})\right)
g_{\rm Ca,L} \overline{g_{\rm Ca,L}}
\left(V-\overline{V_{\rm Ca}}\right)
\quad,
\ee
with
\be{g_Ca,L}
\frac{dg_{\rm Ca,L}}{dt}
&=&
\frac{\sigma_{\rm act}(V,V_{\rm Ca,L},\kappa_{\rm Ca,L})-g_{\rm Ca,L}}{\tau_{\rm Ca,L}}
\nonumber\\
\frac{dh_{\rm Ca,L}}{dt}
&=&
\frac{\sigma_{\rm inact}(V,W_{\rm Ca,L},\lambda_{\rm Ca,L})-h_{\rm Ca,L}}{\theta_{\rm Ca,L}}
\quad.
\ee
For T-type channels all L's in the subscript have to be
replaced by T. The term $1-H(\ldots)$ appears only for L-type channels
and has to be replaced by $1$ for T-type.
While HVA-channels were included in different ways in
other models, LVA-channels were neglected in all other
models.
In the following a set of experiments providing 
opening characteristics of L- and T-type channels are
summarised. The simulations will rely on the single
protein characteristics as provided in \cite{magee95}.

{\bf Single T-type (LVA) channel properties:}
Open T-type channels show a single protein conductivity
of $\overline{g_{\rm Ca,T}}\approx 10pS$ \cite{magee95}
(measured in dendrites of pyramidal neurons in CA1 hippocampus).
They open already for membrane potentials near resting
potential in the range of $-70$ to $-30mV$ with a half opening at
$-55mV$ \cite{dunne01}. The threshold
for opening was also measured in adrenal medulla endothelial
cells (in whole-cell patch-mode)
to be $-50mV$ with $V_{\rm Ca,T}=-30mV$ \cite{vinet99}
(this experiment uses calcium and not barium).
The threshold value of $-50mV$ was also found in \cite{magee95}
with $V_{\rm Ca,T}=-32mV$.
This consistently infers $\kappa_{\rm Ca,T}\approx 10mV$
in order to find low activity at the activation threshold.
Indeed, $\kappa_{\rm Ca,T}=7.0mV$ was found in \cite{magee95}.
For inactivation $W_{\rm Ca,T}=-67mV$
and $\lambda_{\rm Ca,T}=6.5mV$ are reported in \cite{magee95}.
Activation happens on a time scale of 
$\tau_{\rm Ca,T}=10ms$ while inactivation 
is rapid \cite{dunne01} but 
is slower with $\theta_{\rm Ca,T}=18ms$ \cite{vinet99}.

{\bf Single L-type (HVA) channel activation:}
Open L-type channels show a single protein conductivity
of $\overline{g_{\rm Ca,L}}\approx 27pS$ \cite{magee95},
measured in dendrites of pyramidal neurons in CA1 hippocampus. 
In \bcs~a single-channel conductivity of $22pS$ was
found \cite{smith93}. 
Activation is found between -40 and 0mV
(\cite{dunne01}, 
\cite{vinet99} in a whole-endothelial cell measurement, 
and \cite{plant88} in whole-cell measurements of mouse \bcs). 
The activation threshold is higher $-20mV$ in 
CA1-pyramidal cells \cite{magee95}.
The half maximal activation potential
is $V_{\rm Ca,L}=-18mV$ \cite{vinet99} with a width of
$\kappa_{\rm Ca,L}\approx 10mV$.
In CA1-pyramidal neurons measured with barium
$V_{\rm Ca,L}=+9mV$ is found with 
$\kappa_{\rm Ca,L}=6mV$ \cite{magee95}
(however, activation depends on
the barium concentration in the patch).
Channel characteristics have also been measured in
\bcs~with $V_{\rm Ca,L}=-3.8mV$
and $\kappa_{\rm Ca,L}=8.4mV$ 
\cite{smith93}. 
The variability of whole-cell measurements of the currents 
is presumably related to different protein densities in different cell
types rather that to different single-channel characteristics --
see \cite{vinet99} and references in the discussion of 
L-type channels. Another ambiguity comes from the replacement
of calcium by barium for technical reasons, which induces
differences in the measured conductivity.

{\bf No L-type inactivation observed:}
Voltage-dependent inactivation is not observed for L-type 
(\cite{dunne01}, \cite{plant88} in \bcs, 
\cite{vinet99} in endothelial cells, and
\cite{villarroya99} in bovine chromaffin cells (whole-cell measurement),
\cite{magee95} for CA1-pyramidal neurons).
Thus, $\tau_{\rm Ca,L}> 1s$ is used. Note that
the half inactivation potential and the steepness
are irrelevant under these circumstances. Alternatively
one could simply set $h_{\rm Ca,L}=1$.

However, there is an experiment showing inactivation
of L-type channels in cardiac cells upon increased 
calcium levels \cite{hoefer97}. 
Inactivation was characterised by a
Hill-function with Hill-coefficient $n_{\rm Ca,L}=1$
(thus suggesting inactivation to be mediated by the
binding of calcium ions to a single regulating site
per Ca,L-protein) and half maximal inactivation
at $C_{\rm Ca,L}=4\mu M$. This behaviour is 
included in this model for the first time
by the additional
factor $1-H(C,C_{\rm Ca,L},n_{\rm Ca,L})$ in \gl{Ca,L}.

{\bf Sodium conductivity:}
In the same paper \cite{smith93}
it is also shown that the L-type calcium channels exhibit
large conductivity of $44pS$ for sodium, which is not
considered in the simulations.

\subsection{The leakage currents}

The leakage currents in \gln{ion-dynamics}{voltage} are calculated
in the steady state approximation. 
\be{leakage_ss}
- J_{\rm Na}
&=&
\rho_{\rm Na,V} I_{\rm Na,V}^{\rm ss}
+ 2 \rho_{\rm Na,K} I_{\rm Na,K}^{\rm ss} \alpha_{\rm Na,K}
+ \rho_{\rm NCX} I_{\rm NCX}^{\rm ss} \alpha_{\rm NCX}
\nonumber\\
- J_{\rm K}
&=&
\rho_{\rm K,ATP} I_{\rm K,ATP}^{\rm ss}
+ \rho_{\rm K,V} I_{\rm K,V}^{\rm ss}
+ \rho_{\rm K,Ca} I_{\rm K,Ca}^{\rm ss}
-2 \rho_{\rm Na,K} I_{\rm Na,K}^{\rm ss}
\nonumber\\
- J_{\rm Ca}
&=&
\rho_{\rm Ca,L} I_{\rm Ca,L}^{\rm ss}
+ \rho_{\rm Ca,T} I_{\rm Ca,T}^{\rm ss}
- z_{\rm Ca} \rho_{\rm NCX} I_{\rm NCX}^{\rm ss}
+ \rho_{\rm PMCA} I_{\rm PMCA}^{\rm ss}
\quad,
\ee
where the superscript ss denotes the steady state quantities.
The calculated leakage currents guarantee that the cell exhibits
a stable steady state and that the cell returns to its resting state
after stimulation.
In most alternative approaches the leakage currents were not considered
(e.g.~\cite{sherman88,chay97,bertram04}).

The steady state currents are calculated from the steady
state approximation of all kinetic equations for the open
probabilities. These are evaluated on the basis of a set
of properties in the resting state 
$\{V_0,\gamma_0,N_0,K_0,C_0\}$.
\be{currents_ss}
I_{\rm Na,K}^{\rm ss}
&=& 
\overline{I_{\rm Na,K}}
\left(1 - H(K_0,K_{\rm Na,K},n_{\rm Na,K})\right)
\nonumber\\
I_{\rm Na,V}^{\rm ss}
&=& 
\sigma_{\rm act}(V_0,V_{\rm Na,V},\kappa_{\rm Na,V}) \overline{g_{\rm Na,V}}
\left(V_0-\overline{V_{\rm Na}}\right)
\nonumber\\
I_{\rm NCX}^{\rm ss}
&=& 
\overline{I_{\rm NCX}} H(C_0,C_{\rm NCX},n_{\rm NCX})
\nonumber\\
I_{\rm PMCA}^{\rm ss}
&=& 
\overline{I_{\rm PMCA}} H(C_0,C_{\rm PMCA},n_{\rm PMCA})
\nonumber\\
I_{\rm K,ATP}^{\rm ss}
&=& 
\left(1-\sigma_{\rm act}(\gamma_0,\gamma_{\rm K,ATP},\kappa_{\rm K,ATP})\right) 
\overline{g_{\rm K,ATP}}
\left(V_0-\overline{V_{\rm K}}\right)
\nonumber\\
I_{\rm K,V}^{\rm ss}
&=& 
\sigma_{\rm act}(V_0,V_{\rm K,V},\kappa_{\rm K,V}) \overline{g_{\rm K,V}}
\left(V_0-\overline{V_{\rm K}}\right)
\nonumber\\
I_{\rm K,Ca}^{\rm ss}
&=&
\overline{g}_{\rm K,Ca}
\left(V_0-\overline{V_{\rm K}}\right)
H(C_0,C_{\rm K,Ca}(V_0),n_{\rm K,Ca})
\sigma_{\rm act}(V_0,V_{\rm K,Ca},\kappa_{\rm K,Ca})
\nonumber\\
I_{\rm Ca,L}^{\rm ss}
&=& 
\sigma_{\rm act}(V_0,V_{\rm Ca,L},\kappa_{\rm Ca,L}) \overline{g_{\rm Ca,L}}
\left(V_0-\overline{V_{\rm Ca}}\right)
\nonumber\\
I_{\rm Ca,T}^{\rm ss}
&=& 
\sigma_{\rm act}(V_0,V_{\rm Ca,T},\kappa_{\rm Ca,T}) \overline{g_{\rm Ca,T}}
\left(V_0-\overline{V_{\rm Ca}}\right)
\quad,
\ee
where $C_{\rm K,Ca}(V_0)$ is calculated with \gl{dynamic_half_C}.
The intracellular resting concentrations of ions are
used as given in \ta{pars}:
$K_0=95mM$ \cite{atwater78}, 
$N_0=20mM$ \cite{atwater78}, 
$C_0=0.1\mu M$.

%
%

\subsection{Parameter lists}

\begin{table}[ht!]
\scriptsize
\begin{center}
\begin{tabular}{|l||c|c|l|l|} \hline
Variable  & adapted from \cite{sherman88} & full model \ab{bc0221_10}& unit  & reference\\
\hline
use leakage currents& no         & yes       & boolean    & $-$\\
use inactivation    & yes        & yes       & boolean    & $-$\\
PMCA feedback to $V$& no         & yes       & boolean    & $-$\\
\hline
cell radius $R_{\rm cell}$&$6.1$ & $6.1$     & $\mu m$    & \cite{solimena06,straub04}\\
$C_{\rm m}$         & $10$       & $10$      &$fF/\mu m^2$& \cite{gentet00,goepel04}\\
$T$                 & $310$      & $310$     & $K$        & $-$\\
\hline
$\gamma_0$          & $1$        & $1$       & $mM$       & \cite{rorsman03}\\
$\gamma_{\rm stimualtion}$& none & $10$      & $mM$       & \cite{goepel04,beauvois06}\\
\hline
$V_0$               & $-70$      & $-70$     & $mV$       & known\\
$K_0$               & $95$       & $95$      & $mM$       & \cite{atwater78}\\
$K_{\rm ext}$       & $5.7$      & $5.7$     & $mM$       & \gl{nernst}:
                                                           $\overline{V_{\rm K}}=-75mV$\\
$N_0$               & $20$       & $20$      & $mM$       & \cite{atwater78}: $20$,
                                                           \cite{smith88}: $36$\\
$N_{\rm ext}$       & $400$      & $400$     & $mM$       & \gl{nernst}:
                                                           $\overline{V_{\rm Na}}=80mV$\\
$C_0$               & $0.1$      & $0.1$     & $\mu M$    & known\\
$C_{\rm ext}$       & $1.5$      & $1.5$     & $mM$       & \gl{nernst}:
                                                           $\overline{V_{\rm K}}=128mV$\\
use Nernst          & no         & yes       & boolean    & $-$\\
correct $\overline{V_{\rm Ca}}$& no & $78$   & $mV$       & \cite{hille92}:
                                                           $\overline{V_{\rm K}}=50mV$\\
\hline
calcium binding sites $c_0$& $1.0$ & $1.0$   & $mM$       & \cite{sherman88}: $0.3mM$;
                                                         $1.0$ implies $f_{\rm Ca}\approx 0.1\%$\\
$K_c$               & $1$    & $1$   & $\mu M$            & \cite{wasserman81}\\
\hline
\end{tabular}
\end{center}
\caption[]{\sf {\bf General framework of the simulations:}
The parameters as used in the full simulation in \ab{bc0221_10} and
the parameters needed to reproduce
the model \cite{sherman88} are listed. 
Note that the latter differ from the
values used in \cite{sherman88} and are adapted to today's knowledge.
References given without comment support the value used in the full model.
\label{pars}}
\end{table}
%
\begin{table}[ht!]
\scriptsize
\begin{center}
\begin{tabular}{|l||c|c|l|l|} \hline
Variable      & adapted from \cite{sherman88} & full model \ab{bc0221_10} & unit & reference\\
\hline
$\overline{I_{\rm Na,K}}$ & none & $0.00003$ & $pA$       & \cite{maixent93}\\
$K_{\rm Na,K}$            & none & $33.3$    & $mM$       & \cite{chapman83}\\
$n_{\rm Na,K}$            & none & $2$       & \#         & \cite{chapman83}\\
$\tilde{K}_{\rm Na,K}$    & none & $20$      & $mM$       & \cite{chapman83}\\
$\tilde{n}_{\rm Na,K}$    & none & $2$       & \#         & \cite{chapman83}\\
$\alpha_{\rm Na,K}$       & none & $1.5$     & Na/K       & known\\
\hline
$\overline{g_{\rm K,ATP}}$& none & $54$      & $pS$       & \cite{ashcroft84,cook84,dunne01}\\
$\tau_{\rm K,ATP}$        & none & $1$       & $s$        & relatively slow\\
$\gamma_{\rm K,ATP}$      & none & $1.2$ & $mM$ & large current at resting state \cite{kanno02}\\
$\kappa_{\rm K,ATP}$      & none & $6$       & $mM$       & estimated\\
\hline
$\overline{g_{\rm K,V}}$  &$10$  & $10$      & $pS$       & \cite{dunne01}\\
$\tau_{\rm K,V}$          &\gl{tau_K,V}&\gl{tau_K,V}& $s$        & \cite{kelly91}\\
half max $\tau_{\rm K,V}$ & $0.03$ &$0.03$   & $s$    & $8ms<\tau_{\rm K,V}<37ms$ \cite{kelly91}\\
$V_{\rm K,V}$             &$-15$ & $1.0$     & $mV$       & \cite{kelly91}\\
$\kappa_{\rm K,V}$        &$5.6$ & $8.5$     & $mV$       & \cite{kelly91}\\
$\theta_{\rm K,V}$        & none & $400$     & $ms$        & \cite{kelly91} $\le 400ms$\\
$W_{\rm K,V}$             & none & $-25$     & $mV$       & \cite{kelly91}\\
$\lambda_{\rm K,V}$       & none & $7.3$     & $mV$       & \cite{kelly91}\\
\hline
$\overline{g_{\rm K,Ca}}$ &$220$ & $220$     & $pS$       & \cite{dunne01},
                                                           \cite{barrett82}: $f(T)$\\
$C_{\rm K,Ca}$            &$0.001$&\gl{dynamic_half_C}& $mM$ & \cite{barrett82}: $f(V)$\\
$n_{\rm K,Ca}$            &$3$   & $2$       & \#         & \cite{barrett82}\\
$\tau_{\rm K,Ca}$         & none & $100$     & $ms$        & estimated\\
$V_{\rm K,Ca}$            & none & $-40$     & $mV$       & \cite{barrett82}\\
$\kappa_{\rm K,Ca}$       & none & $25$      & $mV$       & \cite{barrett82}\\
\hline
$\overline{g_{\rm Na,V}}$ & none & $14$      & $pS$       & \cite{bezanilla87}\\
$\tau_{\rm Na,V}$         & none &\gl{tau_Na,V}& $s$       & \cite{hille92,goepel99}\\
$V_{\rm Na,V}$            & none & $-35$     & $mV$       & \cite{plant88_PflugArch}\\
$\kappa_{\rm Na,V}$       & none & $8.0$     & $mV$       & \cite{plant88_PflugArch}\\
$\theta_{\rm Na,V}$       & none & $4.6$     & $ms$        & \cite{hille92}\\
$W_{\rm Na,V}$            & none & $-100$    & $mV$       & \cite{plant88_PflugArch,goepel99,dunne01}\\
$\lambda_{\rm Na,V}$      & none & $20$      & $mV$       & \cite{plant88_PflugArch,goepel99,hiriart88}\\
\hline
\end{tabular}
\end{center}
\caption[]{\sf {\bf Properties of single sodium and potassium conducting transmembrane proteins:}
Properties of sodium and potassium conducting membrane proteins.
References given without comment support the value used
in the full model.
A comment of the form $f(X)$ denotes that the reference claim
a dependence on the variable $X$.
\label{pars_NK}}
\end{table}
%
\begin{table}[ht!]
\scriptsize
\begin{center}
\begin{tabular}{|l||c|c|l|l|} \hline
Variable & adapted from \cite{sherman88} & full model \ab{bc0221_10}& unit & reference\\
\hline
$\overline{I_{\rm NCX}}$  & none & $-0.5$    & $fA$       & \cite{juhaszova00}\\
$C_{\rm NCX}$             & none & $1.8$     & $\mu M$    & \cite{blaustein99}\\
$n_{\rm NCX}$             & none & $1$       & \#         & \cite{erler04}\\
$\alpha_{\rm NCX}$        & none & $3$       & Na/Ca      & \cite{blaustein99}\\
\hline
$\overline{I_{\rm PMCA}}$ &$0.01$& $0.01$    & $fA$       & \cite{juhaszova00}\\
$C_{\rm PMCA}$            &$1.8$ & $0.1$     & $\mu M$    & \cite{elwess97}\\
$n_{\rm PMCA}$            &$2$   & $2$       & \#         & \cite{caride01_JBiolChem}\\
$\alpha_{\rm PMCA}$       &$1$   & $1$       & Ca/ATP     & \cite{carafoli91}\\
\hline
$\overline{g_{\rm Ca,L}}$ &$27$  & $27$      & $pS$       & \cite{magee95,smith93}\\
$\tau_{\rm Ca,L}$         &$1$& $6$          & $ms$       & \cite{magee95}: fast\\
$V_{\rm Ca,L}$            &$0.0$ & $0.0$     & $mV$       & \cite{magee95,vinet99,smith93}\\
$\kappa_{\rm Ca,L}$       &$12$  & $12$      & $mV$       & \cite{magee95,dunne01,smith93}\\
$\theta_{\rm Ca,L}$       &$10$  & $10$      & $s$        & \cite{magee95}: no inact.\\
$W_{\rm Ca,L}$            &$100$ & $100$     & $mV$       & \cite{magee95}: no inact.\\
$\lambda_{\rm Ca,L}$      &$10$  & $10$      & $mV$       & \cite{magee95}: no inact.\\
$C_{\rm Ca,L}$            & none & $4$       & $\mu M$    & \cite{hoefer97}\\
$n_{\rm Ca,L}$            & none & $1$       & \#         & \cite{hoefer97}\\
\hline
$\overline{g_{\rm Ca,T}}$ & none & $10$      & $pS$       & \cite{magee95}\\
$\tau_{\rm Ca,T}$         & none & $10$      & $ms$       & \cite{vinet99}\\
$V_{\rm Ca,T}$            & none & $-30$     & $mV$       & \cite{magee95}\\
$\kappa_{\rm Ca,T}$       & none & $7.0$     & $mV$       & \cite{magee95}\\
$\theta_{\rm Ca,T}$       & none & $18$      & $ms$       & \cite{vinet99}\\
$W_{\rm Ca,T}$            & none & $-67$     & $mV$       & \cite{magee95}\\
$\lambda_{\rm Ca,T}$      & none & $6.5$     & $mV$       & \cite{magee95}\\
\hline
\end{tabular}
\end{center}
\caption[]{\sf {\bf Properties of single calcium conducting transmembrane proteins:}
Properties of calcium conducting transmembrane proteins.
References given without comment support the value used in the full model.
\label{pars_Ca}}
\end{table}

\begin{table}[ht!]
\scriptsize
\begin{center}
\begin{tabular}{|l||c|c|l|} \hline
Variable & adapted from \cite{sherman88} & full model \ab{bc0221_10}& unit\\
\hline
$\rho_{\rm K,ATP}$        & none & $0.092$    & $/\mu m^2$ \\
$\rho_{\rm K,V}$          &$0.24$& $8.0$        & $/\mu m^2$ \\
$\rho_{\rm K,Ca}$         &$0.08$& $0.45$     & $/\mu m^2$ \\
$\rho_{\rm Na,K}$         & none & $2000$     & $/\mu m^2$ \\
$\rho_{\rm Na,V}$         & none & $1.15$     & $/\mu m^2$ \\
\hline
$\rho_{\rm NCX}$          & none & $7.5$     & $/\mu m^2$ \\
$\rho_{\rm PMCA}$         &$10000$& $1350$     & $/\mu m^2$ \\
$\rho_{\rm Ca,L}$         &$0.1$ & $0.9$   & $/\mu m^2$ \\
$\rho_{\rm Ca,T}$         & none & $0.1$    & $/\mu m^2$ \\
\hline
\end{tabular}
\end{center}
\caption[]{\sf {\bf Densities of ion-conducting transmembrane proteins:}
The densities of transmembrane proteins as used in the two simulation
set-ups are listed. These parameters were used as fit parameters
for the simulations. Thus, no references are given. Note that the
large PMCA-density is a result of an unrealistic half activation
calcium concentration used in \cite{sherman88} (see also \ta{pars_Ca}).
\label{densities}}
\end{table}

\end{document}